\documentclass[fleqn,usenatbib]{mnras}

\usepackage[T1]{fontenc}
\usepackage{multirow}
\usepackage{booktabs}
\usepackage{pdflscape}

\DeclareRobustCommand{\VAN}[3]{#2}
\let\VANthebibliography\thebibliography
\def\thebibliography{\DeclareRobustCommand{\VAN}[3]{##3}\VANthebibliography}

\usepackage{graphicx}   
\usepackage{amsmath}    
\usepackage{amssymb}    
\usepackage{subfigure}  
\usepackage{comment}
\usepackage{orcidlink}
\usepackage[figuresleft]{rotfloat}
\usepackage{CJKutf8}
\usepackage{newtxtext,newtxmath}

\title{Metallicity Gradient of Barred Galaxies with TYPHOON}

\author[Qian-Hui Chen et al.]{
    Qian-Hui Chen (陈千惠)\orcidlink{0000-0002-4382-1090}$^{1,2,3,4}$,\thanks{E-mail: Qianhui.Chen@anu.edu.au}
	Kathryn Grasha\orcidlink{0000-0002-3247-5321}$^{1,2,9}$\thanks{ARC DECRA Fellow},
	Andrew J. Battisti\orcidlink{0000-0003-4569-2285}$^{1,2}$,
	\newauthor
	Lisa J. Kewley\orcidlink{0000-0001-8152-3943}$^{5,1,2}$,
	Barry F. Madore\orcidlink{0000-0002-1576-1676}$^{6,7}$,
	Mark Seibert\orcidlink{0000-0002-1143-5515}$^{6}$, 
	Jeff A. Rich\orcidlink{0000-0002-5807-5078}$^{6}$ 
	\newauthor
	and
	Rachael L. Beaton\orcidlink{0000-0002-1691-8217}$^{8,6}$\thanks{Hubble Fellow}
	\\
	$^{1}$Research School of Astronomy and Astrophysics, Australian National University, Canberra, ACT 2611, Australia\\
	$^{2}$ARC Centre of Excellence for All Sky Astrophysics in 3 Dimensions (ASTRO 3D), Australia\\
	$^{3}$CAS Key Laboratory for Research in Galaxies and Cosmology, Department of Astronomy, University of Science and Technology of China, Hefei 230026, China\\
	$^{4}$School of Astronomy and Space Sciences, University of Science and Technology of China, Hefei, 230026, China\\
	$^{5}$Institute for Theory and Computation, Harvard-Smithsonian Center for Astrophysics, Cambridge, MA 02138, USA \\
	$^{6}$The Observatories, Carnegie Institution for Science, 813 Santa Barbara Street, Pasadena, CA 91106, USA\\
	$^{7}$Department of Astronomy and Astrophysics, University of Chicago, Chicago, IL 60637, USA\\
	$^{8}$Department of Astrophysical Sciences, 4 Ivy Lane, Princeton University, Princeton, NJ 08544, USA\\
	$^{9}$Visiting Fellow, Harvard-Smithsonian Center for Astrophysics, 60 Garden Street, Cambridge, MA 02138, USA\\
}

\date{Accepted XXX. Received YYY; in original form ZZZ}

\pubyear{2022}

\begin{document}
\begin{CJK}{UTF8}{gbsn}
	\label{firstpage}
	\pagerange{\pageref{firstpage}--\pageref{lastpage}}
	\maketitle
	
	\begin{abstract}
		Bars play an important role in mixing material in the inner regions of galaxies and stimulating radial migration. Previous observations have found evidence for the impact of a bar on metallicity gradients but the effect is still inconclusive. 
		We use the TYPHOON/PrISM survey to investigate the metallicity gradients along and beyond the bar region across the entire star-forming disk of five nearby galaxies. Using emission line diagrams to identify star-forming spaxels, we recover the global metallicity gradients ranging from $-$0.0162 to $-$0.073 dex/kpc with evidence that the galactic bars act as an agent in affecting in-situ star formation as well as the motions of gas and stars. 
		We observe cases with a `shallow-steep' metallicity radial profile, with evidence of the bar flattening the metallicity gradients inside the bar region (NGC~5068 and NGC~1566) and also note instances where the bar appears to drive a steeper metallicity gradient producing `steep-shallow' metallicity profiles (NGC~1365 and NGC~1744).
		For NGC~2835, a `steep-shallow' metallicity gradient break occurs at a distance $\sim$ 4 times the bar radius, which is more likely driven by gas accretion to the outskirt of the galaxy instead of the bar. The variation of metallicity gradients around the bar region traces the fluctuations of star formation rate surface density in NGC~1365, NGC~1566 and NGC~1744.
		A larger sample combined with hydrodynamical simulations is required to further explore the diversity and the relative importance of different ISM mixing mechanisms on the gas-phase metallicity gradients in local galaxies.

	\end{abstract}
	

	\section{Introduction}
	HII regions are clouds of gas in the interstellar medium (ISM) ionized by the ultraviolet radiation of newly formed, massive OB$-$type stars. From the emission line spectrum of HII regions, the chemical composition traces present-day metal abundances in the gas-phase ISM of galaxies. 
	Star formation, stellar radial migration, stellar winds, inflows, outflows, mergers, and accretion processes impact the gas-phase oxygen abundance of the ISM. Although these physical processes interact in a complex manner, chemical evolution models are now able to link the distribution of metals in the ISM with key physical processes that drive galaxy evolution \citep{Kobayashi_2007, Dave_2011, Kobayashi_Nakasato_2011, Torrey_2013, Taylor_2020, Li_2021}. 
	The dependence of the gas-phase metallicity, denoted $\mathrm{12+log(O/H)}$, with galactocentric radius, is typically parameterized by a characteristic slope (the so-called metallicity gradient) and has been the focus of numerous studies over the past few decades  \citep[e.g., ][]{Searle_1971, Lequeux_1979, Vila-Costas_1992, Martin_1994, Tremonti_2004, Mannucci_2010, Genzel_2011, Sanchez-Blazquez_2011,Sanchez_2014, Ho_2015, Kaplan_2016, Sanchez_2020}. 
	The complex interplay between the processes that enhance the production of metals and those that decrease or dilute metals within galaxies places important constraints on galaxy formation \citep{SAlmeida_2014}. 
	Thus, metals act as tracers of the formation and assembly history of galaxies. In this manner, careful determinations of the radial variation of metals across galaxies provide a critical window into the current evolutionary state of galaxies.
	
	Simulations \citep{Matteucci_1989, Boissier_1999, Pilkongton_2012, Gilbson_2013, Ma_2017, Sillero_2017, Tissera_2019, Hemler_2021} show that a negative gas-phase metallicity gradient (i.e. lower metallicity with larger radial distance to the center of galaxy) provides strong evidence for classical ``inside–out'' formation. The inside-out model implies that star formation begins earlier and lasts longer in the metal-rich center than in the metal-poor outskirt of a galaxy. Observations in the local universe confirm that local spiral galaxies typically have negative metallicity gradients \citep[e.g.,][]{Zarisky_1994, Moustakas_2010, Rupke_2010b, refId0, Ho_2017, Poetrodjojo_2019, Grasha_2022}. Environmental factors also impact the chemical history and distribution of metals in galaxies, where the metallicity gradients can be disrupted by large-scale inflows as a result of galaxy interactions \citep{Kewley_2010, Lopez_2015}.
	
	The impact of galactic barred structures on the metal distribution in the ISM is not fully understood, primarily due to limited observations with a sufficient spatial resolution to study HII region scales across the entire disk of galaxies. 
	Stellar bars drive gas inflows  \citep{Jogee_2005}, enhancing the movement of metal-poor gas from the outer disc toward the centre of galaxies. Spiral galaxies are typically defined into three classes, normal spiral (SA), weak bars (SAB) and strong bars (SB) \citep{Vaucouleurs_1991}. 
	Strong bars are typically long and visually distinct, dominating the light distribution \citep{Nair_2010}, whereas weak bars are small and faint \citep{Vaucouleurs_1963}. Bars (SB and SAB) are commonly found in $\sim$60\% of spiral galaxies in the local universe \citep{Knapen_2000, Eskridge_2000, Whyte_2002, Elmegreen_2004, Menendez_Delmestre_2007, Conselice_2014, Willett_2015, Galloway_2017, Erwin_2018, Kelvin_2018, Reddish_2021}, while strong bars (SB) are found in only 30\% of spiral galaxies in the local universe. 
	Bars are known to play an important role in galaxy evolution and angular momentum redistribution as long-lived phenomena \citep{Villa_Vargas_2010, Gadotti_2015, Vera_2016, Geron_2021}. The gravitational torque of bars helps the gas lose angular momentum and therefore increases the galactic inflows towards the central region \citep{Simkin_1980, Weinberg_1985}. Angular momentum redistribution is particularly efficient at resonances. 
	A bar, in conjunction with spiral arms, can cause a rapid migration of stars through the disk \citep{Matteo_2013}. During the formation and the phase of strong bar activity (t $\textless$ 2 Gyr), stars in the disk are significantly redistributed, with the highest probability of migration at the bar resonances \citep[][]{Minchev_2010}.
	\citet{Fraser_McKelvie_2019} presents a radial mixing mechanism that would manifest itself observationally as weaker stellar age and metallicity gradients within the bar region.

	Detailed analysis of the formation of bars and the impact of bars on spirals, which is imprinted on the distribution of the gas-phase metallicity, may facilitate our understanding of quenching in star-forming galaxies.
	Previous observations with long-slit and integral-field spectroscopy (IFS) analyse the effect of radial gas flows induced by the bars and the subsequent impacts on metallicity gradients \citep{Roberts_1979, Friedli_1994}. 
	Mixing processes such as outflows/inflows are expected to affect the global metallicity gradients, causing a decrease in the gas-phase metallicity gradients with time \citep{Friedli_Benz_1995}.
	The results above suggest a relation between the slopes of the global metallicity abundance gradient and the strength of bars \citep{Martin_1994}. 
	In one scenario, the bar dilutes rich gas within the central region of the galaxy and suppresses star formation \citep[][]{Wang_2012, Wang_2020}, resulting in a radial metallicity gradient that is weaker and flatter when compared with the galactic disc. However, an alternate solution could be that a young bar may energize the star formation process. Such a scenario would be observable with a steep radial metallicity gradient \citep{Alonso-Herrero_2001,Ellison_2011,Lin_2020}. Due to the complex manner in which a bar may impact the mixing of the ISM and quenching processes of star formation, prior observations often present conflicting evidence regarding whether or not barred spiral galaxies systematically exhibit different metallicity gradients than non-barred galaxies \citep{Vila-Costas_1992,Martin_1994,Kaplan_2016}. 
	
	One reason for conflicting evidence from past studies is that there have been relatively few studies that have reported the effect of bars on the measured radial gradient of oxygen abundance or the broken metallicity radial profiles. This is primarily due to the lack of observational coverage of the entire star-forming disks at resolutions necessary to resolve down to HII region scales in order to constrain the resolved emission maps of a galaxy \citep{Roy_1997, Sanchez_2011, Seidel_2016, Wei+2020}. 
	Recently, \citet{Sanchez_2018} characterised the oxygen abundance radial profiles of 102 spiral galaxies with Multi Unit Spectroscopic Explorer \citep[MUSE;][]{Bacon_2014}, finding both steep-shallow and shallow-steep metallicity profiles. They did not address the reason for the location of the broken metallicity profiles. Our work will explore the possible impacts of bars on broken metallicity profiles.
	
	In this paper, we investigate the effect of bars on the metallicity gradients in five local barred galaxies. We use the wide-field spectrograph survey TYPHOON/PrISM \footnote{\href{https://typhoon.datacentral.org.au/}{https://typhoon.datacentral.org.au/}}, which aims to produce highly spatially resolved spectrophotometric data of nearby galaxies. The large field of view of TYPHOON (18-arcmin-long slit with 1.65'' width) allows us to observe the entire disk of nearby star-forming galaxies in an IFU-like manner, at an unprecedented spatial resolution of $\sim$ 4$-$5 pc in the closest galaxies \citep{Sturch_madore_2011} with an average spatial resolution of $\sim$ 50-100 pc. 
	Differences in the measured metallicity gradients between the central barred region and the star-forming disc could be evidence that bars play an important role in radial migration and material inflows \citep{Garcia_2022}. In this work, we will compare the break positions of the barred galaxies with 0.5 $R_e$ and 1.5 $R_e$ to aid in the investigation of the physical driver for the location of the break radii observed in the metallicity profiles.
	
	This paper is organized as follows. Section \ref{sec:obs_reduc} describes our observation and data reduction. In Section \ref{sec:data}, we describe how we measure the metallicity gradient. In Section \ref{sec:discussion}, we discuss the various situations that may give rise to the observed variation in the observed metallicity gradients. In Section~\ref{sec:conclu} we summarise the conclusions of our study. Luminosity distances are adopted from \citet{Leroy_2019}, assuming H$_0$ = 70~km~s$^{-1}$Mpc$^{-1}$ and a flat cosmology with $\Omega_m$ = 0.27.

	\section{Observations and Data Reduction}\label{sec:obs_reduc}
	\subsection{Observations}\label{sec:obs}
	TYPHOON is a survey using the Progressive Integral Step Method (PrISM) also known as “step-and-stare” or “stepped-slit” technique to observe 44 of the closest and largest galaxies in the southern hemisphere. Data were obtained using the 2.5m du Pont telescope at the Las Campanas Observatory in Chile. 
	The galaxies were observed using a long-slit ($18'$ $\times$ $1.65''$) aperture progressively moved by precisely one slit width orthogonal to the long axis, step by step, in order to build up a dispersed-image data cube.  We only observe when the seeing conditions are better than 1.65$''$ to avoid slit loss.
	The imaging spectrograph of TYPHOON, the Wide Field re-imaging CCD (WFCCD), is configured to have a resolving power of approximately R $\approx$ 850 at 7000~\AA\ and R $\approx$ 960 at 5577~\AA. 
	A 3D output datacube is produced after the observations, recording spatial and spectral information. More detailed information about the TYPHOON/PrISM survey is in \citet{Ho_2017}, \citet{GrashaIAU_2022}, and Seibert et al. (In prep.). 
	
	Prior studies utilising the TYPHOON data have investigated the metallicity gradient of HII regions within NGC~1365 \citep{Ho_2017} and NGC~2997 \citep{Ho_2018}, and the contribution of diffuse ionised gas with spatial resolution to emission line measurements \citep{Poetrodjojo_2019}.
	A recent work \citep{Grasha_2022} presents the localized ISM study including metallicity in HII regions with six TYPHOON spiral galaxies, finding a tight relation between local physical conditions and their localized enrichment of the ISM.
	
	The TYPHOON data are reduced using a standard long-slit data reduction procedure (Seibert et al. In prep.). Our spectra cover the wavelength range of 3650$\sim$8150~\AA\ with a flux calibration accuracy of 2\% \citep{Ho_2017}. The reduced 2-D spectra are later tiled together to form 3-D data cubes with spectral and spatial samplings of $1.5$\AA\ and $1.65''$, respectively. In summary, the pixel size is 1.65"/pixel. For reference, we show sample spectra from different positions of NGC~1365 (red circles in Fig~\ref{fig:whitelight}) after reduction is shown as black lines in Fig~\ref{fig:spaxel}.
	
	Our sample includes five nearby and face-on barred galaxies with reduced data cubes at the time of publication: NGC 1365, NGC 1566, NGC 1744, NGC 2835 and NGC 5068 (Table \ref{tab:info}). 
	We choose face-on galaxies with low to moderate inclination (i $\textless$ 75\textdegree) to limit the effects of extinction and line-of-sight confusion \citep[e.g., ][]{Emsellem_2022}.
	For these galaxies, the slit was placed along the north-south direction and moved from east to west after 600 seconds of integration. The total integration time for each galaxy is: 133,800s for NGC~1365, 118,200s for NGC~5068, 94,800s for NGC~2835, and 75,000s for NGC~1744. The 3-D output datacubes have spatial resolutions on the individual target galaxies ranging from 41 to 145 parsec/pixel (Table \ref{tab:info}). The high spatial resolution helps ensure that the measured metallicity gradient results are not contaminated by diffuse ionised gas \citep[DIG; ][]{Poetrodjojo_2019}. 
	The DIG is hotter and has a lower density in comparison to HII regions \citep{Collins_2001}. 
	This increases the observed metal emission line fluxes from DIG and enhances the metal lines relative to the Balmer hydrogen emission (e.g., [SII]/ H$\alpha$; \citeauthor{blanc_2009} \citeyear{blanc_2009}). Additionally, DIG emission is shifted towards the LINER and AGN regions of the BPT diagram \citep{Baldwin_1981}.
	The contamination of DIG, which serves to effectively flatten the measured metallicity profiles \citep{Poetrodjojo_2019}, makes the broken metallicity gradient analysis unreliable in surveys with low angular resolution. We thus advise caution when comparing our gradients and breakpoints with surveys having spatial resolution larger than the scale of HII regions.

	\begin{table*}
		\footnotesize
		\newcommand{\tabincell}[2]{\begin{tabular}{@{}#1@{}}#2\end{tabular}}
		\centering
		\scriptsize
		\begin{tabular}{ccccccccccc}
			\hline
			Galaxy & Morphology & R.A & Dec & \tabincell{c}{Inclination\\ (degrees)} & \tabincell{c}{P.A.\\ (degrees)} & \tabincell{c}{Distance\\ (Mpc)} & \tabincell{c}{logM$_{*}$ \\ (M$_{\odot}$)} & \tabincell{c}{$R_{25}$\\ (arcmin)} & \tabincell{c}{Number of\\ Spaxels}& \tabincell{c}{Spatial resolution\\ (parsec/pixel)}\\
			\hline
			NGC 1365 & SB(s)b & 03h33m36.371s & -36d08m25.45s & 35.7 & 49.5 & 18.1 & 10.75 & 5.61 & 2512 & 144.79\\
			NGC 1566 & SAB(s)bc & 04h20m00.42s & -54d56m16.1s & 49.1 & 44.2 & 17.9 & 10.67 & 4.16 & 2638 & 143.19\\
			NGC 1744 & SB(s)d & 04h59m57.80s & -26d01m20.0s & 69.9 & -88 & 9.9 & 9.23 & 4.06 & 669 & 79.19\\
			NGC 2835 & SB(rs)c & 09h17m52.91s & -22d21m16.8s & 56.2 & -91.3 & 10.36 & 9.67 & 3.30 & 2436 & 82.87\\
			NGC 5068 & SAB(rs)cd & 13h18m54.81s & -21d02m20.8s & 27.3 & 137 & 5.16 & 9.36 & 3.62 & 3979 & 41.28\\
			\hline
		\end{tabular}
		\caption{Properties of the target galaxies. Columns list the galaxy name, RC3 morphology, right ascension, declination, inclination angle, 
		position angle, distance, log(M$_{*}$) and $R_{25}$ which is defined as the 25 mag arcsec$^2$ B-band isophote. 
		We reject regions with signal-to-noise ratios (S/Ns) $\leq$ 3 in any of the diagnostic lines (H$\beta$ (4861\AA), [OIII]$\lambda$5007, [NII]$\lambda\lambda$6548,6583, H$\alpha$ (6563\AA), and [SII]$\lambda\lambda$6716,6731, see Sec~\ref{sec:signal_to_noise}), and also those with line ratios inconsistent with photoionization based on their [NII]/H$\alpha$ and [OIII]/H$\beta$ line ratios \citep[e.g., excited by shocks or AGNs;][]{Ka03}, see Sec~\ref{sec:bpt}). The last column shows the spatial resolution in the unit of parsec per pixel. We adopt the stellar mass described in \citet{Leroy_2019} and the rest of the physical parameters are from the NASA extragalactic database (NED) which are used to deproject the galaxies (Section~\ref{sec:deprojection}). }
		\label{tab:info}
	\end{table*}
	
	\subsection{Deprojection of the Galaxy Disks}\label{sec:deprojection}
	To ensure we are free from the effects of inclination within each galaxy, we deproject every pixel before we calculate its distance to its galaxy centre. Here we follow the method in \citet{Grasha_2017} which can be briefly summarized as the following two steps:
	
	\begin{enumerate}
		\item[(1)] Rotate the whole galaxy according to the position angle $\theta$ until the major-axis is horizontal.
		\begin{gather}
			x' = x \cos\theta + y \sin\theta\\
			y' = -x \sin\theta +y \cos\theta
		\end{gather}
		where $x'$, $y'$ are the rotated coordinates, $x$, $y$ are the initial coordinates and $\theta$ is the position angle (Table~\ref{tab:info}).
		
		\item[(2)] Deproject for the line-of-sight inclination as:
		\begin{gather}
			x'' = x'  \\
			y'' = \frac{y'}{cos i}
		\end{gather}
		where $x'$, $y'$ are the coordinates after rotation, $x''$, $y''$ are the deprojected coordinates and $i$ refers to inclination angle from NED dataset\footnote{\href{http://ned.ipac.caltech.edu/}{http://ned.ipac.caltech.edu/}}.
		
	\end{enumerate}
	
	All distances reported in this paper are de-projected galactocentric distances calculated with respect to the deprojected coordinates positions if $x''$ and $y''$.

	\subsection{Data Reduction}\label{sec:data_reduction}
	
	\subsubsection{Spectral Fitting with LZIFU}\label{sec:lzifu}
	We adopt {\sc LZIFU} \citep{Ho_2016, Ho+2016} as the tool to measure the emission line fluxes in the TYPHOON data cubes. {\sc LZIFU} works in two steps, 
	\begin{enumerate}
	\item Modelling and removing the continuum on a spaxel-to-spaxel basis using {\sc PPXF} \citep{Cappellari_Emsellem_2004, Cappelari_2017}. {\sc MIUSCAT} simple stellar population models \citep{Vazdekis_2012} are adopted. 
	\item Appling Levenberg-Marquardt least-square method to fit emission lines as Gaussians. For the emission lines, we model five emission lines simultaneously using single Gaussian component for each emission line --- H$\beta$ (4861\AA), [OIII]$\lambda$5007, [NII]$\lambda\lambda$6548,6583, H$\alpha$ (6563\AA), and [SII]$\lambda\lambda$6716,6731. We tie together the velocity and velocity dispersion of all the lines and fix the flux ratios of [OIII]λλ4959/5007 and [NII]λλ6548/6583 to those given by quantum mechanics.
	\end{enumerate}
	The final outputs of {\sc LZIFU} are emission line flux with corresponding error maps. We perform all analyses at the native resolution of each galaxy (Tab~\ref{tab:info}) for a spaxel-by-spaxel study.

	\subsubsection{Signal to Noise S/N Cut}\label{sec:signal_to_noise}
	For all analysis in this work, we apply a signal-to-noise ratio $\geq$ 3 limit for the following six emission lines: H$\alpha$, [OIII]$\lambda$5007, H$\beta$, [NII]$\lambda$6583 and [SII]$\lambda \lambda$6716,31. 
	The signal is defined as the measured total flux of the best-fit emission line while the noise is the square root of the variance datacube, which considers several sources of error including Poisson noise, sky standard deviation etc, over the wavelength of the corresponding emission line.
	The total number of spaxels with S/N $\geq$ 3 in each galaxy are given in Table \ref{tab:info} and shown in Fig~\ref{fig:ha}.

	\subsubsection{Extinction Correction}\label{sec:extinction}
	We correct all emission line fluxes for extinction using the \citet{Fitzpatrick_1999} dust extinction function for the Milky Way. We follow the method in \citet{Poetrodjojo_2019} as:
	\begin{equation}
		\mathrm{E(B-V) = \frac{\log_{10}\left(\frac{(H\alpha / H\beta)_{obs}}{(H\alpha / H\beta)_{int}}\right)}{0.4(k(H\beta)-k(H\beta))} }
	\end{equation}
	where $\mathrm{(H\alpha / H\beta)_{obs}}$ is the observed Balmer ratio and $\mathrm{(H\alpha / H\beta)_{int}}$ is the intrinsic ratio of 2.86 for case B recombination \citep{Osterbrock_1989}. Assuming a typical R$_v$ value of 3.1 to determine the $k$ values at each wavelength, we obtain maps of E(B--V) which are then applied to calculate the intrinsic line flux \citep{Calzetti_2000} as:
	\begin{equation}
		\mathrm{F_{intrinsic} = F_{observed} \times 10^{0.4\, k_{\lambda}\ \times\ E(B-V)}.}
	\end{equation}
	
\begin{landscape}
	\begin{figure}
		\centering
		
		\subfigure{%
			\includegraphics[width=0.42\textwidth,height=0.46\textwidth]{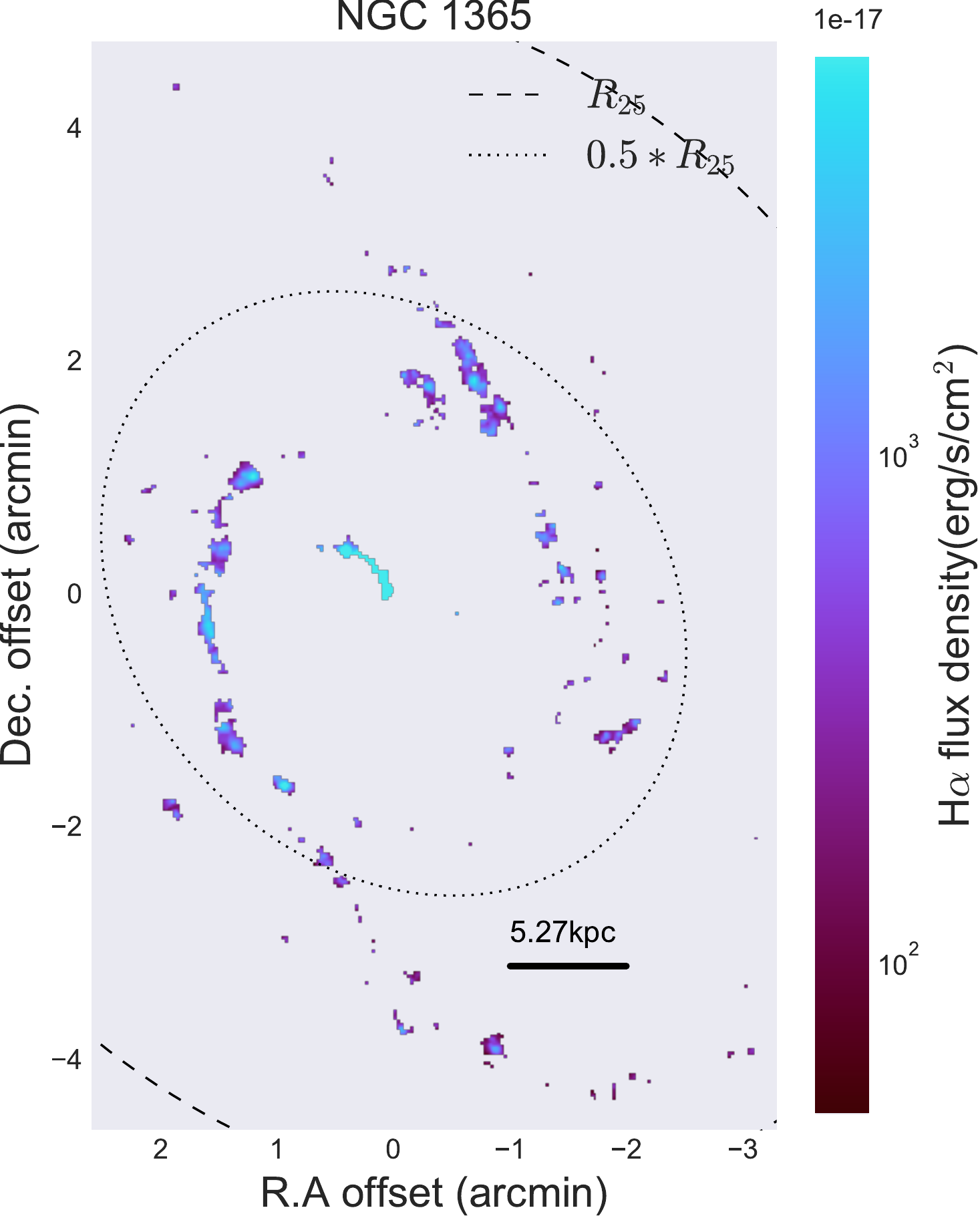}
		}
		\subfigure{%
			\includegraphics[width=0.42\textwidth,height=0.46\textwidth]{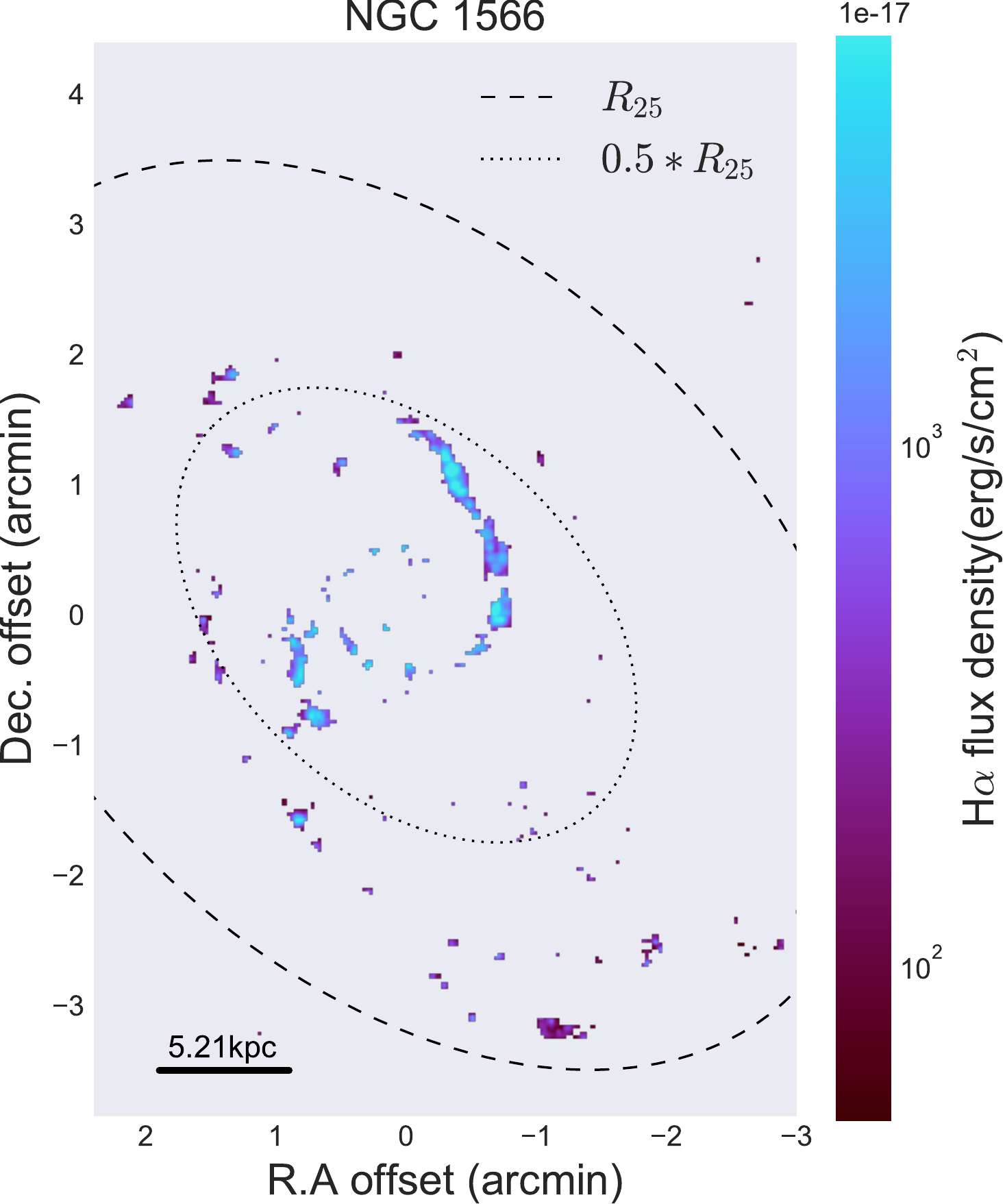}
		}
		\subfigure{%
			\includegraphics[width=0.42\textwidth,height=0.46\textwidth]{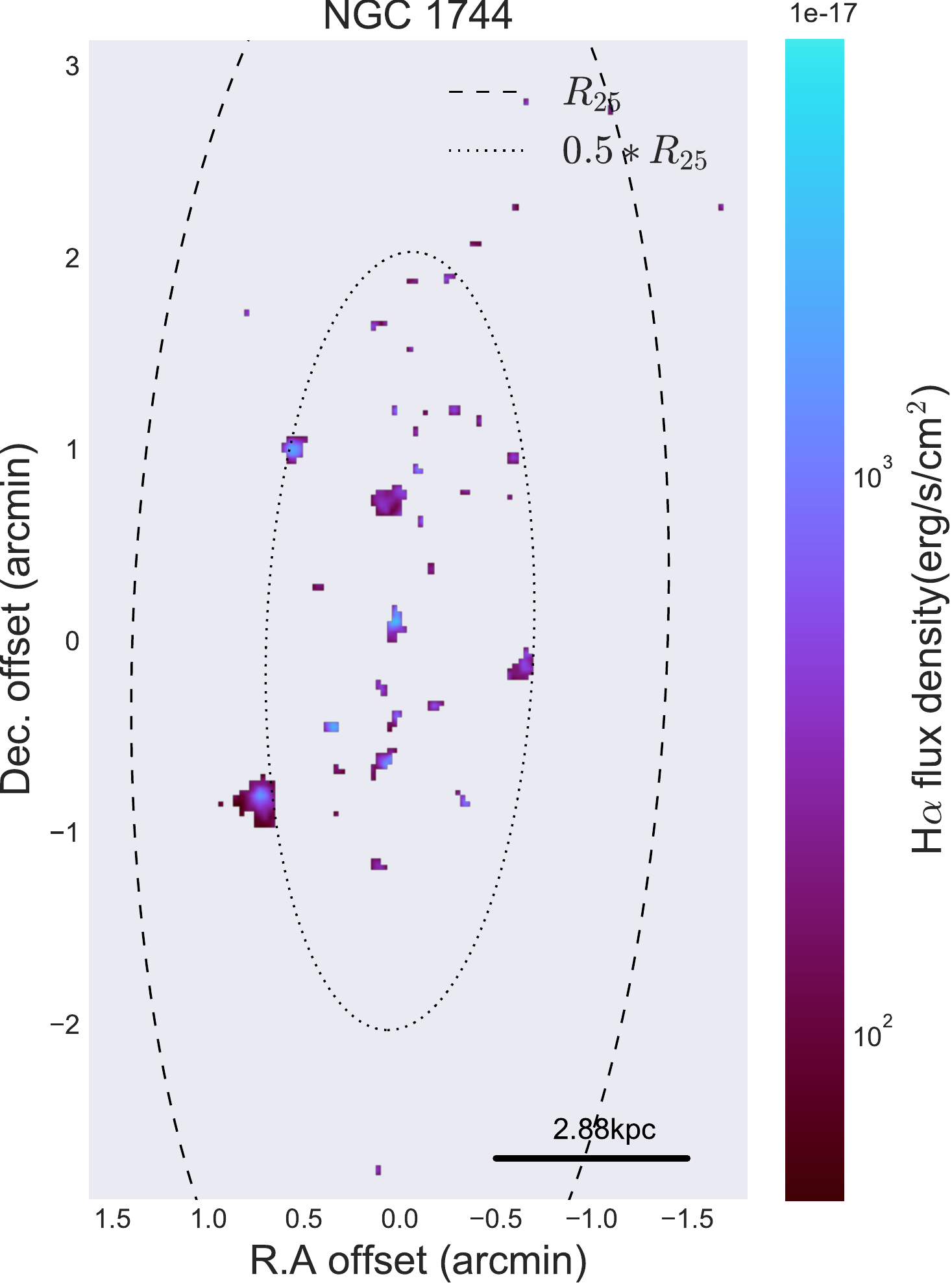}
		}
		\subfigure{%
			\includegraphics[width=0.42\textwidth,height=0.46\textwidth]{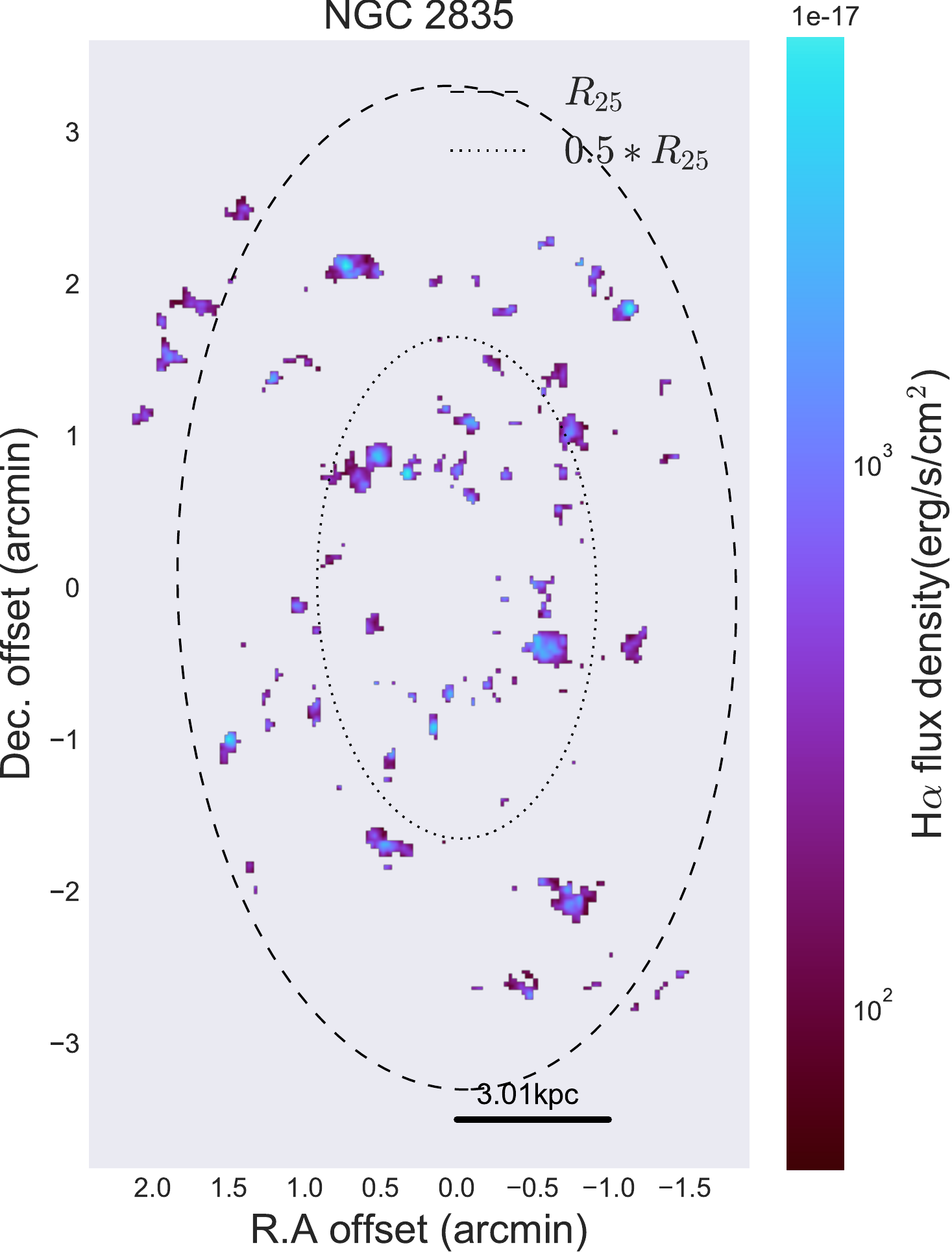}
		}
		\subfigure{%
			\includegraphics[width=0.42\textwidth,height=0.46\textwidth]{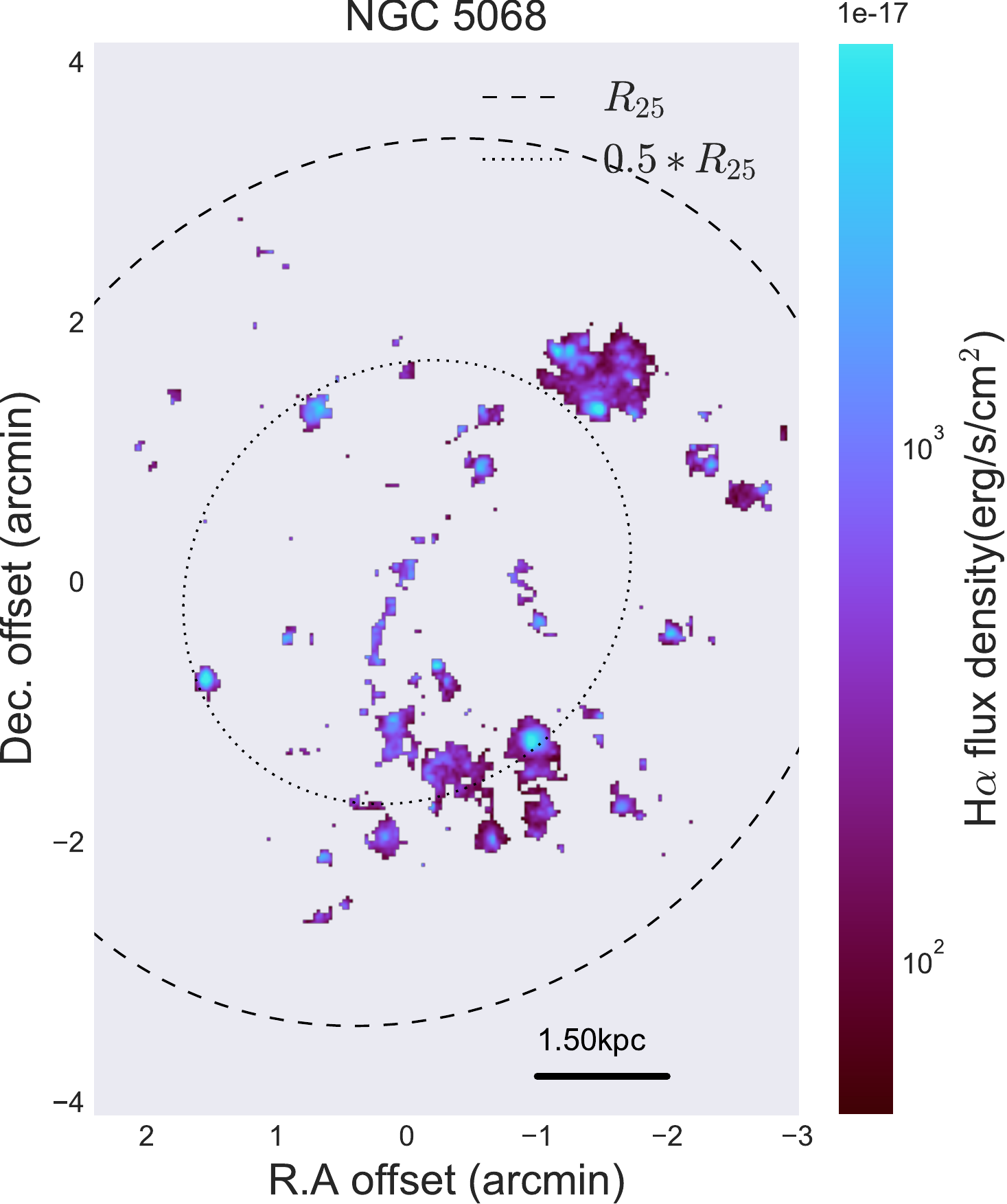}
		}
	\caption{ H$\alpha$ maps after applying the S/N limit $\textgreater$ 3 on all of the diagnostic lines (H$\beta$, [OIII]$\lambda$5007, H$\alpha$, [NII]$\lambda$6583 and [SII]$\lambda \lambda$6716,31}, see Sec~\ref{sec:signal_to_noise}), and extinction corrections (Sec~\ref{sec:extinction}). The black dashed line shows the $R_{25}$ radius ellipse while the dotted line shows the $R_{25}/2$ ellipse. The solid lines in the bottom right are scale bars corresponding to $\mathrm{1~arcmin}$. All of the galaxies show luminous H$\alpha$ regions (light yellow) along the bars and spiral arms.
		\label{fig:ha}
	\end{figure}
	
	\end{landscape}

	Figure~\ref{fig:ha} shows the H$\alpha$ flux maps and R$_{25}$ radius ellipses for all five galaxies in this study. Only spaxels with S/N $\textgreater$ 3 are shown (Section~\ref{sec:signal_to_noise}). All of the galaxies show luminous H$\alpha$ regions (light yellow) along the bars and spiral arms. In this work, we adopt the assumption that R$_{25} \approx$ 3.6 R$_e$ \citep{Williams_2009} for better comparison with \citet{Sanchez_2018} in Section~\ref{sec:prework}.
	
	\section{data analysis}\label{sec:data}
	
	\subsection{BPT Diagrams}\label{sec:bpt}
	In order to separate different types of emission line objects according to their excitation mechanism (i.e., starburst or active galactic nuclei; AGN), we adopt the BPT classification diagram introduced by \citet{Baldwin_1981}. The BPT diagram is a high-excitation [O III]$\lambda$5007/H$\beta$ line ratio versus low-excitation [N II]$\lambda$6584/H$\alpha$ line ratio diagnostic diagram, when combined with the \citet{Ke01} and \citet{Ka03} demarcation lines, allows for the classification of emission line ratio measurements according to the ionizing source. The \citet{Ke01} line identifies the upper bound to pure star formation-driven ionisation sources from theoretical modeling; galaxies lying above this line must be dominated by another ionising source such as an AGN, shocks, or LINERS (low-ionization nuclear emission-line regions). The \citet{Ka03} is an empirical line from data fitting of Sloan Digital Sky Survey; galaxies lying below the \citet{Ka03} are driven by photoionisation sources (i.e., pure star formation).
	The region between these two lines may contain a mixture of ionizing sources of star formation and hard components (e.g. AGN, shocks, or diffuse ionised gas). 
	
\begin{landscape}
	\begin{figure}
		\centering
		
		\subfigure{%
			\includegraphics[width=0.38\textwidth]{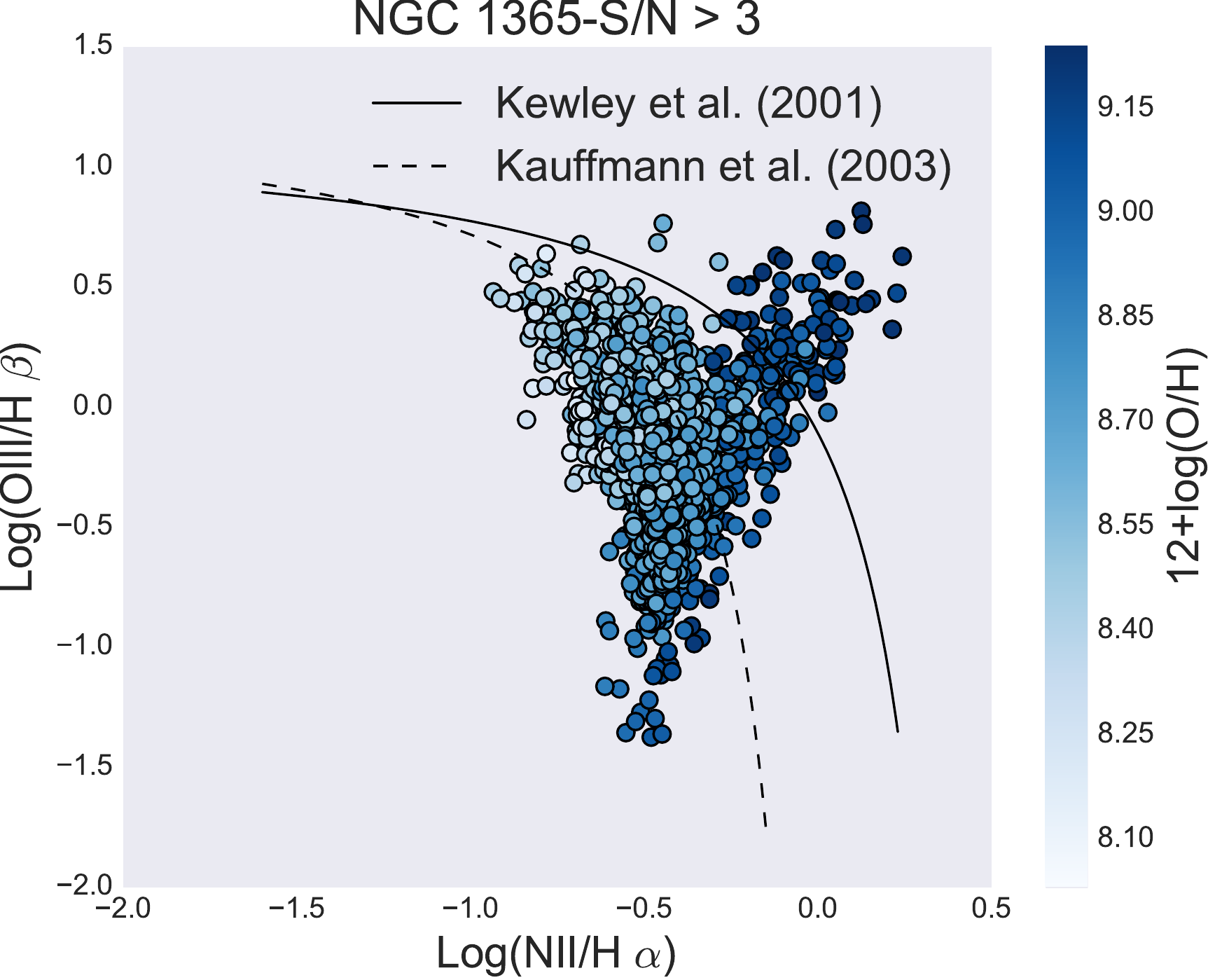}
		}
		\subfigure{%
			\includegraphics[width=0.38\textwidth]{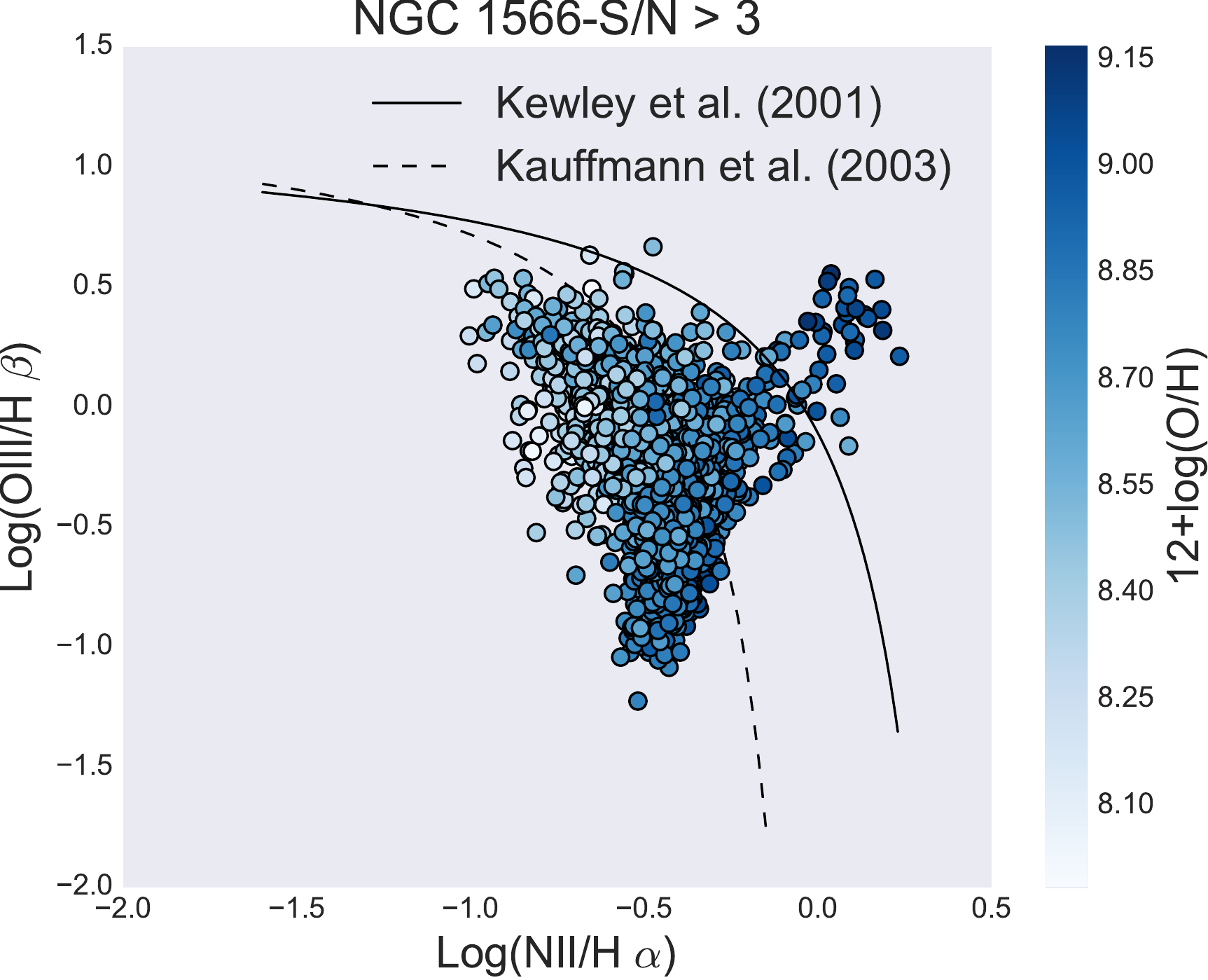}
		}
		
		\subfigure{%
			\includegraphics[width=0.38\textwidth]{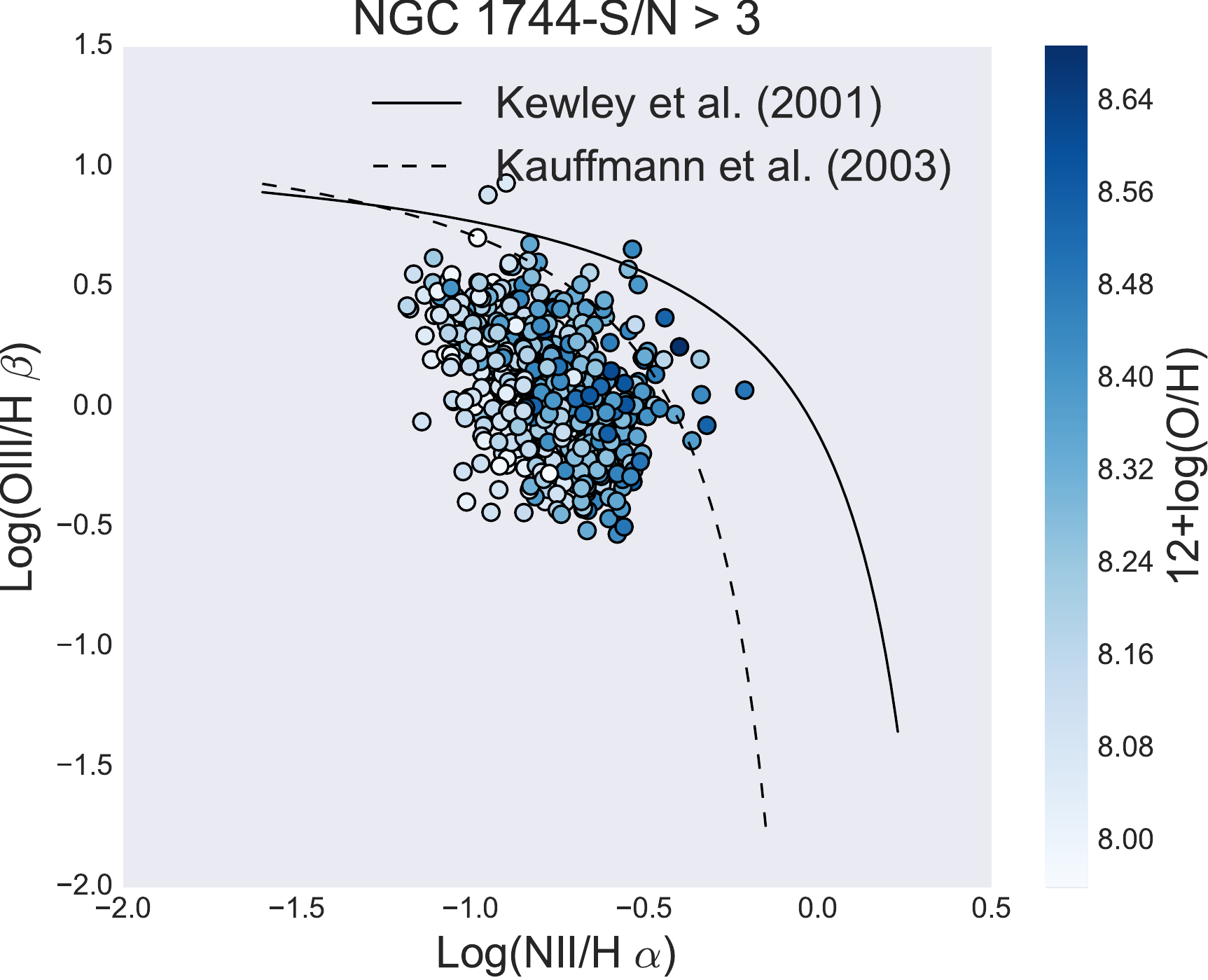}
		}
		\subfigure{%
			\includegraphics[width=0.38\textwidth]{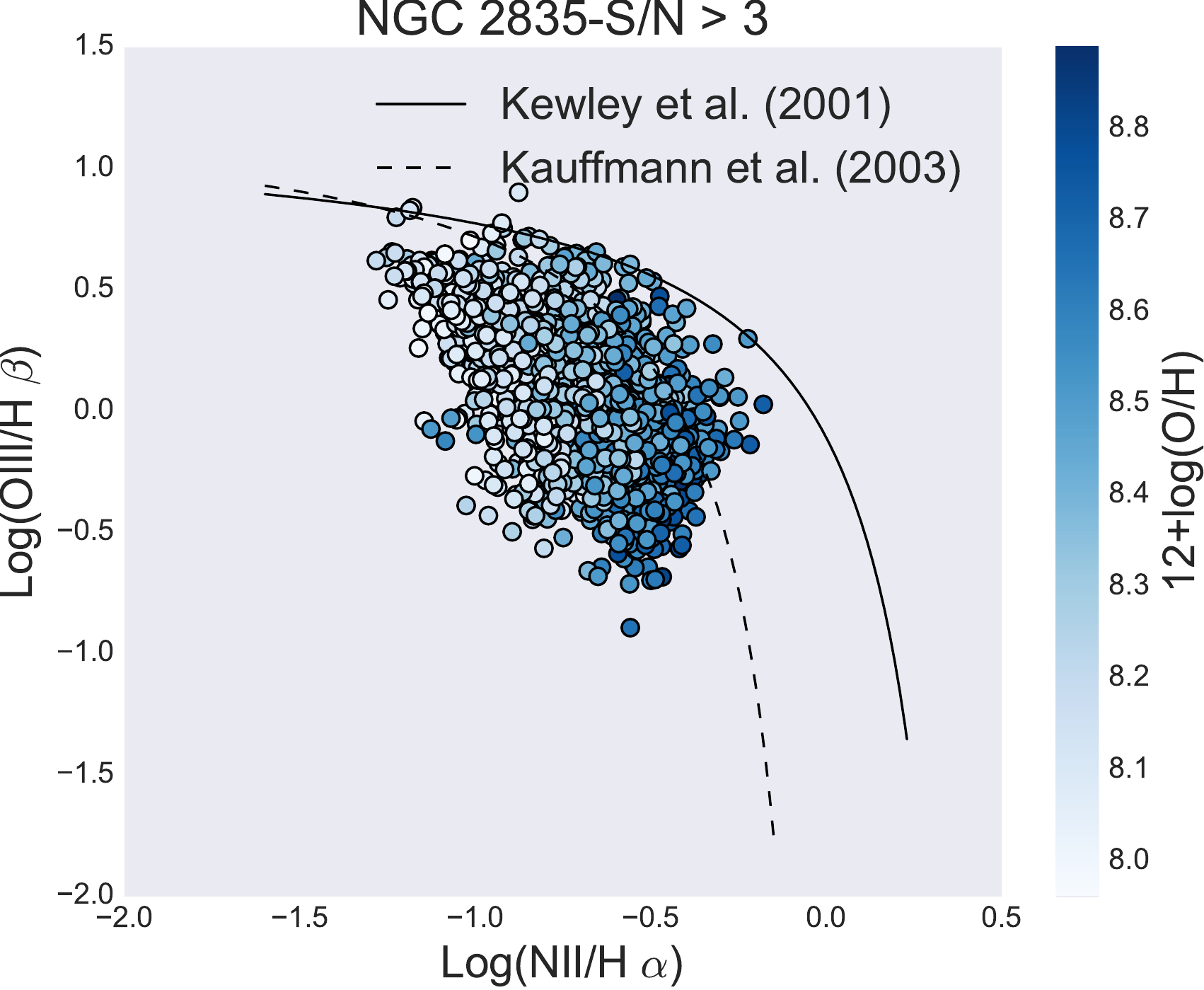}
		}
		\subfigure{%
			\includegraphics[width=0.38\textwidth]{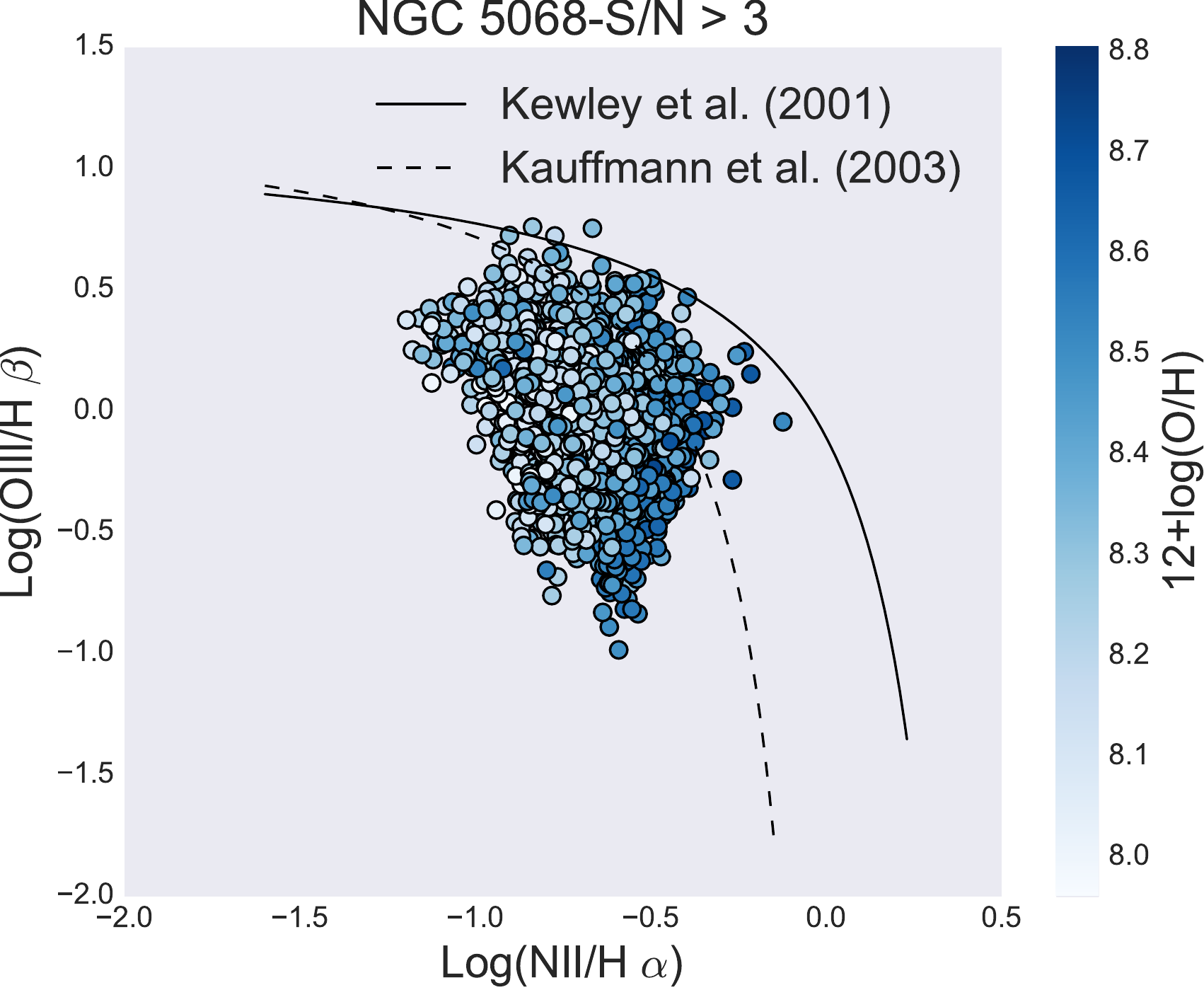}
		}

		\caption{BPT diagrams of our galaxies. The classification lines come from \citet{Ka03} (dashed line) and \citet{Ke01} (solid line). The spaxels are colored according to their metallicity derived from the N2S2 diagnostic (Section~\ref{sec:N2S2}). We exclude spaxels over the \citet{Ka03} demarcation line in all analyses} of metallicity gradients to exclude spaxels that are powered by sources other than pure star formation.
		\label{fig:bpt}
	\end{figure}
	\end{landscape}

	Figure~\ref{fig:bpt} shows the BPT diagrams for the galaxies in this study colored by metallicity. To ensure the spaxels we consider are driven only by star formation, we ignore all spaxels above the pure star-forming \citet{Ka03} line in the rest of the analysis in this work.
	
	\subsection{Metallicity Measurements}\label{sec:N2S2}
	There are numerous diagnostics to determine the gas-phase metallicity due to different methods often yielding discrepant oxygen abundances \citep[see][]{Kewley_Ellison_2008}.
	
    In this work, we measure the metallicity following the N2S2-N2H$\alpha$ indicator \citep{Dopita_2016} (hereafter, N2S2) which uses the H$\alpha$, [NII]$\lambda$6484 and [SII]$\lambda\lambda$6717,31 emission lines. All of these four emission lines are well observed by TYPHOON, which is most sensitive to optical red wavelengths. This diagnostic is sensitive to metallicity primarily through the [NII]/[SII] ratio. Unlike the N2S2 diagnostic, N2O2, R$_{23}$ and O$_3$N$_2$ are degenerate with ionisation parameter, requiring the [OII]3727,29 lines to break the degeneracy. Because the [OII]$\lambda\lambda$3727,29 lines are at the blue sensitivity of TYPHOON dataset and are barely detected in individual spaxels, we are unable to determine the spaxel-by-spaxel metallicity with the N2O2, R$_{23}$ and O$_3$N$_2$ diagnostic. The recent TYPHOON work by \citet{Grasha_2022} analyse gas-phase metallicity on HII-region basis instead of at the level of individual spaxel. Summing up the composite flux of every individual HII region enables them to obtain reliable [OII]$\lambda\lambda$3727,29 meausrements in order to use the N2O2 diagnostic.
	
	We note that with the inclusion of the [SII] doublet lines, the N2S2 diagnostic is sensitive to both metallicity and the fractional contribution from DIG \citep{Shapley_2019, Kumari_2019}. The high spatial resolution of TYPHOON data allows the individual star-forming regions to be identified and separated from the surrounding DIG. We therefore are not concerned about DIG contamination given the resolution of our spaxels at the scale of individual HII regions. The contamination of DIG to star formation only becomes an issue for studies with a resolution that is unable to resolve individual HII regions \citep[$\sim$kpc;][]{Poetrodjojo_2019}.
	
\begin{landscape}
	\begin{figure}
		\centering
		\subfigure{%
			\includegraphics[width=0.4\textwidth,height=0.43\textwidth]{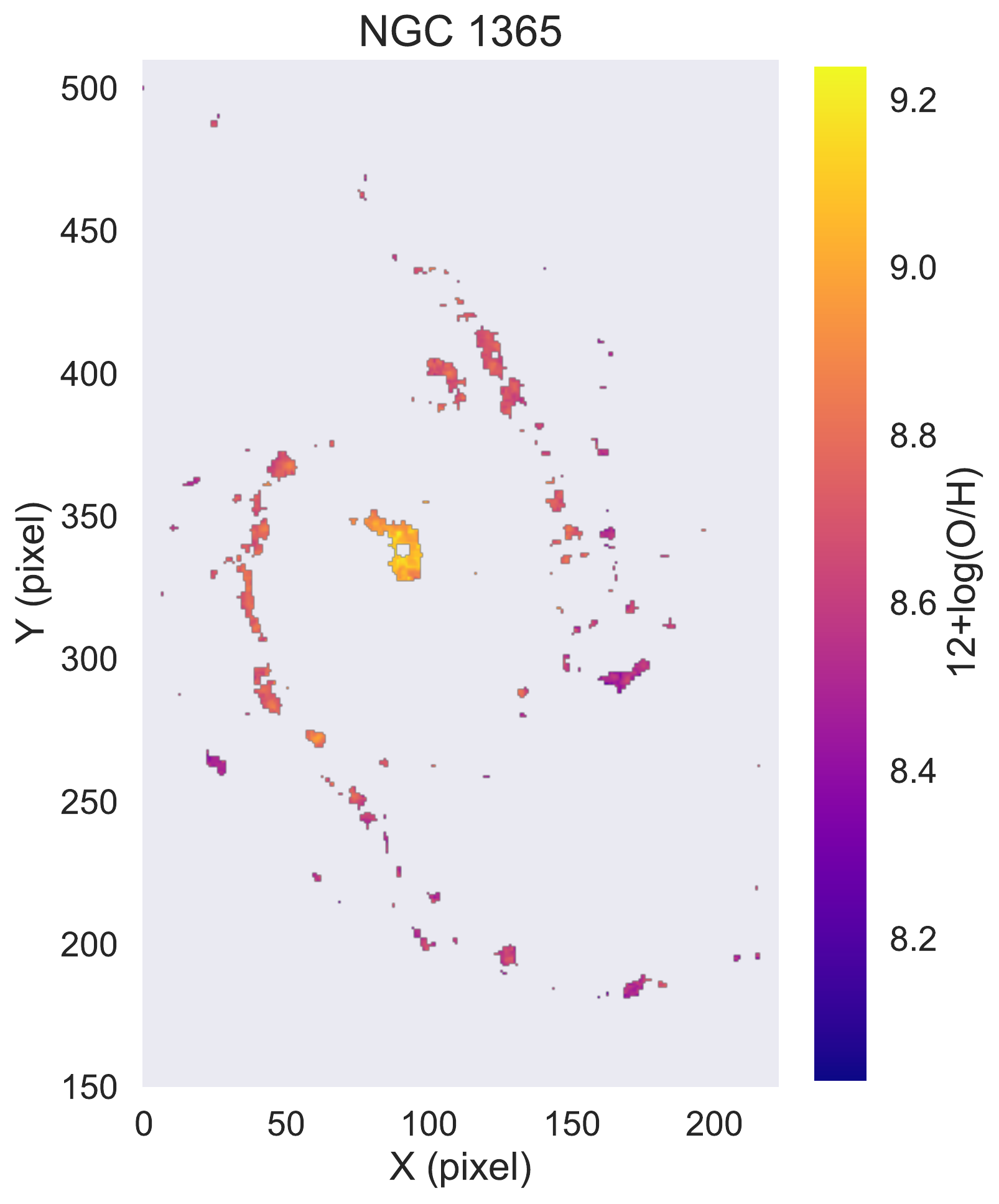}
		}
		\subfigure{%
			\includegraphics[width=0.4\textwidth,height=0.43\textwidth]{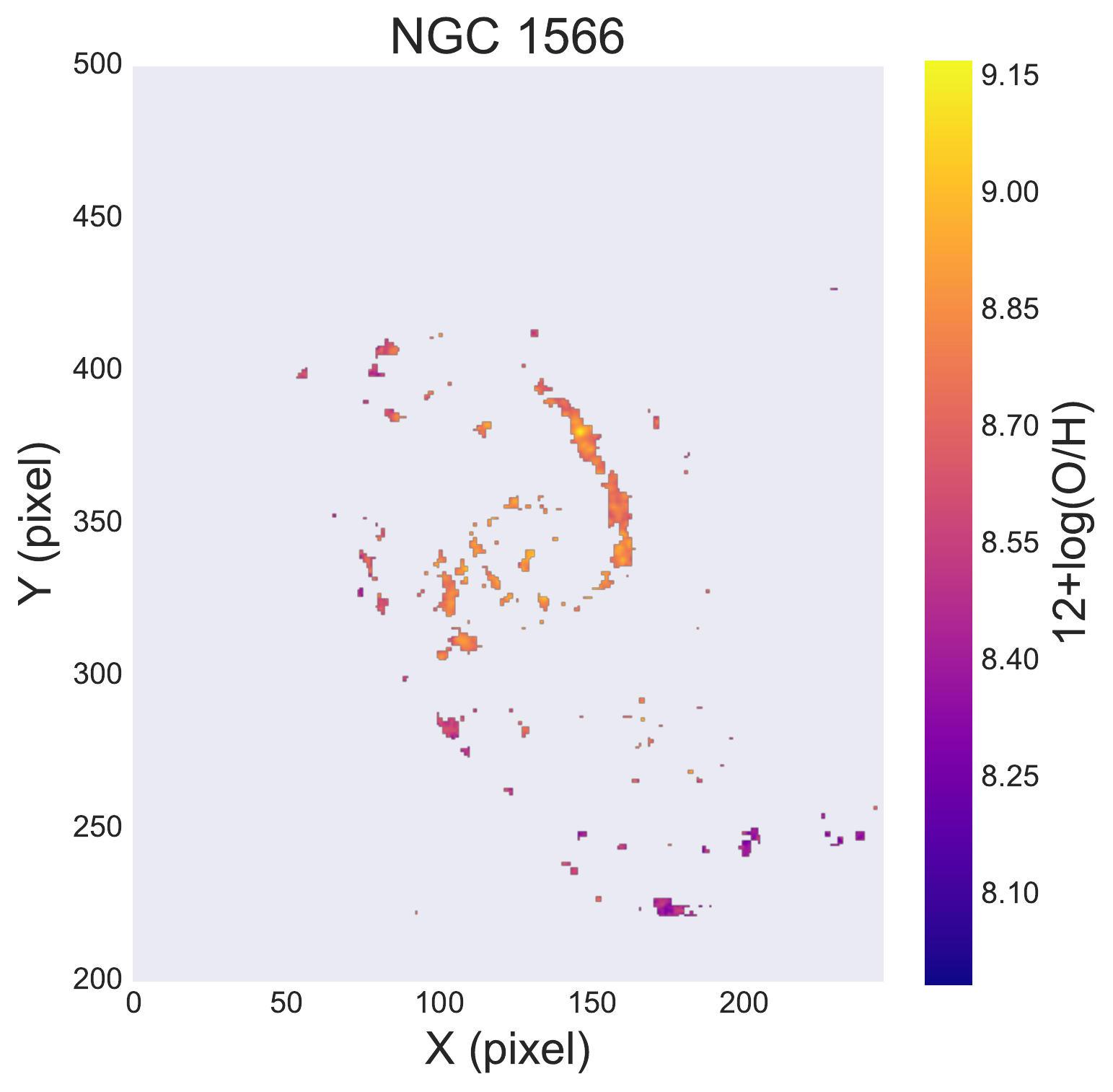}
		}
		
		\subfigure{%
			\includegraphics[width=0.4\textwidth,height=0.43\textwidth]{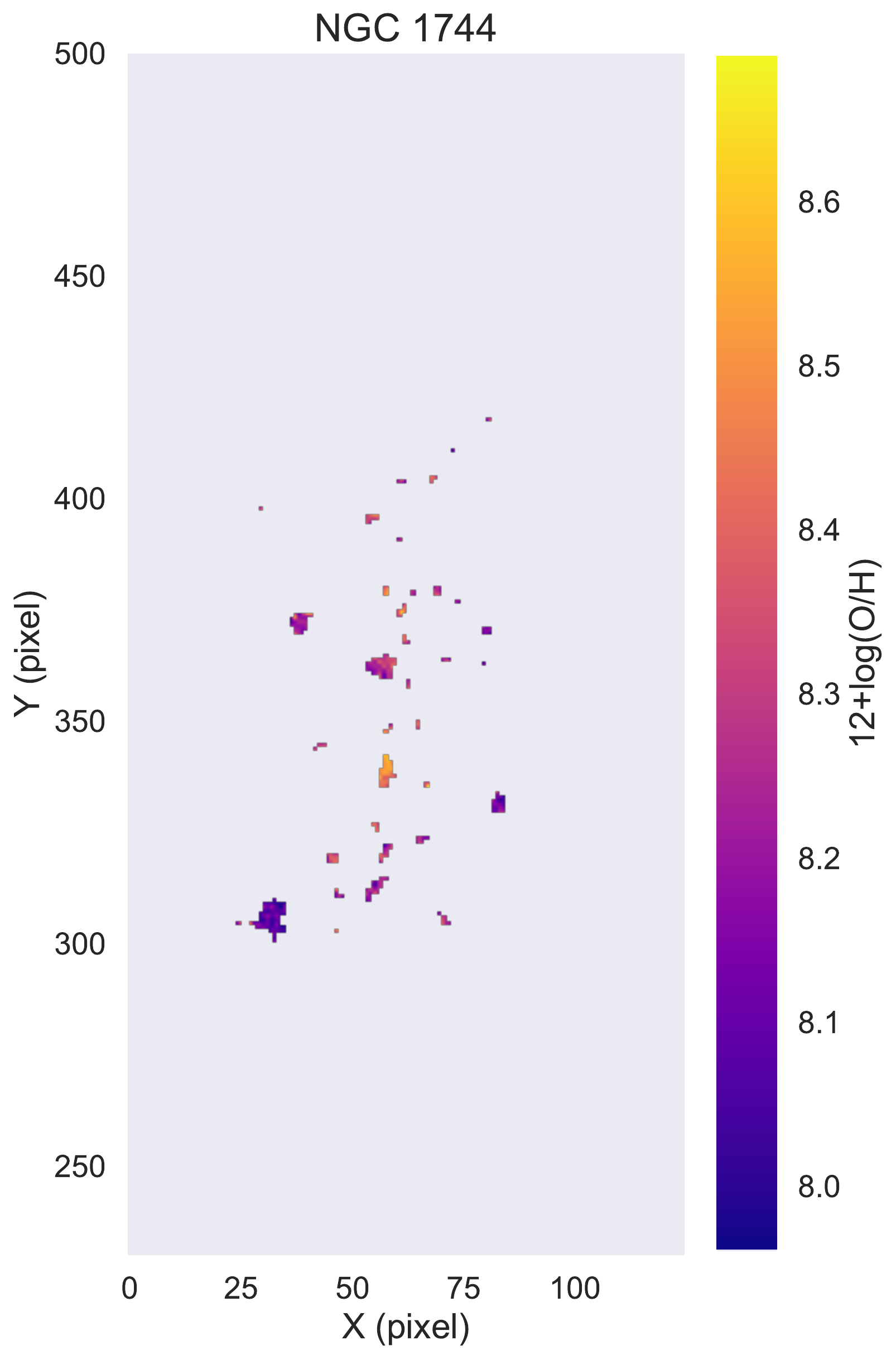}
		}
		\subfigure{%
			\includegraphics[width=0.4\textwidth,height=0.43\textwidth]{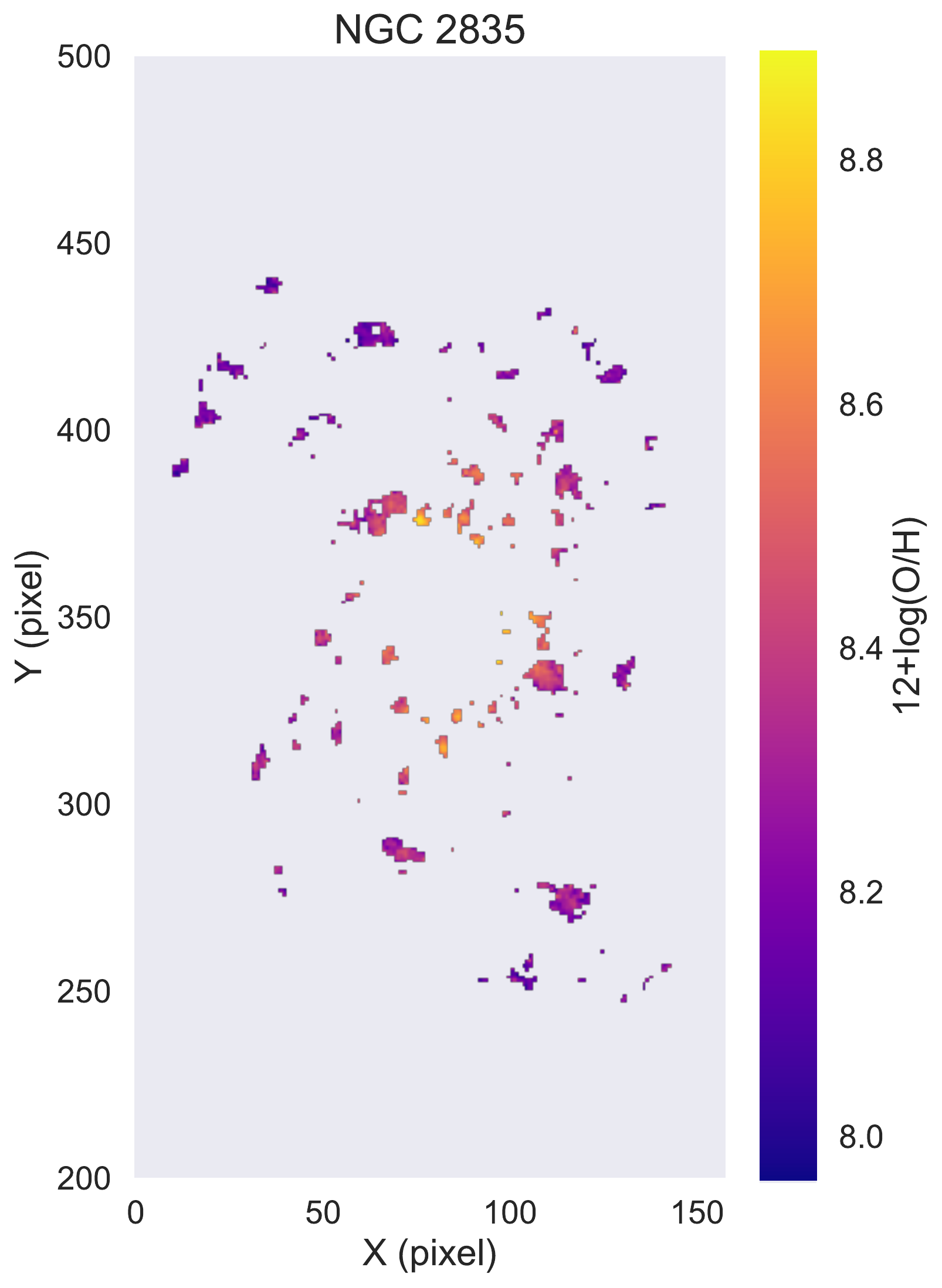}
		}
		\subfigure{%
			\includegraphics[width=0.4\textwidth,height=0.43\textwidth]{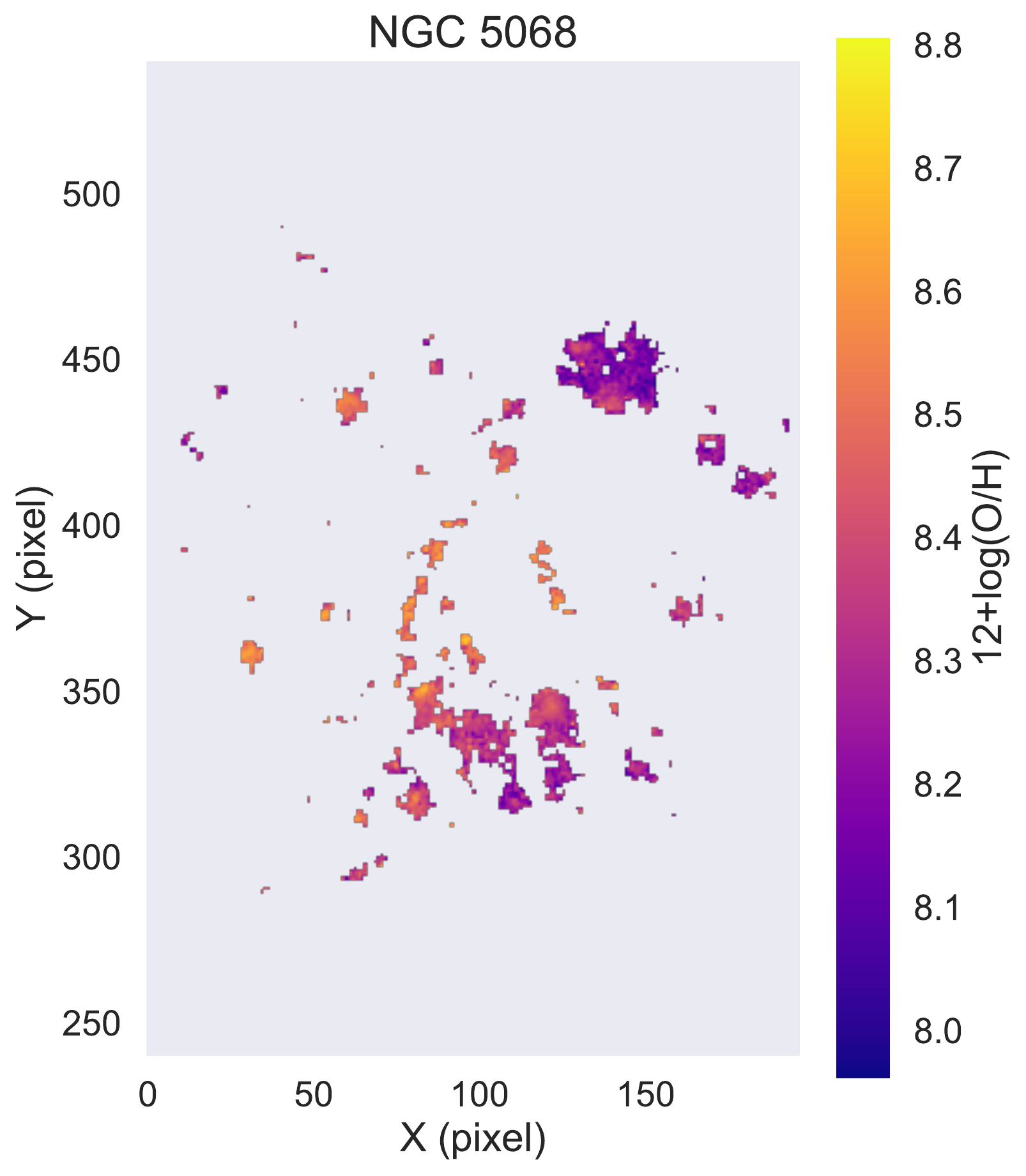}
		}
		\caption{Metallicity maps of our samples, derived using the N2S2 diagnostic \citep{Dopita_2016}. Higher metallicity is} observed in the inner regions of all the galaxies.
		\label{fig:zmap}
	\end{figure}
	\end{landscape}

	The N2S2 metallicity diagnostic \citep{Dopita_2016} is calculated as:
	\begin{gather}
		y = \log [NII]/[SII] + 0.264 \log[NII]/H\alpha \\
		12+\log(O/H) = 8.77+y + 0.45(y+0.3)^5  
	\end{gather}
	over the range of 12+log(O/H) = 7.96 \textasciitilde\, 9.24. Fig~\ref{fig:zmap} shows the metallicity maps derived using the N2S2 metallicity diagnostic. Every galaxy shows a clear negative radial gradient with larger metallicity values at smaller radii compared with the lower metallicity values recovered at larger galactocentric radii. We do not see any asymmetric in the metallicity maps except for NGC~1566 which shows lower ($\sim$ 0.15 dex) metallicity in the outer region of the southern spiral arm than in the northern arm.
	
	\subsection{Gas-phase Oxygen Abundance Gradients}
	We present radial profiles of the oxygen abundance distribution with both a single linear fit (Section~\ref{sec:singleline}) and a double linear fit (Section~\ref{sec:doubleline}) to constrain any possible effect of the central bar on the central metallicity distribution. 
	The comparison of single- and piecewise-linear fits for each individual galaxy are discussed in Sec~\ref{sec:doubleline}. 
	
	\subsubsection{Single linear fit}\label{sec:singleline}
\begin{landscape}
	\begin{figure}
		\centering
		
		\subfigure{%
			\includegraphics[width=0.45\textwidth]{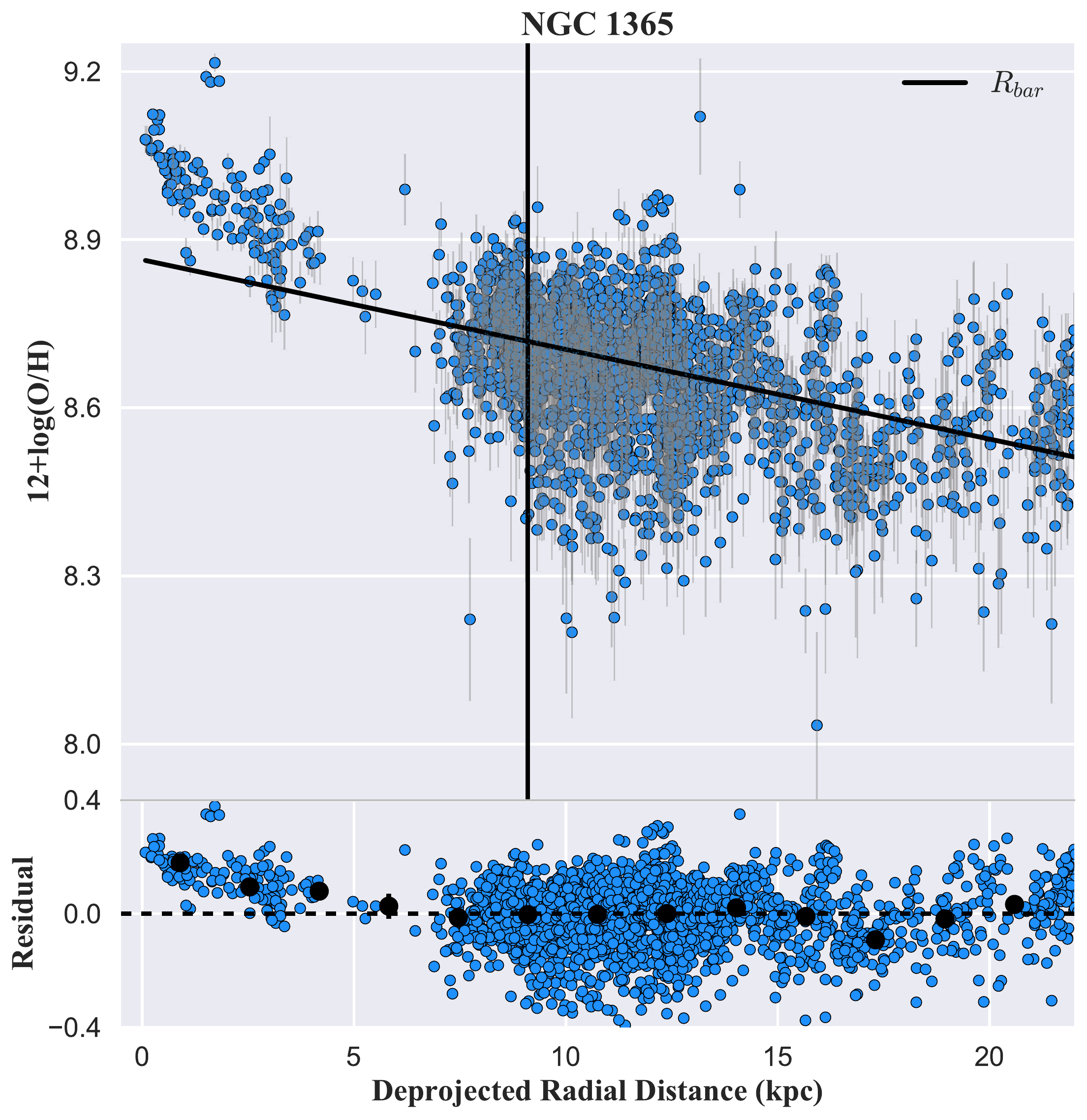}
		}
		\subfigure{%
			\includegraphics[width=0.45\textwidth]{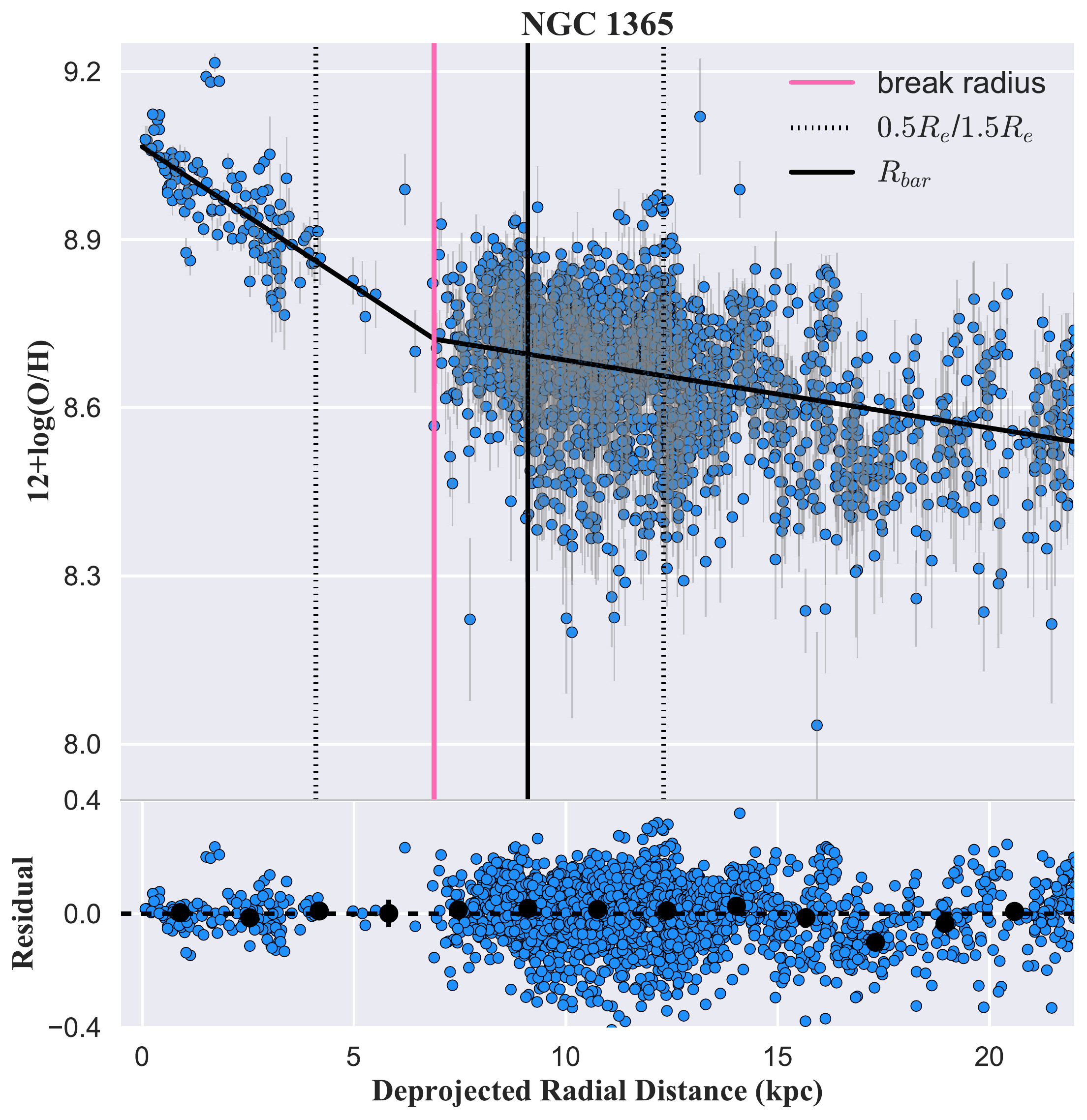}
		}
		\subfigure{%
			\includegraphics[width=0.35\textwidth]{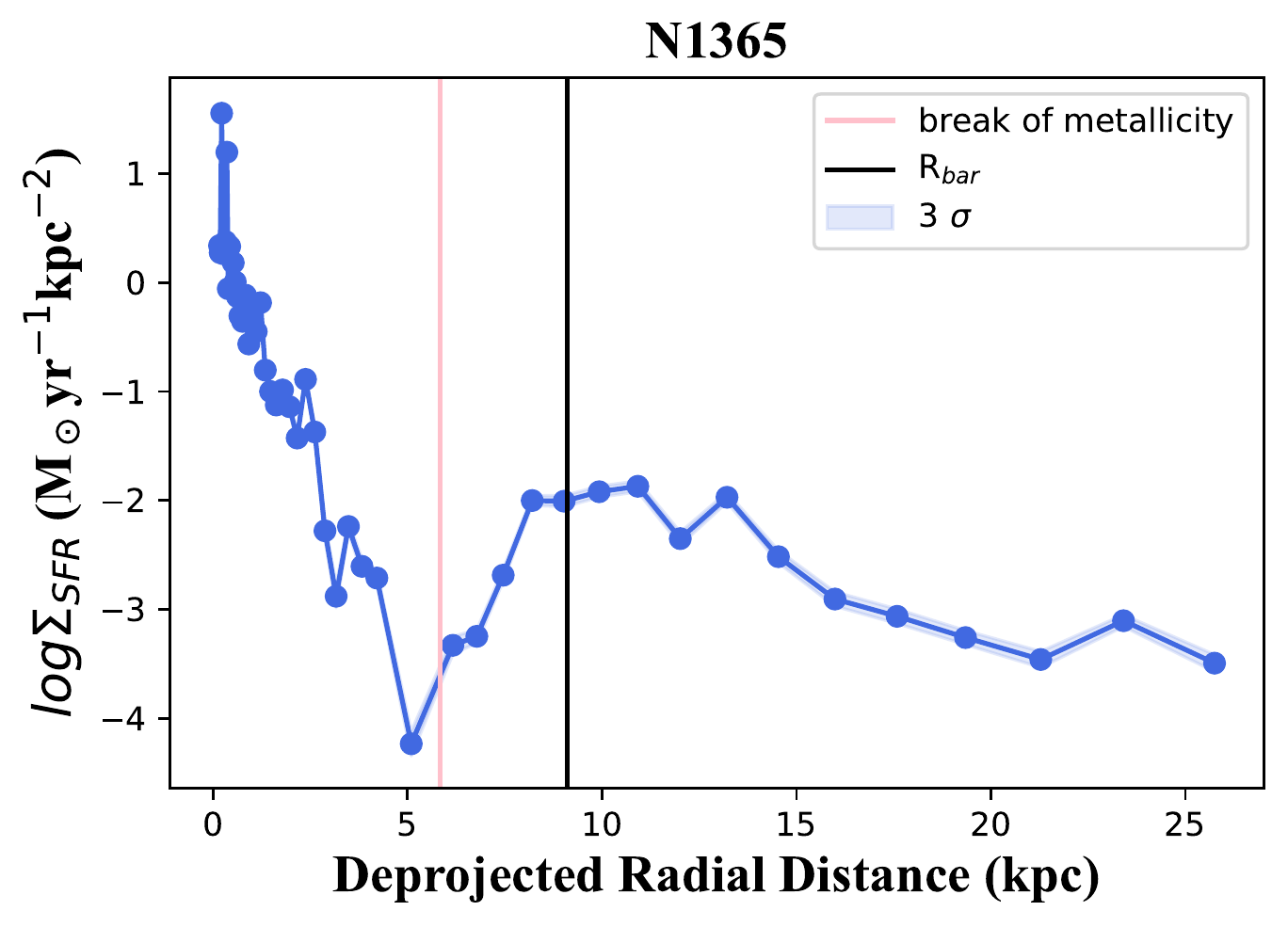}
		}

		\caption{{\bf {Left panel}}: Oxygen abundance radial distribution. Gray line shows 1$\sigma$ errors for oxygen abundance. Black diagonal line shows the best linear fit to the data. The vertical black solid lines in all panels indicate the end of bars (hereafter R$_\mathrm{bar}$, Appendix~\ref{sec:barend}) as well as in the middle and left panel}. NGC~1365 shows a negative metallicity radial gradient. The residual between the data and the best fit is plotted in the bottom sub-panel, with black scatters as the medians of each 1~kpc bin. 
		{\bf {Middle panel}}: Oxygen abundance gradient fitted with a double linear fit. The data and errorbar are the same as the left panel. Black diagonal line shows the best double-linear fit. Pink solid lines represent the break radius, which is determined by the least square method (Section~\ref{sec:doubleline}. Vertical dotted lines mark the location of 0.5 R$_e$ and 1.5 R$_e$. 
		The residual between data and best fit is shown in the bottom sub-panel as the left panel, with medians of each 1~kpc bin overplotted as the black scatters.
		A double linear fit better describes the radial distribution of metallicity than single linear fits in our study. The observed broken metallicity gradients in these barred galaxies may be attributed to the effects of bars on mixing the ISM. 
		{\bf {Right panel}}: star formation rate (SFR) surface density radial profile $\Sigma_{\mathrm{SFR}}$ as a function of deprojected radial distance. SFRs are derived from H$\alpha$ luminosity. Vertical pink lines are break radius, the same as the middle panel. The shadow of the lines represents 1$\sigma$ of SFR surface density, calculated from the uncertainty of H$\alpha$ and adding the 2\% uncertainty of flux calibration (see Sec~\ref{sec:obs}). 
		We observe changes in SFR gradients near the break radius of metallicity gradients in all the galaxies.
		\label{fig:Ograd}
	\end{figure}
	\end{landscape}

\begin{landscape}
	\begin{figure}
		\centering
		\subfigure{%
			\includegraphics[width=0.45\textwidth]{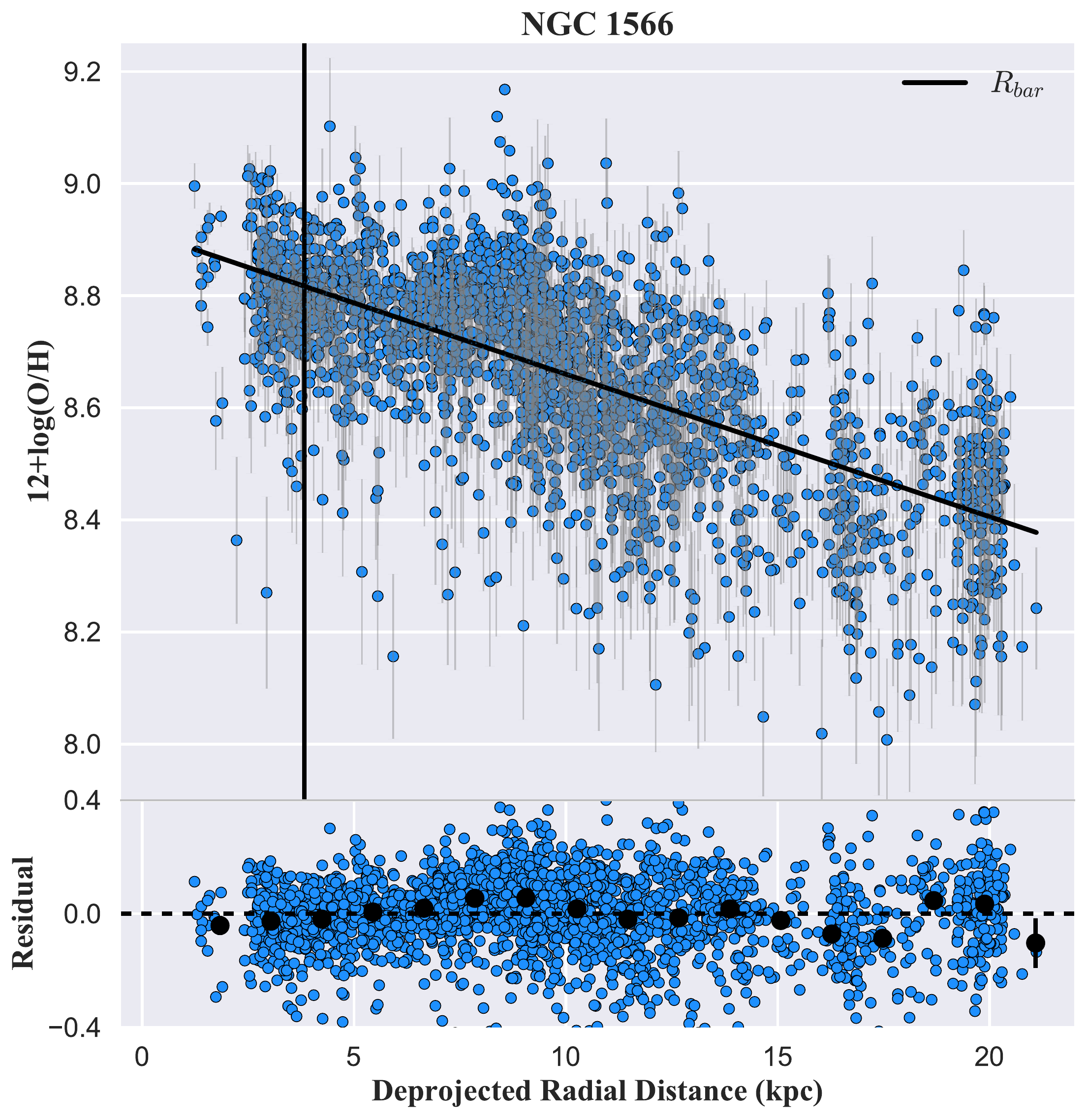}
		}
		\subfigure{%
			\includegraphics[width=0.45\textwidth]{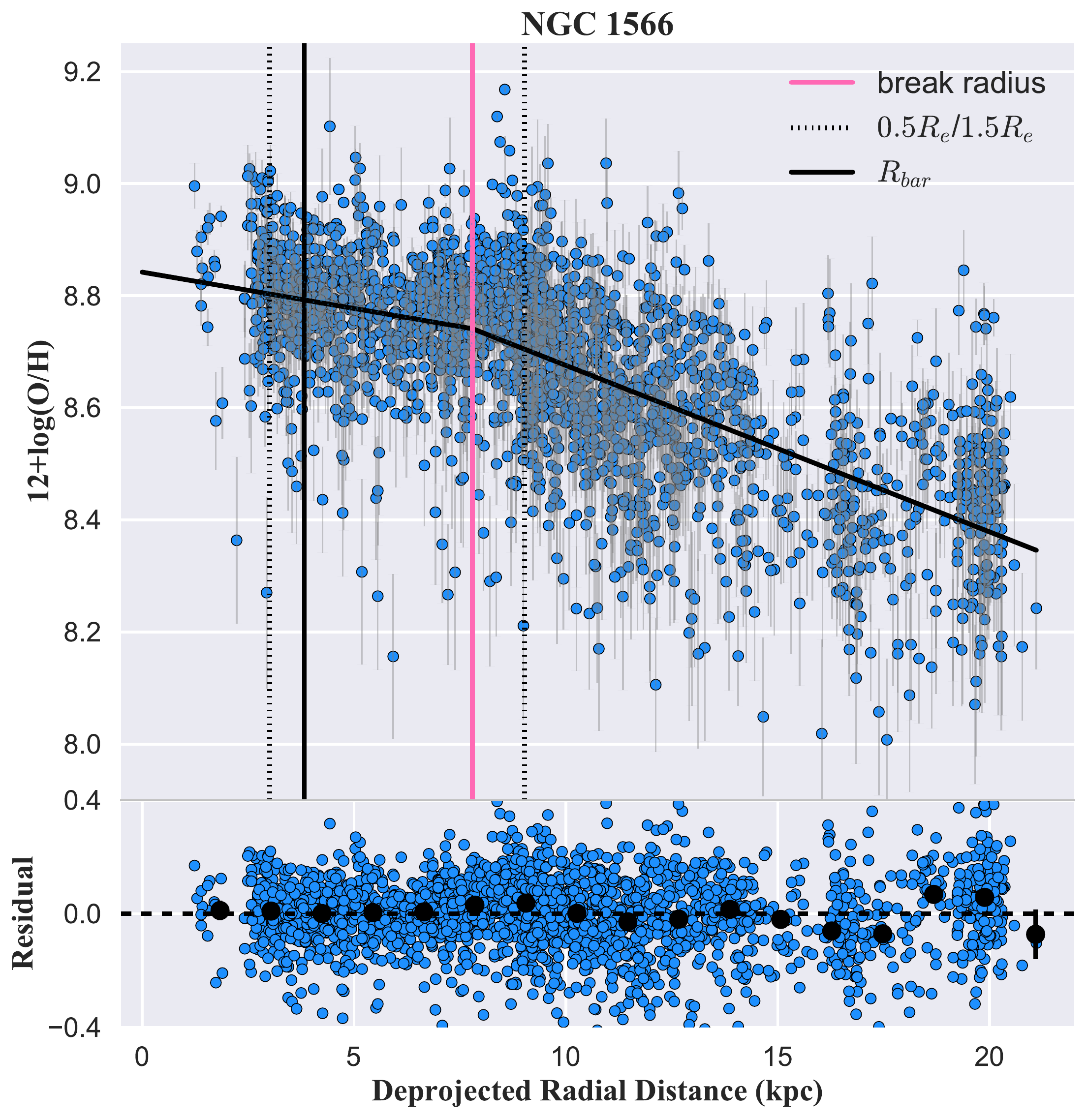}
		}
		\subfigure{%
			\includegraphics[width=0.35\textwidth]{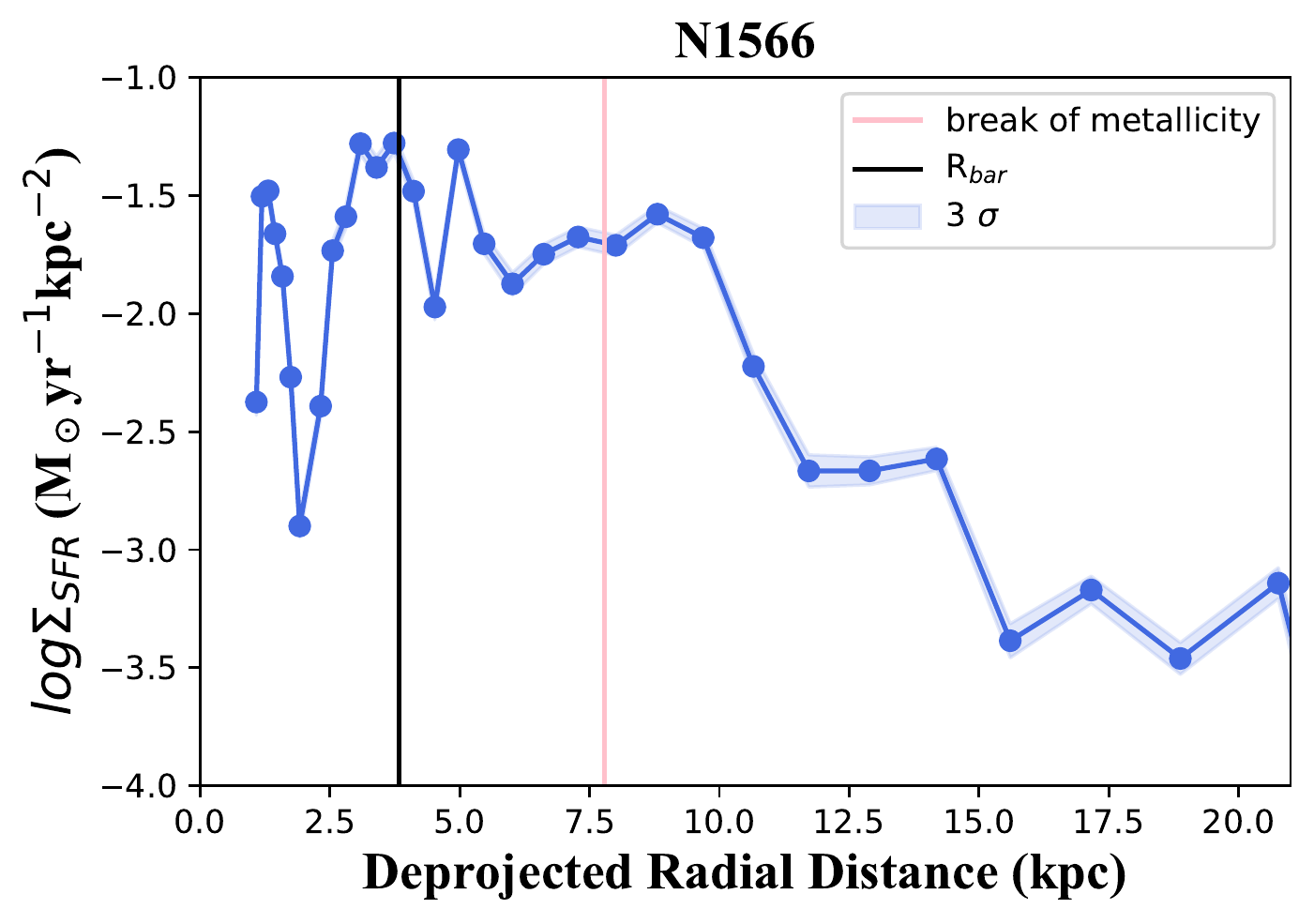}
		}
		
		\subfigure{%
			\includegraphics[width=0.45\textwidth]{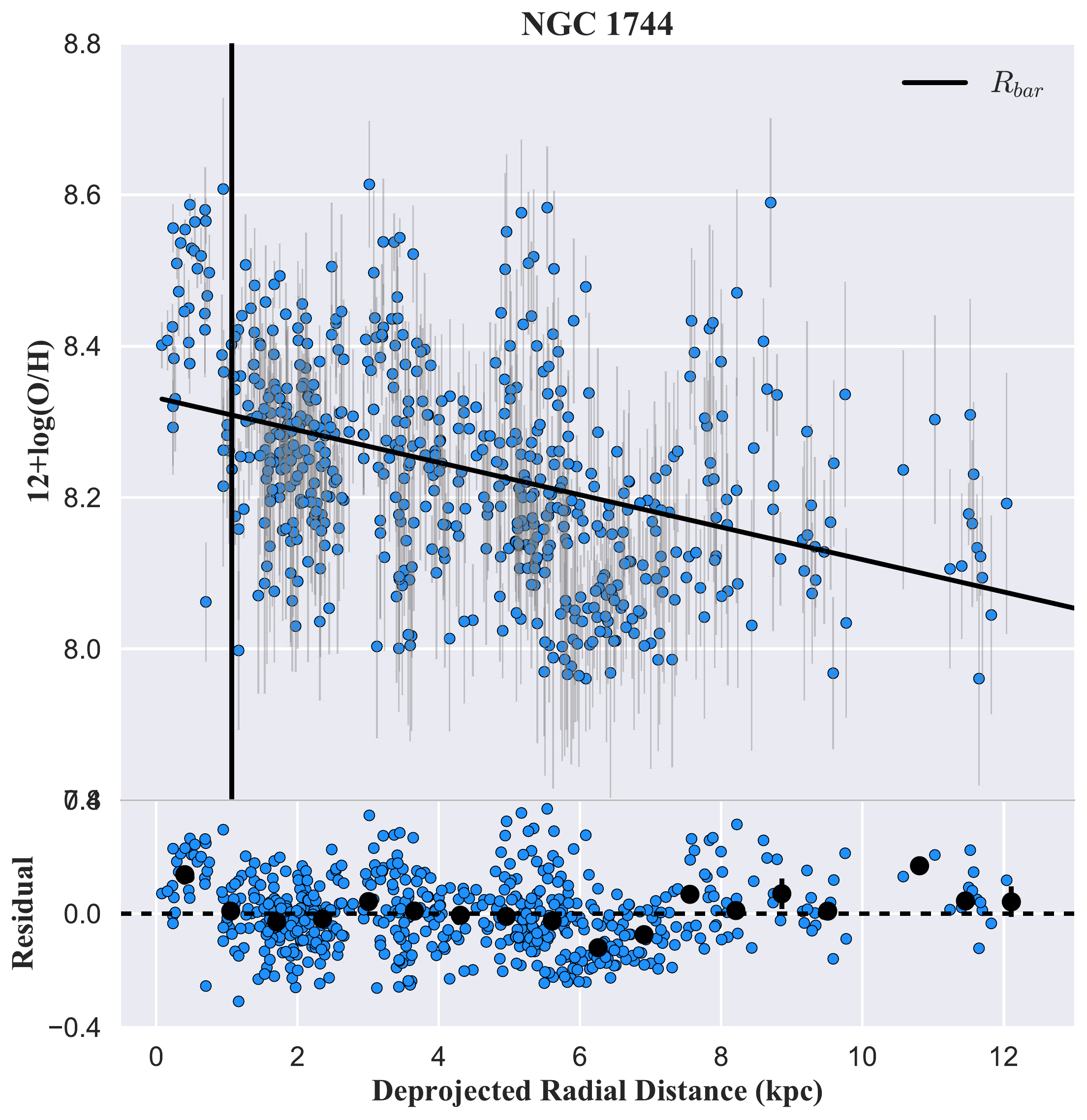}
		}
		\subfigure{%
			\includegraphics[width=0.45\textwidth]{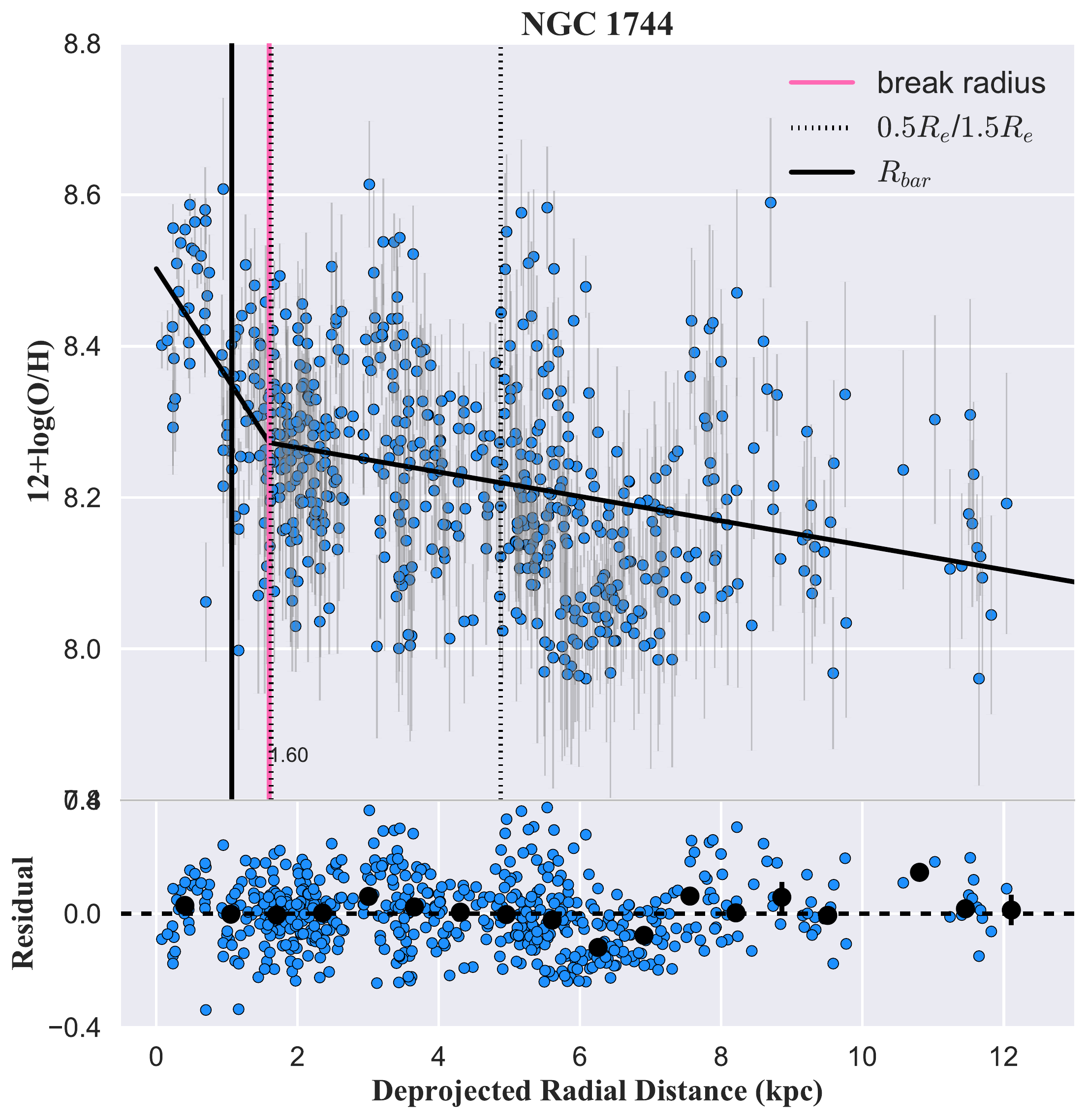}
		}
		\subfigure{%
			\includegraphics[width=0.35\textwidth]{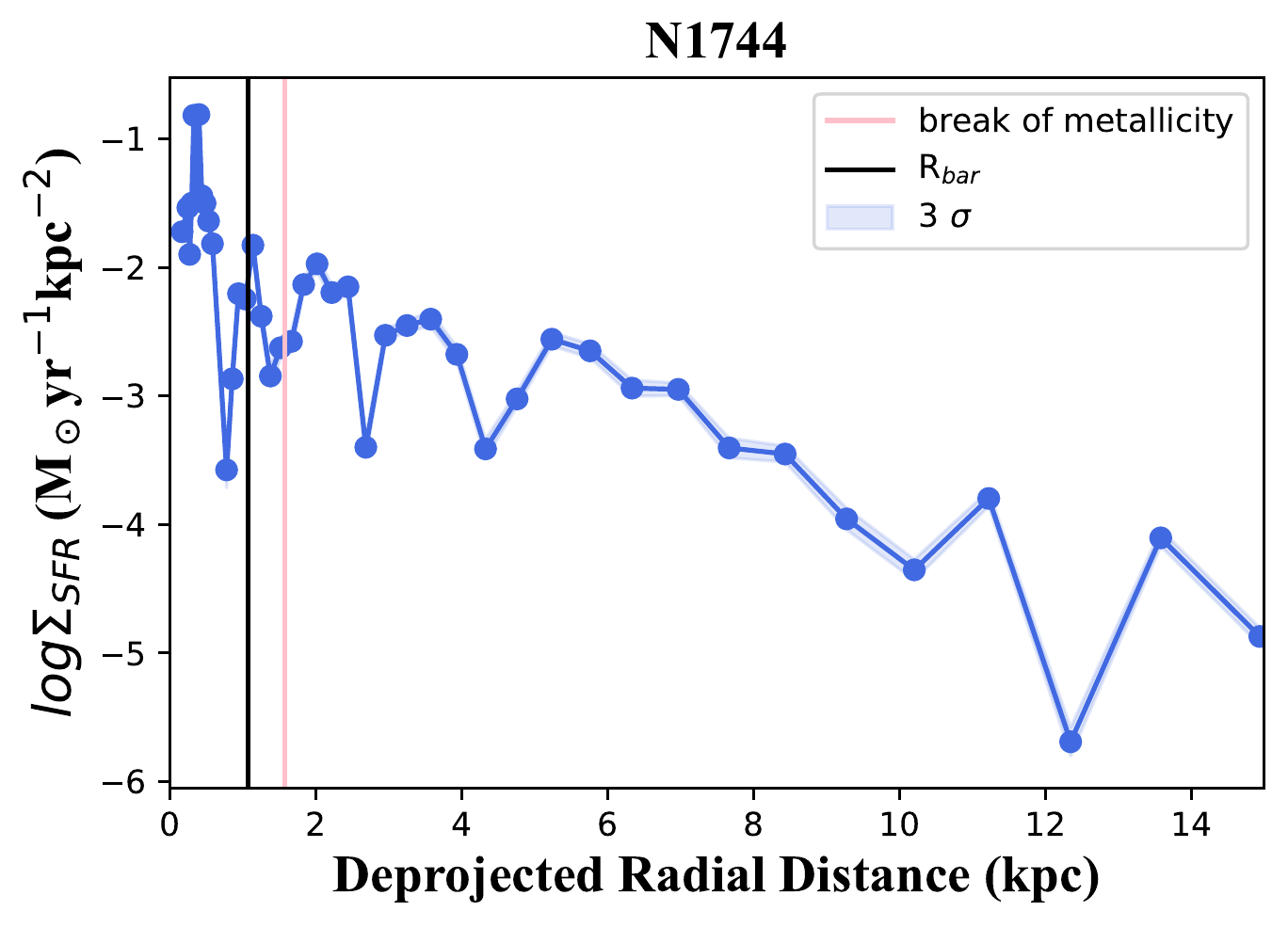}
		}

		\caption{Similar to Fig~\ref{fig:Ograd} but for NGC~1566 and NGC~1744.}
		\label{fig:Ograd2}
	\end{figure}
	\end{landscape}

\begin{landscape}
	\begin{figure}
		\centering
	\subfigure{%
			\includegraphics[width=0.45\textwidth]{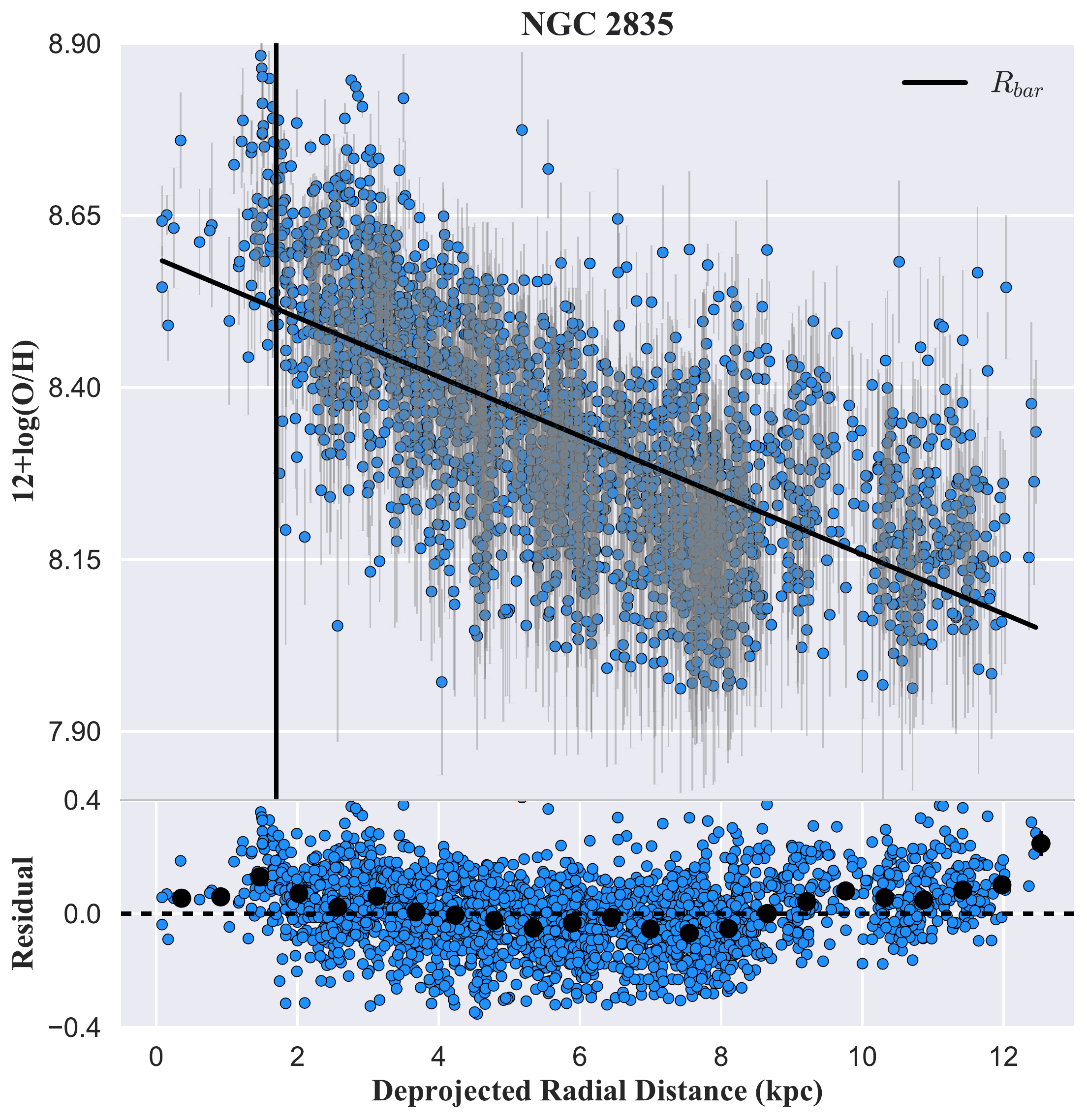}
		}
		\subfigure{%
			\includegraphics[width=0.45\textwidth]{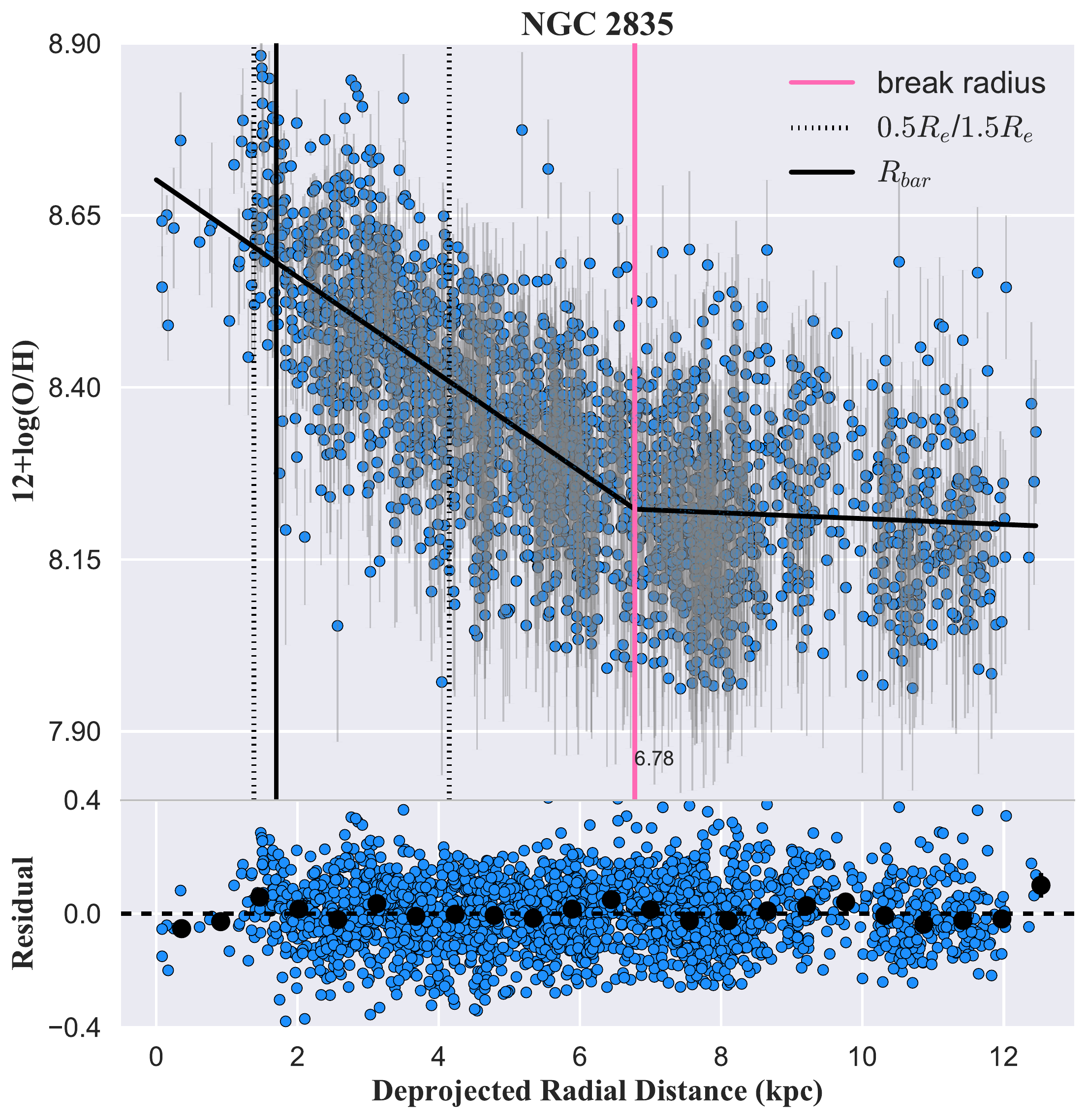}
		}
		\subfigure{%
			\includegraphics[width=0.35\textwidth]{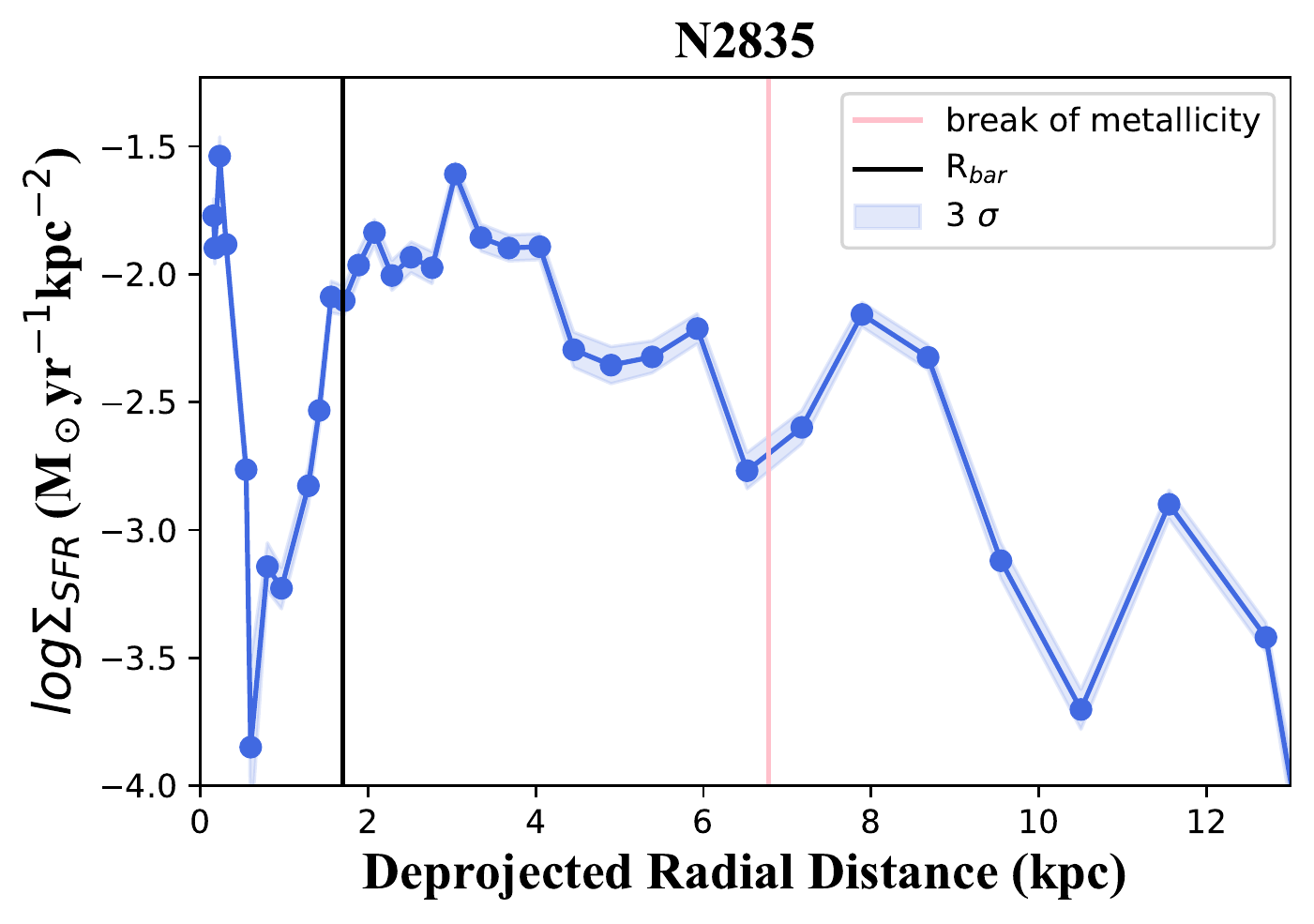}
		}
		
		\subfigure{%
			\includegraphics[width=0.45\textwidth]{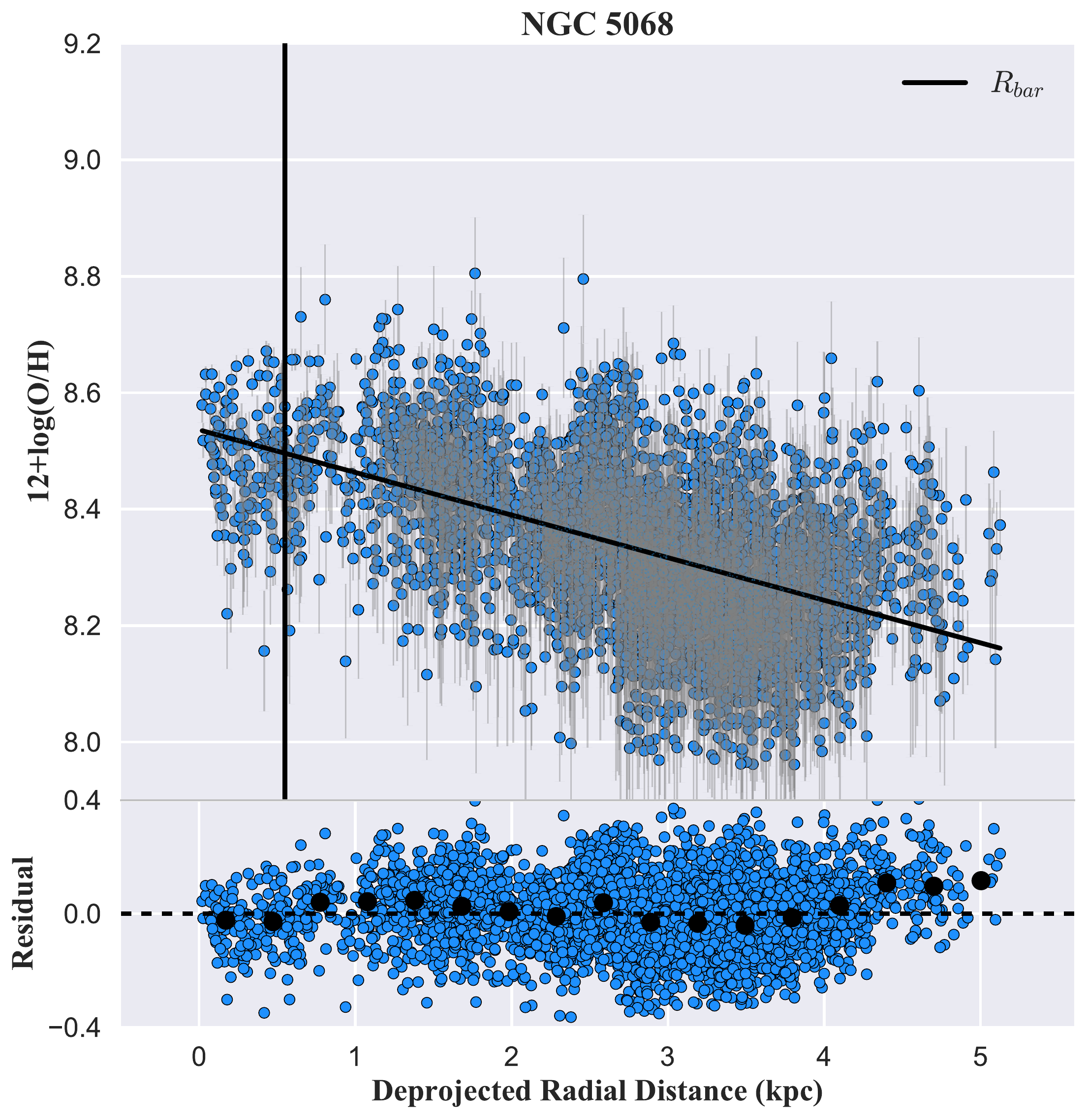}
		}
		\subfigure{%
			\includegraphics[width=0.45\textwidth]{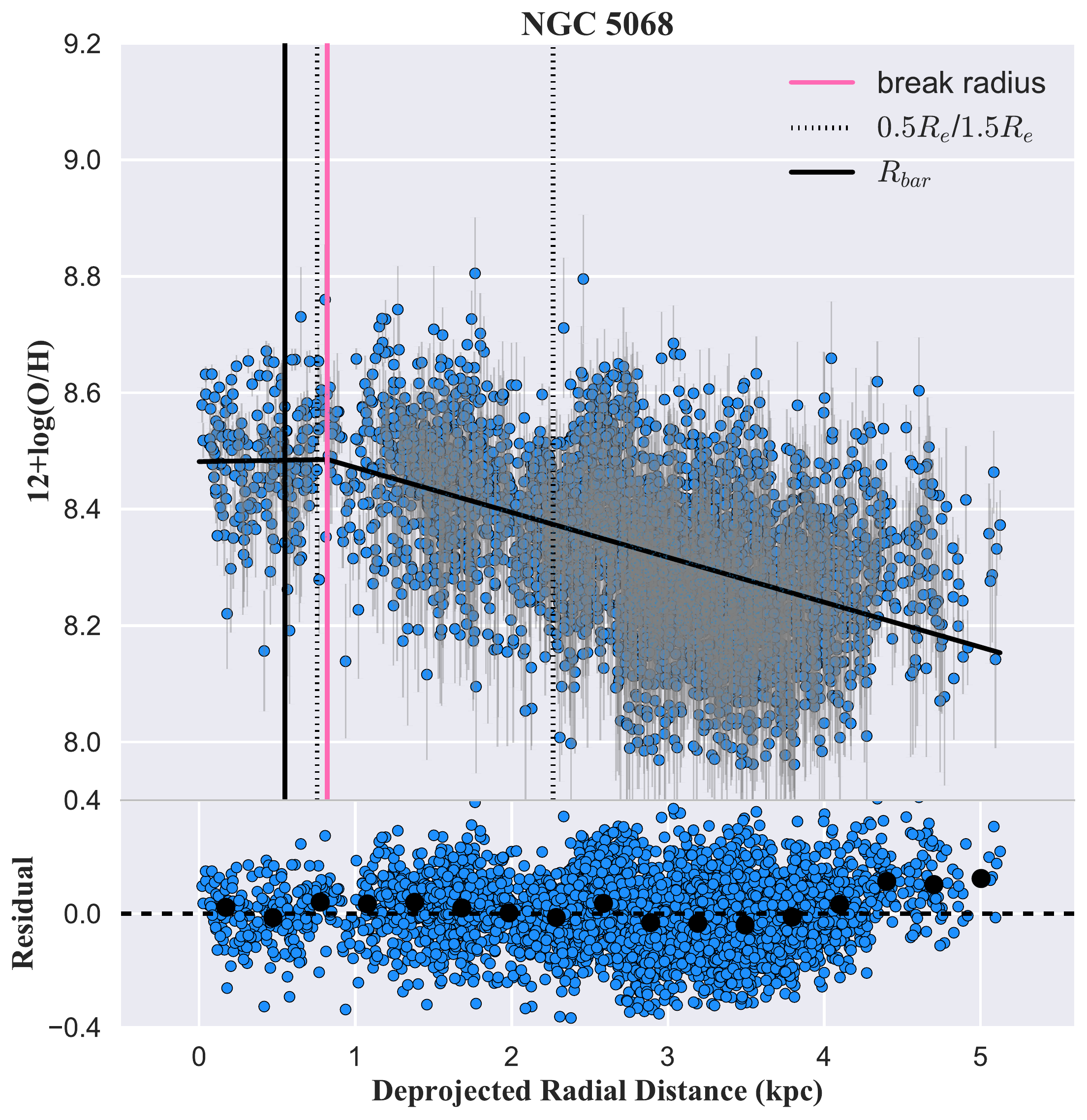}
		}
		\subfigure{%
			\includegraphics[width=0.35\textwidth]{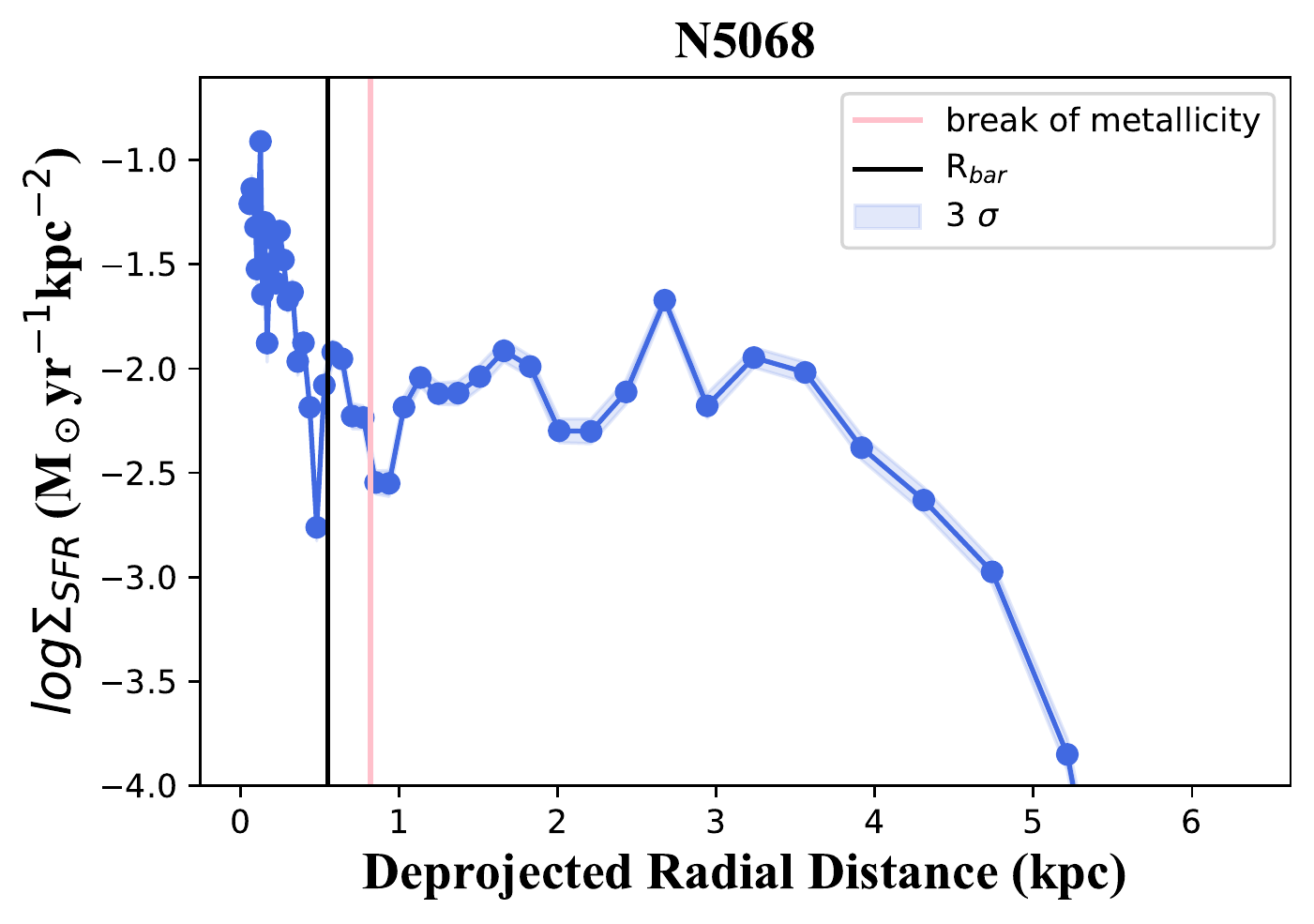}
		}
		\caption{Similar to Fig~\ref{fig:Ograd} but for NGC~2835 and NGC~5068.}
		\label{fig:Ograd3}
	\end{figure}
	\end{landscape}

	For each galaxy, we perform the least squares linear fit to all spaxels above S/N $\textgreater$ 3 to measure the metallicity gradient. The single linear best fit is shown as a black solid line in left column of Fig~\ref{fig:Ograd}, Fig~\ref{fig:Ograd2} and Fig~\ref{fig:Ograd3}. The reported gradients with a single linear fit and associated 1$\sigma$ errors are reported in Table~\ref{tab:gradient}. All galaxies in this study show a decreasing metallicity with increasing radial distance. The global metallicity gradients, determined by single linear fits, are shallow and range from $-$0.0162 to $-$0.073 dex/kpc. The shallow metallicity gradients we recover are consistent with those found by \cite{Seidel_2016}, who found the metallicity gradients to be flatter in barred galaxies than in unbarred galaxies.

	As seen in the left column of Fig~\ref{fig:Ograd}, Fig~\ref{fig:Ograd2} and Fig~\ref{fig:Ograd3}, most of our galaxies  can be well-described with a single linear function, albeit with significant scatter. The residuals between data and best single-linear fits fluctuate around 0 (dash line in the bottom panels of the left and middle columns of Fig~\ref{fig:Ograd}, Fig~\ref{fig:Ograd2} and Fig~\ref{fig:Ograd3}.
	In the next section, we perform piecewise linear fits to metallicity to investigate whether the observed data can be better described with piecewise compared to single-linear fits.
	
	\subsubsection{Piecewise linear fit}\label{sec:doubleline}
	In Sec~\ref{sec:singleline}, we find a deficiency in the single-linear fits of the metallicity gradients as the residuals near bar radius deviate from 0, especially for NGC~1365, NGC~1744 and NGC~2835.
	The deviation of the metallicity gradients from a linear fit in the bar regions of these galaxies is of note as this may be attributed to the impact of the bars on the metal distribution in the gas. To better explore the impact of bars on the oxygen abundance gradients, we apply piecewise linear fits to the observed metallicity distribution. The break radius, inner slope, outer slope and intercept are all included as parameters in the fit, determined by least chi-squares. The piecewise linear best fits are presented in the middle column of Fig~\ref{fig:Ograd}, Fig~\ref{fig:Ograd2} and Fig~\ref{fig:Ograd3} and listed in Table~\ref{tab:gradient}.
	
	We find, in general, smaller residuals (scatters in the bottom panels of the left and middle columns of Fig~\ref{fig:Ograd}, Fig~\ref{fig:Ograd2} and Fig~\ref{fig:Ograd3} using piecewise linear fits especially near the break radii, albeit the marginal improvement in NGC~1566 and NGC~5068. Our results indicate that the piecewise linear function is a more suitable model to describe metallicity gradients of our barred galaxies, with a marked improvement in the residuals near R$_\mathrm{bar}$. We classify the broken metallicity profiles into shallow-steep (NGC~1566 and NGC~5068) and steep shallow profiles (NGC~1365, NGC~1744 and NGC~2835), further discussed in the following paragraphs. 
	
	For NGC~1365, for a single linear fit to metallicity gradient (Sec~\ref{sec:singleline}), we measure a slope of $-$0.0162 $\pm$ 0.0005 dex/kpc (left panel of Figure~\ref{fig:Ograd}). However, the single linear fit poorly constrains the individual spaxel measurements, showing an excess in positive residuals within the bar region.
	We re-do the analysis with a two-component fit to the metallicity gradient of $-$0.055 dex/kpc inside the bar region, significantly steeper than the gradient in the disc region ($-$0.0123 dex/kpc) with a break occurring at 5.8~kpc (middle panel of Figure~\ref{fig:Ograd}). The break of metallicity gradient occurs near the edge of the bar region (7.1~kpc), identified manually from optical images (Appendix~\ref{sec:barend}). 
	The smaller residuals observed for the piecewise linear function (Figure~\ref{fig:Ograd}) suggests that this piecewise function constrains the data better than a single linear fit for NGC~1365.
	Furthermore, this steep-shallow metallicity radial gradient in NGC~1365 is consistent with previous observations, see Appendix~\ref{sec:individual_galaxy}.

	We measure a linear fit for the metallicity gradient of NGC 1566 as $-$0.0254 $\pm$ 0.0006 dex/kpc (left column of Figure~\ref{fig:Ograd2}). When fitted with a piecewise function, we find that NGC~1566 is better constrained with a shallow-steep profile (middle column of Figure~\ref{fig:Ograd2}), showing residuals within R$_\mathrm{bar}$ (inner 10~kpc) to be more consistent with smaller scatter centred around zero.
	The metallicity gradient inside the bar region ($-$0.013 dex/kpc) is flatter than the gradient in the disc region ($-$0.0297 dex/kpc) with a break at 7.79~kpc.

	In NGC~1744, we fit a single component to measure the global metallicity gradient of $-$0.021 $\pm$ 0.002 dex/kpc (left column of Figure~\ref{fig:Ograd2}). All spaxels within the R$_\mathrm{bar}$ region (inner 1.07 kpc) lie above the single linear fit, highlighting the inability for a single fit in describing the metallicity gradient of NGC~1744.
	For the piecewise linear fit, we recover a steep-shallow metallicity profile. The lower residuals within R$_\mathrm{bar}$ suggest that the piecewise fit is a better descriptor for the metallicity gradient, especially within R$_\mathrm{bar}$.
	We measure a significantly steeper metallicity gradient ($-$0.15 dex/kpc) inside the bar region with a break at 1.58~kpc with a shallower metallicity gradient beyond the break of $-$0.016~dex/kpc (middle column of Figure~\ref{fig:Ograd2}). To our knowledge, this is the first reported metallicity gradient for this galaxy.

	We measure a global metallicity gradient in NGC 2835 of $-$0.0431 $\pm$ 0.0009 dex/kpc from a single component fit (left column of Fig~\ref{fig:Ograd3}).
	The piecewise fit shows a steep-shallow metallicity profile in NGC~2835 with a residual distribution centered around zero throughout the galaxy (middle bottom panel of Fig~\ref{fig:Ograd3}).
	We find an outer flattening at 6.78 kpc in the double-linear fitting, with a steep metallicity gradient inside the break of $-$0.071~dex/kpc and $-$0.004~dex/kpc beyond the break radius at 6.78~kpc. This break radius is significantly larger than the bar boundary of 1.33~kpc \citep{Elmegreen_1996}.
	This suggests that the observed break in the metallicity gradient is highly unlikely to be bar-driven and instead quite possibly the result of inflows driven by possible past interactions, which is hinted by the faint extended disc observed in ESO Digitized Sky Survey \footnote{\href{https://archive.eso.org/dss/dss}{https://archive.eso.org/dss/dss}}. The observed HI distribution from Very Large Array shows a disturbed profile that is significantly deviated from the centre of NGC~2835 \citep[][]{Condon_1987}, which supports the occurrence of prior interactions.

	In NGC~5068, we measure a negative metallicity gradient of $-$0.073 $\pm$ 0.002 dex/kpc (left column of Figure~\ref{fig:Ograd3}). NGC~5068 is marginally better described with a shallow-steep piecewise-linear function (middle column of Figure~\ref{fig:Ograd3}), as the medians of residuals are closer to zero with a piecewise linear fit than a single linear fit, for the regions within R$_\mathrm{bar}$.
	We find an inner flattening ($-$0.005 dex/kpc) of the metallicity gradient with a break occurring at 0.82~kpc which lies near the end of bar (0.55~kpc). We measure a metallicity gradient beyond the break radius of $-$0.077 dex/kpc.

	The broken metallicity gradients of barred galaxies may be attributed to the effect of bars, together with spiral arms, on mixing ISM and radial migration. Prior studies have revealed that bars play an important role in radial migration as bars are an efficient mechanism that can create large scale radial gas flows \citep{Athanassoula_2003}. The bar-driven gas-flows would mix the gas and flatten the metallicity gradient inside the bar region \citep{Sanchez_2011, Seidel_2016}. A bar may also induce large scale gas inflows, triggering starbursts within the centre of the galaxy, producing  steep-shallow metallicity gradient. As time progresses, enriched gas from outward flows penetrates the outer disk and the bar mixes the ISM inside the bar region. As a result, the higher metal content in the central bar region of the galaxy is diluted. 
	
	Another potential influence of bar-driven inflow is suppressing the gas consumption within the central region of barred galaxies \citep{Sheth_2005,Geron_2021}. The bar-driven inflows expedite the quenching process \citep{Spinoso_2017, James_2018, Newnham_2020}, which would result in an observable  shallow-steep metallicity profile truncated near the end of a bar. A reliable explanation is that the gas can become too dynamically hot for star formation through an increase in velocity dispersion and/or shear, which would quench star formation \citep{Zurita_2004}. This would result in a  shallow-steep metallicity profile truncated near the central bar region. The mixing processes of ISM combined with the depression of star formation result in a flattened metallicity gradient within the bar region.
	
	The theories above may explain the inability for a single linear fit to explain the inner and disk regions of barred galaxies. The observed deviations from a linear trend may signify the existence of a perturbing event, such as a recent inflow of metal-poor gas from merger events \citep{Mihos_1996,Bustamante_2018} or a galactic-scale gas mixing, which potentially redeposits metal-rich gas \citep{Belfiore_2015,Belfiore_2017} especially within the inner central bar region.
	
	We note that the high spatial resolution of TYPHOON allows the individual star-forming regions to be separated from surrounding DIG (see Sec~\ref{sec:obs}). For surveys having a lower spatial resolution, such as MaNGA and SAMI with $\sim$ 1 kpc/pixel resolution, the broken metallicity gradient analysis might not be reliable for those survey data.

	\begin{table*}
		\centering
		\scriptsize
		\begin{tabular}{|c|c|c|c|c|c|c|c|c|c|c|}
			\hline
			\multirow{2}*{Galaxy} & \multicolumn{2}{|c|}{Global Gradient} && Break Radius && \multicolumn{2}{|c|}{Inner Gradient} & \multicolumn{2}{|c|}{Outer Gradient} & $\Delta(\mathrm{metallicity})$\\
			\cmidrule(r){2-3} \cmidrule(r){7-8} \cmidrule(r){9-10}
			~ & (dex kpc$^{-1}$) & (dex R$_{25} ^{-1}$) && kpc && (dex kpc$^{-1}$) & (dex R$_{25} ^{-1}$) & (dex kpc$^{-1}$) & (dex R$_{25} ^{-1}$) & (dex kpc$^{-1}$) \\
			\hline
			NGC 1365 &  -0.0162 (0.0005) & -0.479 (0.015) && 5.84 (0.79) && -0.055 (0.008) & -1.625 (0.236) &  -0.0123 (0.0005) & -0.363 (0.015) & -0.0427\\
			NGC 1566 & -0.0254 (0.0006) & -0.550 (0.013) && 7.79 (0.59) && -0.013 (0.015) & -0.282 (0.325) & -0.0297 (0.0009) & -0.643 (0.020)  & 0.0167\\
			NGC 1744 & -0.021 (0.002) & -0.246 (0.023) && 1.58 (0.20) && -0.15 (0.03) & -1.75 (0.35) & -0.016 (0.002) & -0.187 (0.023) & -0.134\\
			NGC 2835 & -0.0431 (0.0009) & -0.429 (0.009) && 6.78 (0.14) && -0.071 (0.002) & -0.71 (0.02) & -0.004 (0.002) & 0.04 (0.02) & -0.067\\
			NGC 5068 & -0.073 (0.002) & -0.397 (0.011) && 0.82 (0.23) && -0.005 (0.040) & -0.027 (0.217) & -0.077 (0.002) & -0.418 (0.011) & 0.072\\
			\hline
		\end{tabular}
		\caption{Oxygen abundance gradient of global (single-linear fit; left column of Figure~\ref{fig:Ograd}), the inner and outer region of each galaxy (double-linear fit; middle column of Figure~\ref{fig:Ograd}) accompanying 1$\sigma$ uncertainties. The third column lists the break radius determined from double-linear least-square fitting and the associated 1$\sigma$ uncertainty (i.e. 1 standard deviation error). The final column lists the difference between inner and outer best-fitted metallicity gradient (see~Sec.\ref{sec:sfr}). }
		\label{tab:gradient}
	\end{table*}

	\section{Discussion}\label{sec:discussion}
	Our galaxy samples present two types of metallicity gradients including shallow-steep profiles (NGC~1566 and NGC~5068) and steep-shallow profiles (NGC~1365, NGC~1744 and NGC~2835). In this section, we will discuss possible mechanisms that give rise to the observed variety of metallicity gradient profiles.

	\subsection{Comparison with previous work}\label{sec:prework}
	\citet{Sanchez_2018} report a shallow-steep metallicity in spiral galaxies with a typical truncation near 0.5 R$_e$ (vertical dotted lines in Figure~\ref{fig:Ograd}, Figure~\ref{fig:Ograd2} and Figure~\ref{fig:Ograd3}). They also reported the break radius of steep-shallow metallicity is near $\sim$ 1.5 $R_e$ (vertical dotted lines in Figure~\ref{fig:Ograd}) but the location of the steep-shallow break in the metallicity profile varies over a wide range of radial distance in their study.
	
	We find that the shallow-steep transition of metallicity indeed does occur near 0.5 R$_e$ for a single system NGC~5068, consistent with the findings of \citet{Sanchez_2018}. However, we do not find this signature in all galaxies; the break radius of shallow-steep (7.79 kpc) of NGC~1566 is located at over twice of distance of 0.5 R$_e$ ($\sim$ 3 kpc) as typically found by \citet{Sanchez_2018}. None of the galaxies in this study display a steep-shallow metallicity occurring at 1.5 R$_e$; this is consistent with the wide range of steep-shallow truncated position as reported in \citet{Sanchez_2018}. We do caution that the \citet{Sanchez_2018} dataset does not focus on barred galaxies as we do in this study. We additionally note that the \cite{Sanchez_2008} sample uses observations from MUSE, with a lower median spatial resolution of $\sim$460~pc, substantially larger than the typical size of HII regions\citep[10-200pc;][]{Azimlu_2011}. We remind the readers that the inability to resolve sub-HII region scales may influence the measured metallicity values and gradients due to the contamination from DIG emission \citep{ZhangK_2017,Poetrodjojo_2019}.

	\subsection{Star Formation Rate}\label{sec:sfr}
	Previous works \citep[][]{Huang_1996,Sakamoto_1999,Ellison_2011,Catalan-Torrecilla_2017} report higher SFR in the bar regions of spiral galaxies due to a bar serving as a highway, fueling cool gas to the centre of a galaxy. The resulting reinvigorated star formation within a barred will enhance the metallicity within the centre of a galaxy with respect to the region outside the bar. However, gas-phase metallicity is a complex balance between in-situ star formation, accretion/outflows, and ISM mixing \citep[][]{Sharda_2021,Wang_2022}. Additionally, the consumption of accreted metal-poor gas leads to a reduction of SFR and a higher gas-phase metallicity \citep[][]{Sanchez-Almeida_2019}. Thus a comprehensive understanding of how a galactic bar impacts the SFR process and gas-phase metallicity requires a combined analysis of the SFR and metallicity inside and beyond the bar region.
	
	To further investigate how galactic bars impact the gas-phase metallicity distribution, we derive the star formation rate surface density radial profiles $\Sigma_{\mathrm{SFR}}$ for each galaxy (right column of Figure~\ref{fig:Ograd}). 
	We use the {\sc IRAF-ellipse} package to calculate the H$\alpha$ intensity map and convert the H$\alpha$ luminosity to a SFR \citep{Kennicutt_1998}. Geometric parameters for \textit{IRAF} are the same as reported in Tab~\ref{tab:info}. The vertical pink lines in H$\alpha$ are the break radius of the metallicity profile (Section~\ref{sec:doubleline}). 
	
	$\Sigma_{\mathrm{SFR}}$ is higher and has a steeper gradient within the break radius than the outer region for NGC~1365, NGC~1744 and NGC~5068.
	This supports the stimulating effect of bars on star formation. When material mixes and stimulates radial migrations, the bar is able to fuel the center of galaxies and increase the SFR in the inner region. As a result, this gives rise to an enhanced metallicity within the inner region of NGC~1365 and NGC~1744. 
	
	In the interaction scenario for bar formation, NGC~1365 develops a gas-rich bar. With angular momentum redistribution via the spiral arms, the bar can induce large-scale radial inflows of interstellar gas with intense star formation occurring along the bar and spiral arms. A starburst is triggered in the nuclear ring as gas condenses and collapses in the centre. The vigorous star formation in the bar brings an enrichment of oxygen abundance inside the bar region. With the fuel progressively consumed, the star formation is limited to the ring and the spiral arms. As suggested by the limited population of HII regions in the bar \citep{Ho_2017}, the star formation rate is weak near the end of the bar but intense in the galaxy centre. The increasing SFR gradient inside this barred region supports that the bar of NGC~1365 facilitates the star formation in the center (Sec.~\ref{sec:sfr}).
	
	Similar to NGC~1365, the steep-shallow metallicity gradient for NGC~1744 may also result from enhanced star formation within the bar. \citet{Ryder_1993} report an enhancement in the HII regions across the bar and into the northern/western arms. Based on the observations of these bright HII regions, star formation activity is still actively occurring within the mentioned regions, driving the observed rise in the metallicity inside the bar region.

	The enhanced $\Sigma_{\mathrm{SFR}}$ within the break radius of NGC~1566 is consistent with the observed bright HII regions along the star-forming ring at 1.7~kpc \citep{smajic_2015}. 
	Well-defined spiral arms are clearly observed driving the gas inflows into the inner Lindblad resonance (ILR) of the nuclear bar \citep{Bosma_1992, Combes_2014}. 
	\citet{Aguero_2004} noted that some emission regions in the arms ($\sim$ 6.5kpc) have brighter H$\alpha$ luminosity than in the nucleus region. 
	Based on the low stellar density inside the bar region of NGC~1566 \citep{Gouliermis_2017} and the young age of the stellar population in the bar \citep[1 $\sim$ 10 Myr;][]{Shabani_2018}, we hypothesise that NGC~1566 is at the early stage of forming a central bar. Indeed, most stars and star clusters in NGC~1566 are at the age of 1 Myr $\sim$ 1 Gyr \citep{Grasha_2017,  Gouliermis_2017}. This is indicative of a bar younger than 1~Gyr in NGC~1566.  
	Additionally, the spiral arms are driving gas inflows, fueling the nuclear bar. This is evident by the brighter H$\alpha$ regions in the arms at $\sim$ 6.5~kpc\citep{Aguero_2004} than in the nucleus indicates that star formation is triggered in spiral arms. This scenario is consistent with the up-lifted metallicity around $\sim$ 6.5~kpc and the break location at 7.79~kpc. This results in a shallow-steep metallicity profile.
	 The discussion on the association of metallicity profile and $\Sigma_{\mathrm{SFR}}$ profile of NGC~1566 can be further studies by comparing multi-wavelength works to investigate the evolution of the central bar.

	For NGC~2835, $\Sigma_{\mathrm{SFR}}$ remains elevated out to $\sim$ 8 kpc, far beyond the end of the bar and approaching the edge of the observations. This supports the scenario for gas accretion occurring in the outskirt of NGC~2835. The steep-shallow metallicity gradient truncated at large galactocentric radii may be attributed to past interactions in the outskirt of a galaxy \citep{Werk_2010, Werk_2011}. In merging events, gas accretion from the surrounding intergalactic medium mixes metals in the outskirt, resulting in an observed outer flattening in the metallicity gradient \citep{Rupke_2010a, Kewley_2010}. \citet{Collacchioni_2020} finds higher gas accretion rates can result in more negative metallicity gradients within 1 R$_e$. The nearly flat metallicity gradient ($-$0.004 dex/kpc) at large radius ($\sim$0.68 R$_{25}$, i.e. 6.78~kpc) of NGC~2835 and lower $\Sigma_{\mathrm{SFR}}$ outside $\sim$ 8 kpc is consistent with the expectation of merging and accretion scenarios.

	NGC~5068 has a steeply increasing $\Sigma_{\mathrm{SFR}}$ from the bar to the center of the galaxy. An inner steep metallicity gradient is expected from the $\Sigma_{\mathrm{SFR}}$ profile. However, it has an inner flattening in metallicity gradient. Combining our observations hydrodynamical simulations may provide more constraints on the effect of the bar within NGC~5068 \citep[see, e.g.,][]{Sanchez_2008, Bournaud_2010, Sormani_2015, Khoperskov_2018}.

	\section{Conclusions}\label{sec:conclu}
	We investigate the resolved metallicity gradients in five nearby barred galaxies using the TYPHOON IFU survey. The high spatial resolution of TYPHOON provides the ability to explore the gas-phase metallicity gradient on an individual spaxel basis, uncontaminated by DIG emission. The large field of view (18' on one side) allows a measurement of the metallicity gradient out to the outer regions of the star-forming disk (up to 26~kpc). We adopt the N2S2 \citep{Dopita_2016} diagnostic to measure the gas-phase metallicity and constrain the radial distribution of the metallicity.
	The radial gradient of each galaxy is fitted by a single and double linear least-square fitting routine to constrain a break in the metallicity gradient, if present. The relative differences between the metallicity gradients and the breakpoints for the multiple component fits enable us to better investigate the role of the bar in radial migration and material mixing. Our data shows two scenarios on how the bar potentially can impact the metallicity gradients:
	
	\begin{enumerate} 
		\item Metallicity gradients described with an inner steeper and outer shallow gradient. NGC~1365, NGC~2835, and NGC~1744 are best-characterised with a ``steep-shallow'' metallicity gradient.
		\item Metallicity gradients described with an inner shallower and outer steep gradient. NGC~1566 and NGC~5068 are best-characterised  with a ``shallow-steep'' metallicity gradient. 
	\end{enumerate}
	
	Using the H$\alpha$ map, we measure the star formation rate surface density $\Sigma_{\mathrm{SFR}}$ to investigate the effect of galactic bars on star formation and the resulting impact on the gas-phase metallicity distribution. Bars can stimulate star formation activities, indicated by inner steep $\Sigma_{\mathrm{SFR}}$ profiles. This scenario is observed in NGC~1365 and NGC~1744, both of which show "steep-shallow" metallicity gradients. Spiral arms might also play a role in driving gas inflows together with a bar. As a result, bars can also flatten metallicity gradients within the bar region. The flatting within the inner region is seen in the metallicity distribution of NGC~1566. The "steep-shallow" $\Sigma_{\mathrm{SFR}}$ of NGC~5068 is consistent with the central star-forming ring \citep{smajic_2015} while the metallicity of NGC~5068 is observed as "shallow-steep" which indicates a young bar mixing ISM in the galaxy center.
	Our work suggests that SFR radial profiles are a powerful tool when studying the effect of bars on metallicity.
		
	For NGC~2835, the break far beyond the end of the bar in the metallicity gradient is not bar-driven but possibly the consequence of gas accretion to the outskirt of the galaxy \citep{Kewley_2010,Lopez_2015}. The break in the metallicity profile of NGC~2835 is at a distance that is $\sim$4 times greater than the R$_\mathrm{bar}$. This result suggests that the exact impact of galactic features on metallicity gradients are complex and can result from numerous scenarios, influenced by not only bars but also from mixing and in/outflows from intergalactic medium. Further study with larger samples is expected to provide more constraints on this issue.

	The diverse gradients in our barred galaxies indicate that there cannot be one common mechanism of the impact of a bar on the metallicity gradient. Mixing as well as outflows likely play an important role that dominate in some cases. Comparisons with cosmological hydrodynamical simulations are the next step necessary to help explain the diversity in the metallicity gradients, and the physical drivers that control the break locations that are observed in metal distribution of the ISM.

	\section*{Acknowledgements}
	We would like to thank the anonymous referee for their constructive feedback that helped improve the quality of this paper.
	
	This paper includes data obtained with the du Pont Telescope at the Las Campanas Observatory, Chile, as part of the TYPHOON programme, which has been obtaining optical data cubes for the largest angular-sized galaxies in the Southern hemisphere. We thank past and present Directors of The Observatories and the numerous time assignment committees for their generous and unfailing support of this long-term programme.
	
	This work is supported by the Strategic Priority Research Program of Chinese Academy of Sciences (No. XDB 41000000), the National Key R\&D Program of China (2017YFA0402600, 2017YFA0402702), the NSFC grant (Nos. 11973038 and 11973039), and the Chinese Space Station Telescope (CSST) Project. 
	
	This research has made use of NASA's Astrophysics Data System Bibliographic Services (ADS). 
	This research made use of Astropy,\footnote{\href{http://www.astropy.org}{http://www.astropy.org}} a community-developed core Python package for Astronomy \citep{astropy13, astropy18}. 
	This research has made use of the NASA/IPAC Extragalactic Database (NED) which is operated by the Jet Propulsion Laboratory, California Institute of Technology, under contract with NASA.
	
	KG is supported by the Australian Research Council through the Discovery Early Career Researcher Award (DECRA) Fellowship DE220100766 funded by the Australian Government. 
	KG is supported by the Australian Research Council Centre of Excellence for All Sky Astrophysics in 3 Dimensions (ASTRO~3D), through project number CE170100013. 
	KG acknowledges support from Lisa Kewley's ARC Laureate Fellowship (FL150100113). 
	
	We thank Hong-Xin Zhang and Guangwen Chen for helpful comments on the manuscript.

	\section*{Data Availability}
	The TYPHOON team is planning for a public data release on \href{https://datacentral.org.au/}{Data Central} in 2023.
	The TYPHOON data and the code in this article will be shared on reasonable request to the corresponding author.

	\bibliographystyle{mnras}
	\bibliography{ms_bargradients_mnras} 
	
	\appendix
    \section{White-light images with sample spectra}
\begin{figure*}
		\centering
		
		\raggedright
		\subfigure{%
			\includegraphics[width=0.63\textwidth,height=0.71\textwidth]{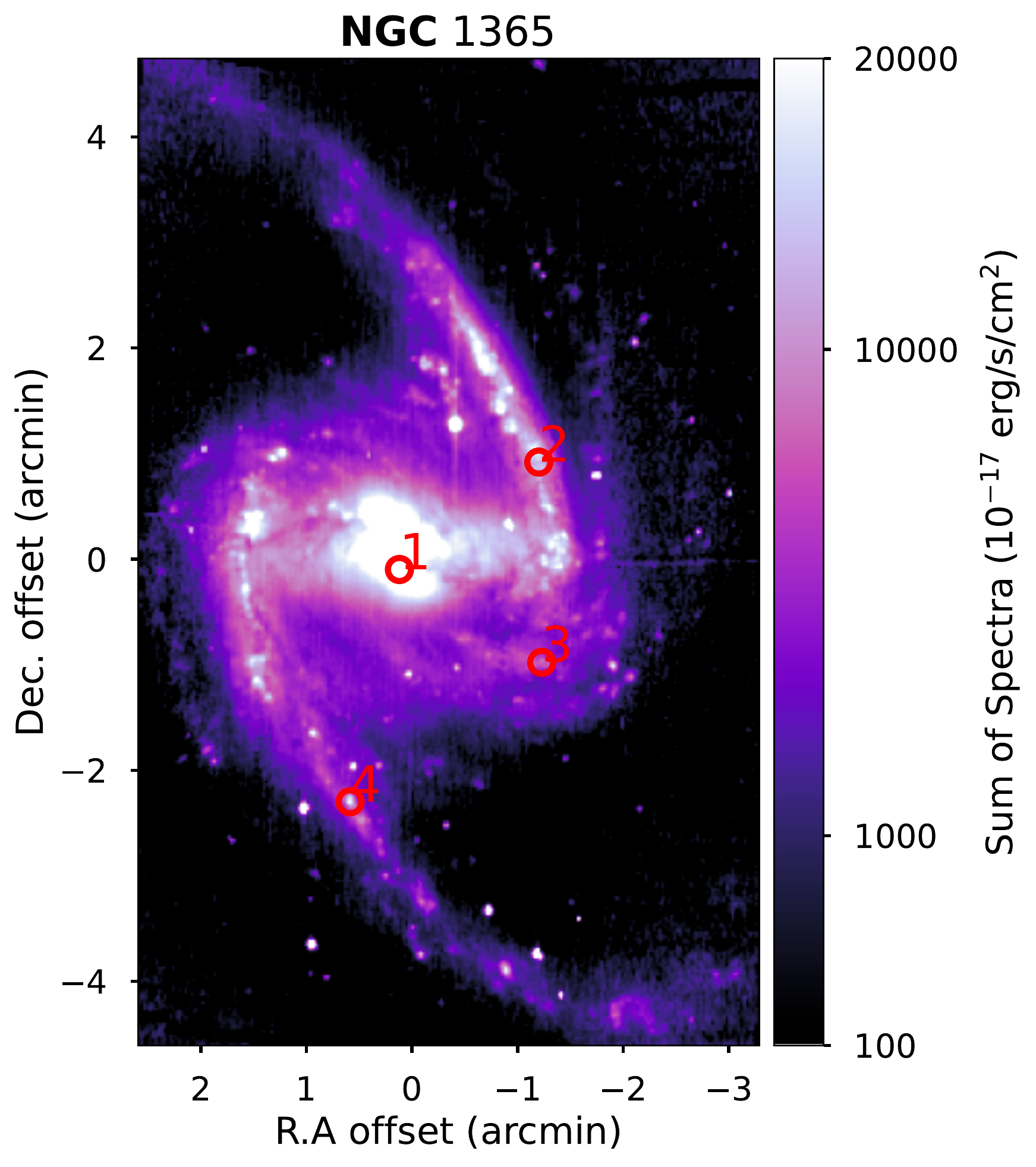}
		}
		
		\raggedleft
		\vspace{-13cm}
		\subfigure{%
			\includegraphics[width=0.35\textwidth,height=0.38\textwidth]{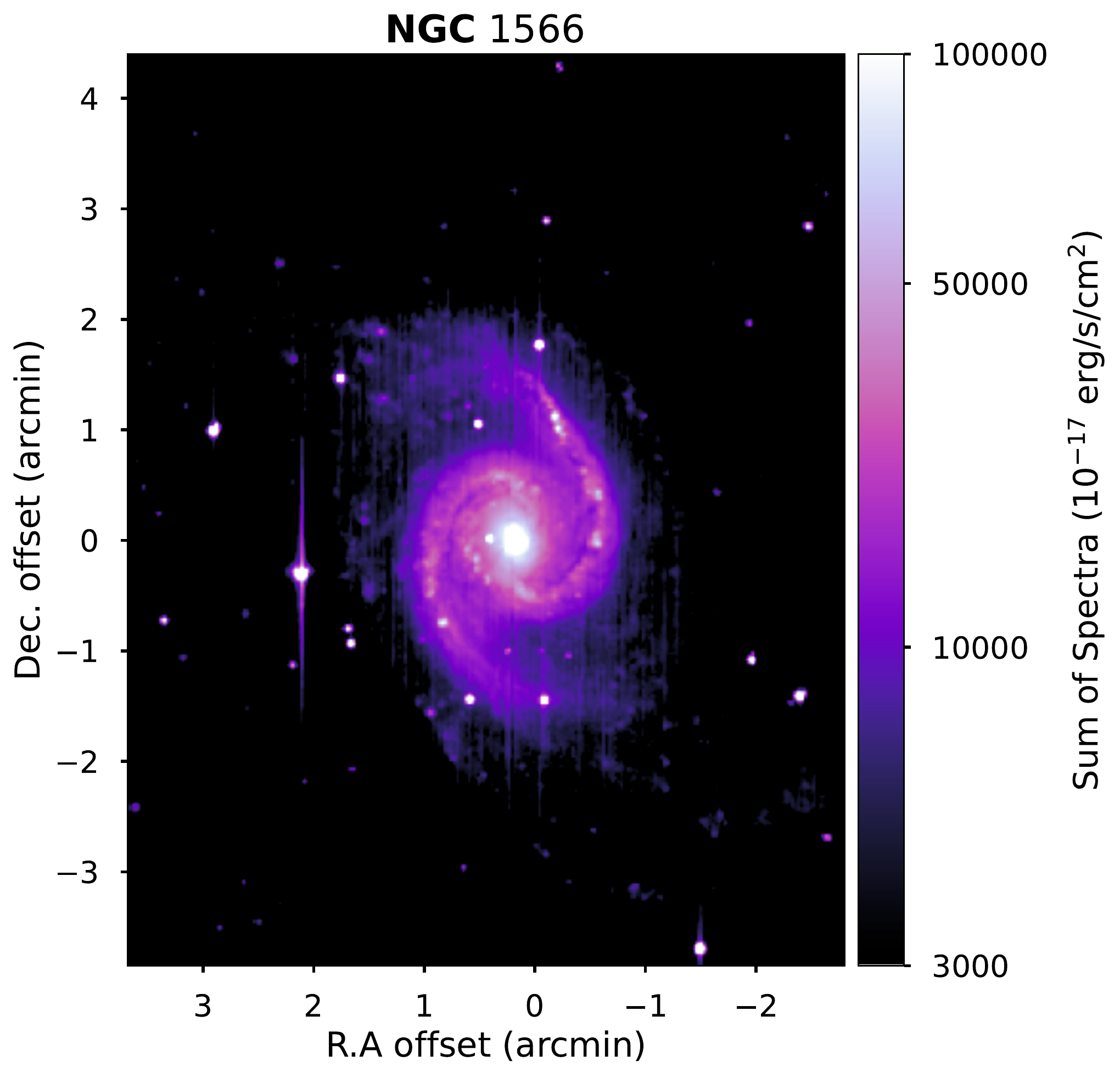}
		}
		
		\raggedleft
		\subfigure{%
			\includegraphics[width=0.35\textwidth,height=0.38\textwidth]{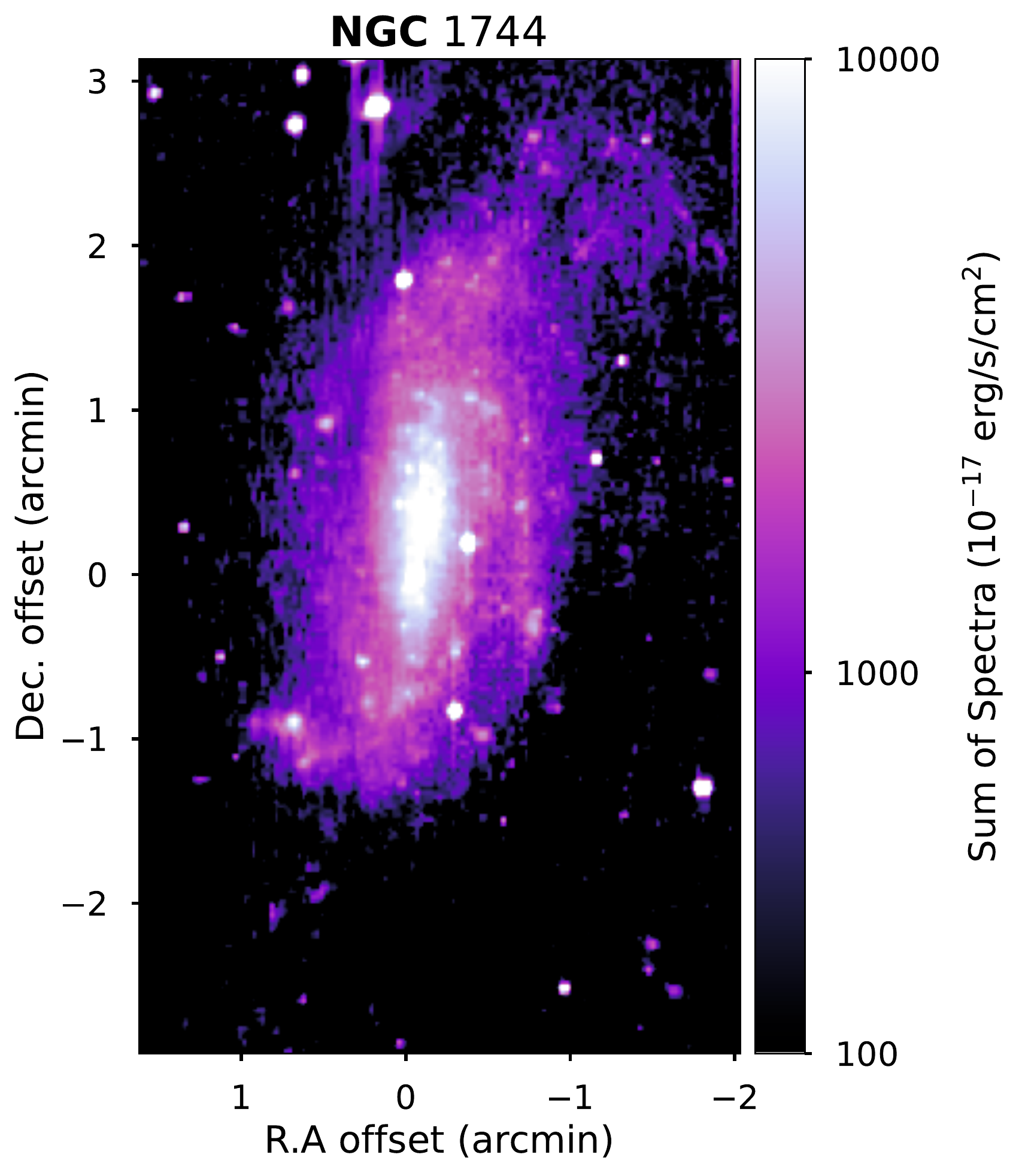}
		}
		
		\centering
		\subfigure{%
			\includegraphics[width=0.4\textwidth,height=0.44\textwidth]{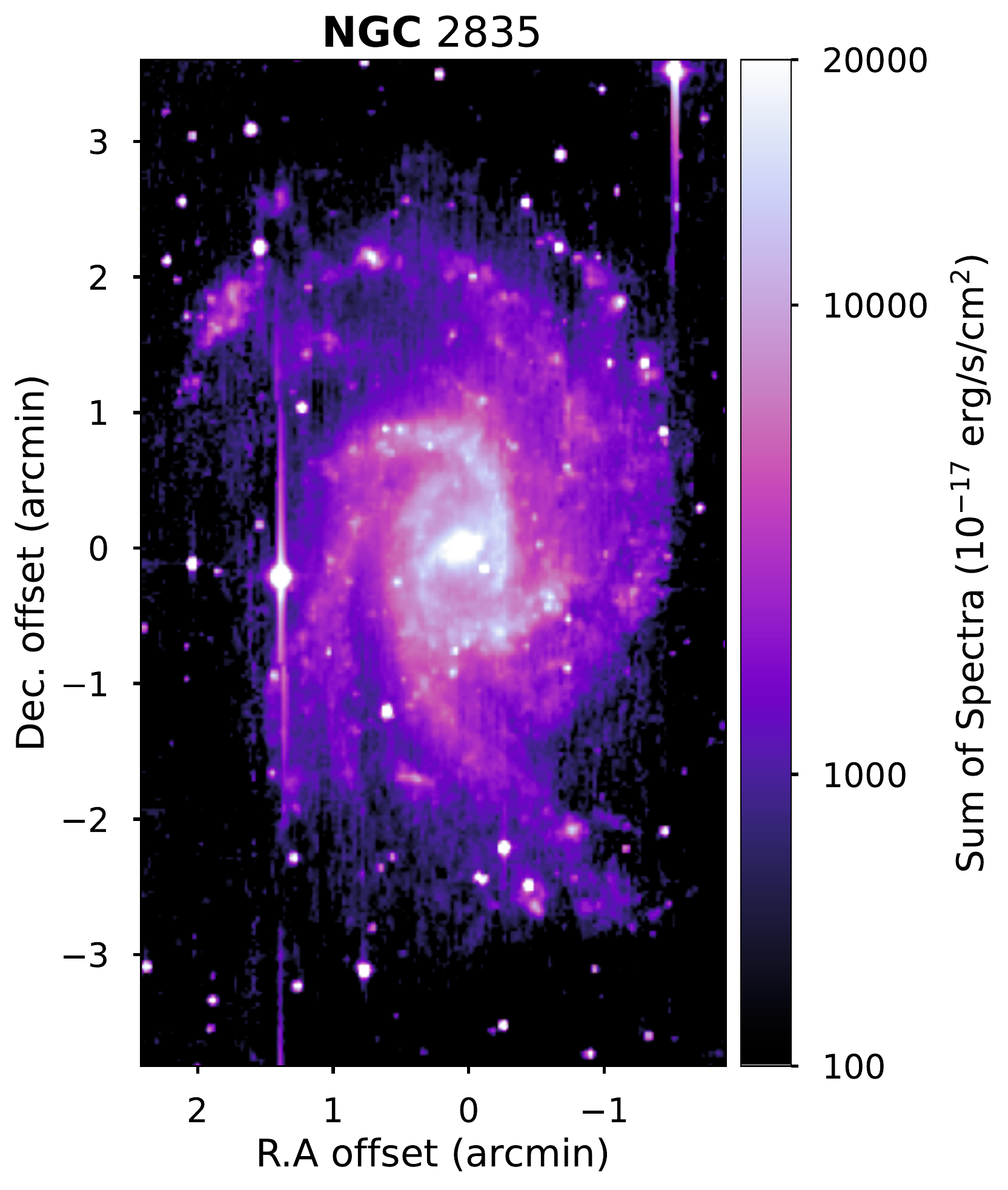}
		}
		\subfigure{%
			\includegraphics[width=0.4\textwidth,height=0.44\textwidth]{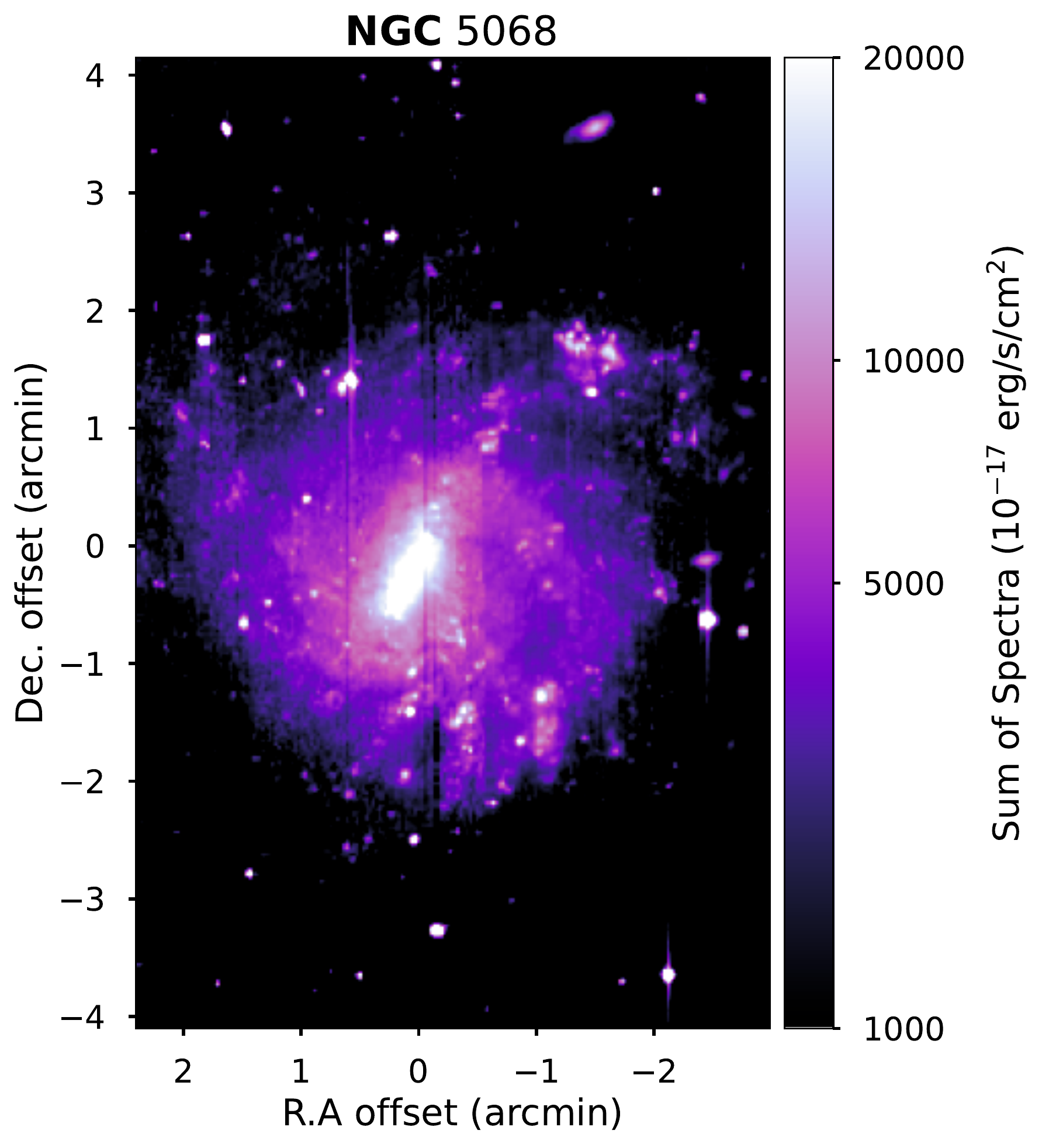}
		}
		\caption{White-light images.}
		\label{fig:whitelight}
	\end{figure*}

\begin{figure*}
	    \centering
	    \includegraphics[width=1\textwidth]{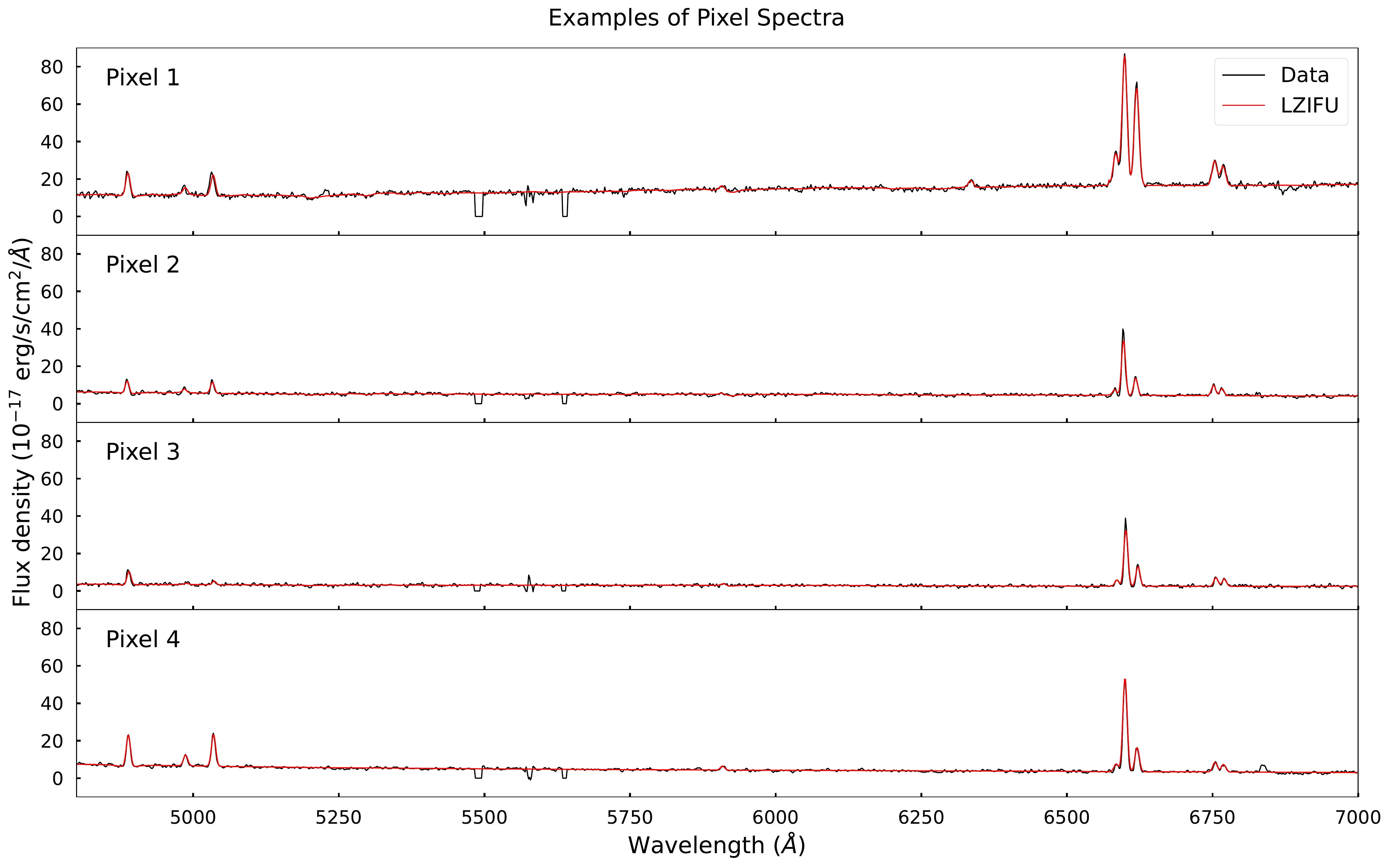}
	    \caption{Sample spectra from four different locations within NGC~1365, marked as red circles in Fig~\ref{fig:whitelight}. The black lines are the observed spectra while the red lines are the best fits from LZIFU (Sec~\ref{sec:data_reduction}).}
	    \label{fig:spaxel}
	\end{figure*}

    By summing the observed spectra along the wavelength, we obtain the white-light images as shown in Fig~\ref{fig:whitelight}. The combination of Fig~\ref{fig:ha} and Fig~\ref{fig:whitelight} demonstrates the data quality of TYPHOON. In Fig~\ref{fig:spaxel}, we present the spectra from four different locations within NGC~1365, marked as red circles in Fig~\ref{fig:whitelight}. The black lines are the observed spectra while the red lines are the best fits from LZIFU (Sec~\ref{sec:data_reduction}).
    
    {}
	
	\section{Definition of Length of Bars}\label{sec:barend}
\begin{figure*}
		\centering
		
		\subfigure{%
			\includegraphics[width=0.37\textwidth]{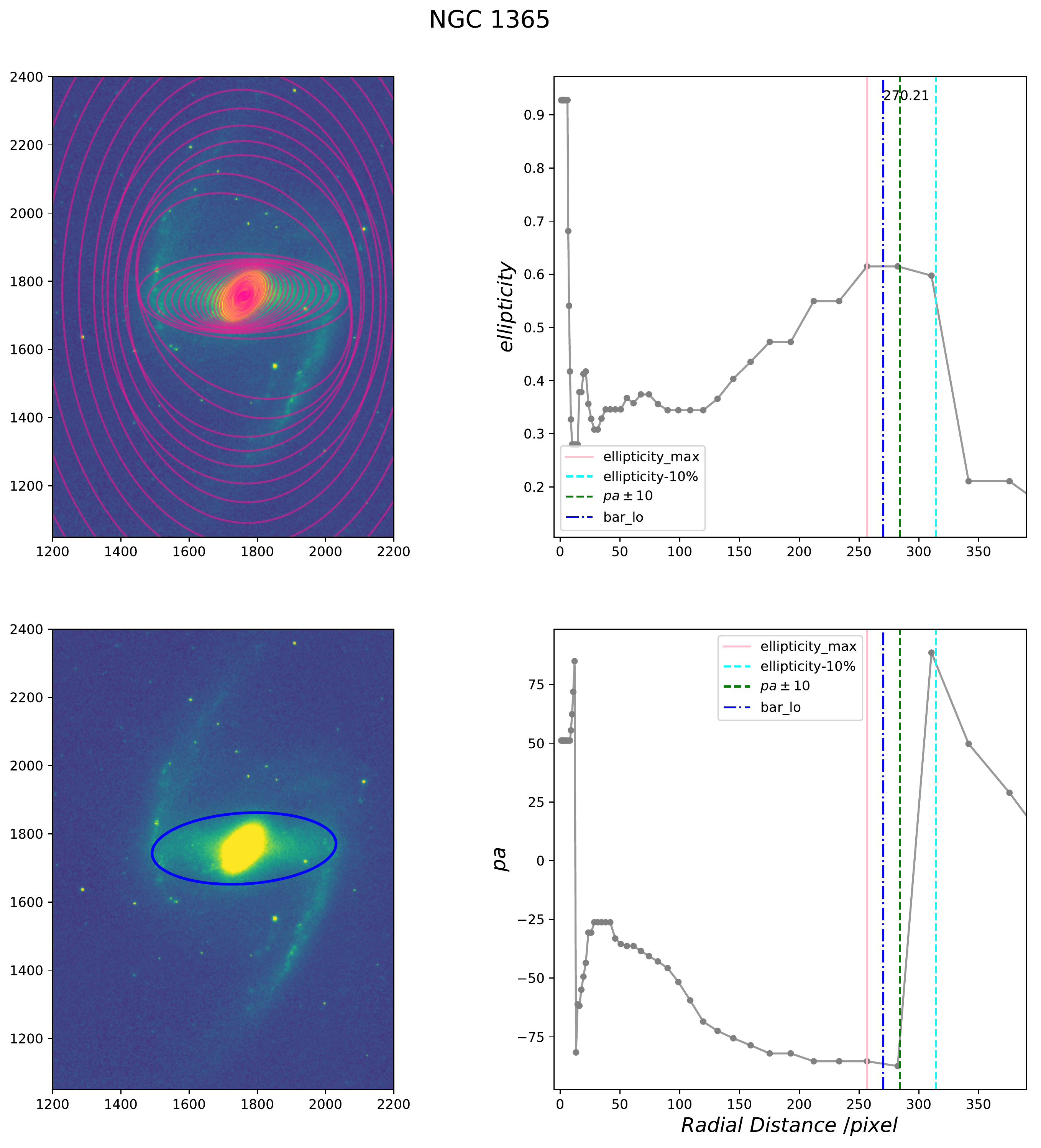}
		}
		\subfigure{%
			\includegraphics[width=0.37\textwidth]{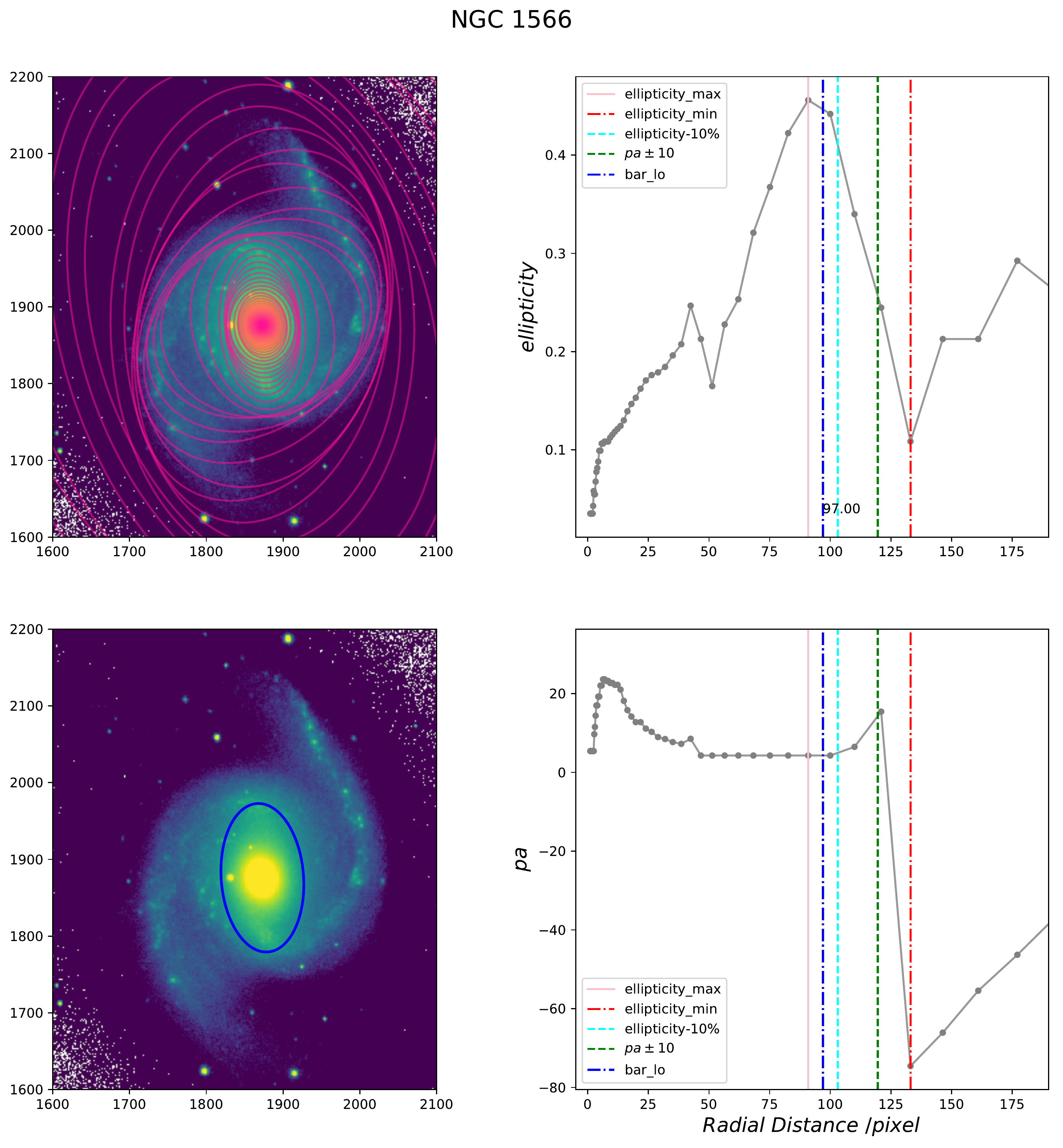}
		}
		\subfigure{%
			\includegraphics[width=0.37\textwidth]{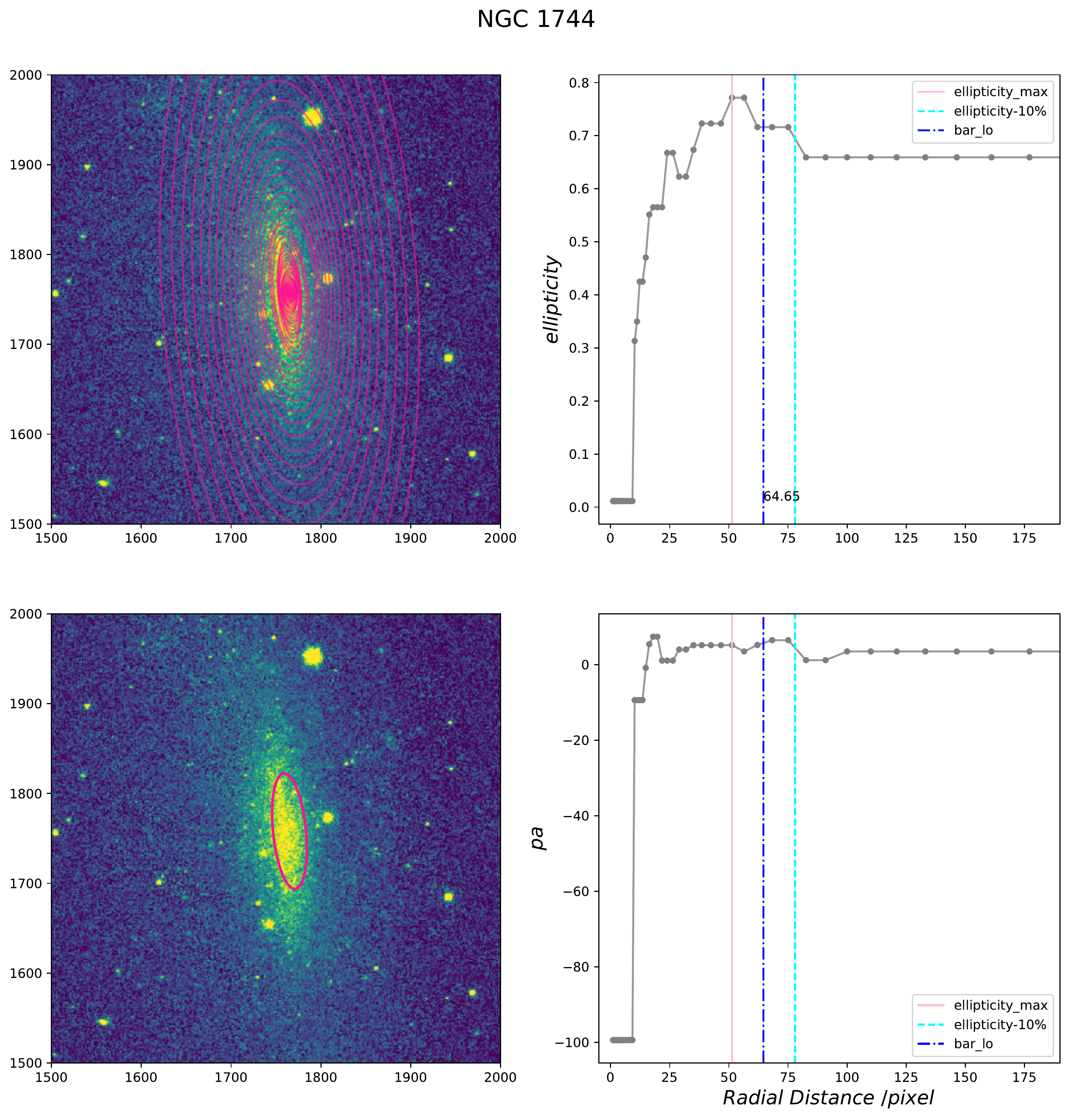}
		}
		\subfigure{%
			\includegraphics[width=0.37\textwidth]{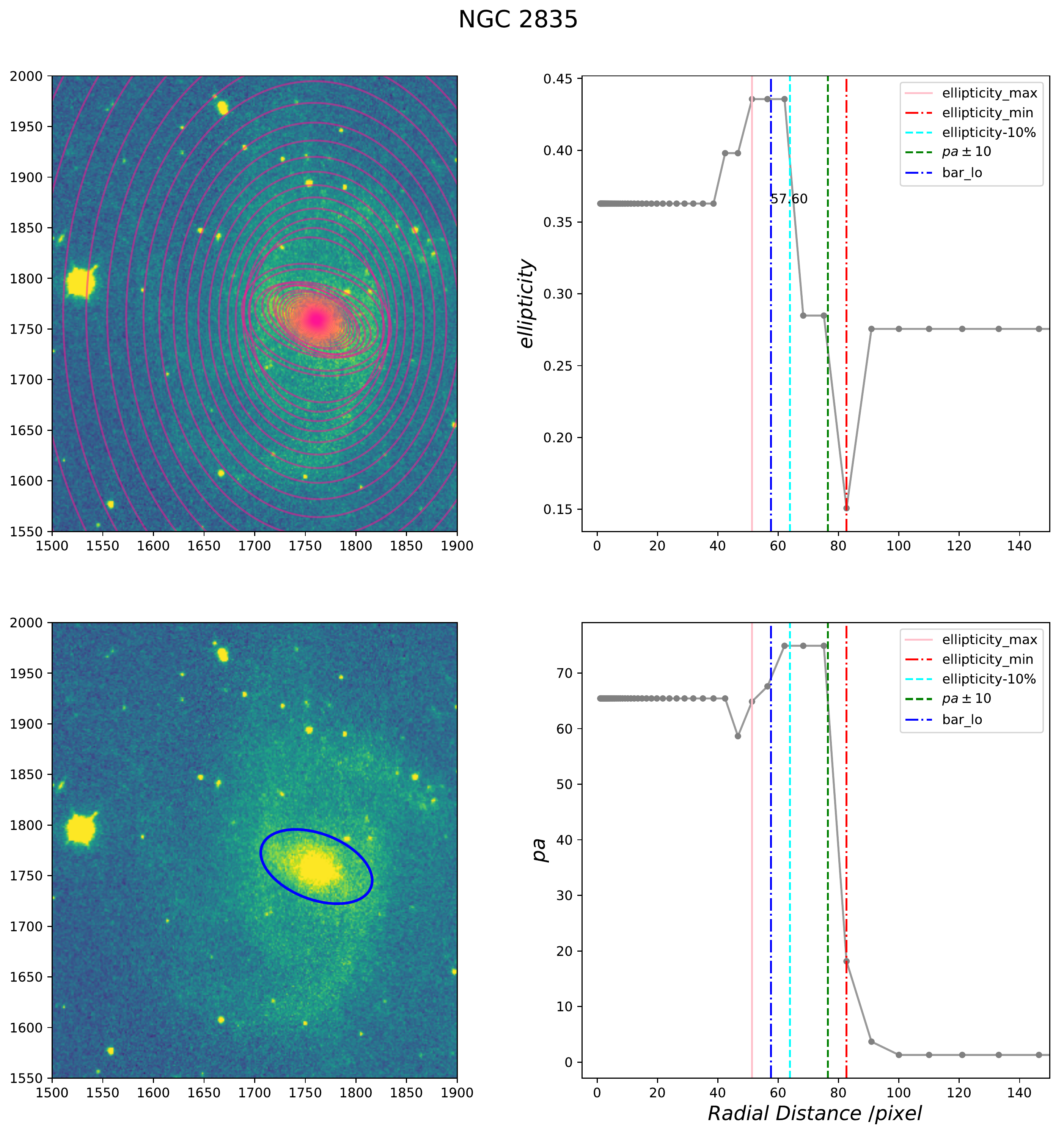}
		}
		\subfigure{%
			\includegraphics[width=0.37\textwidth]{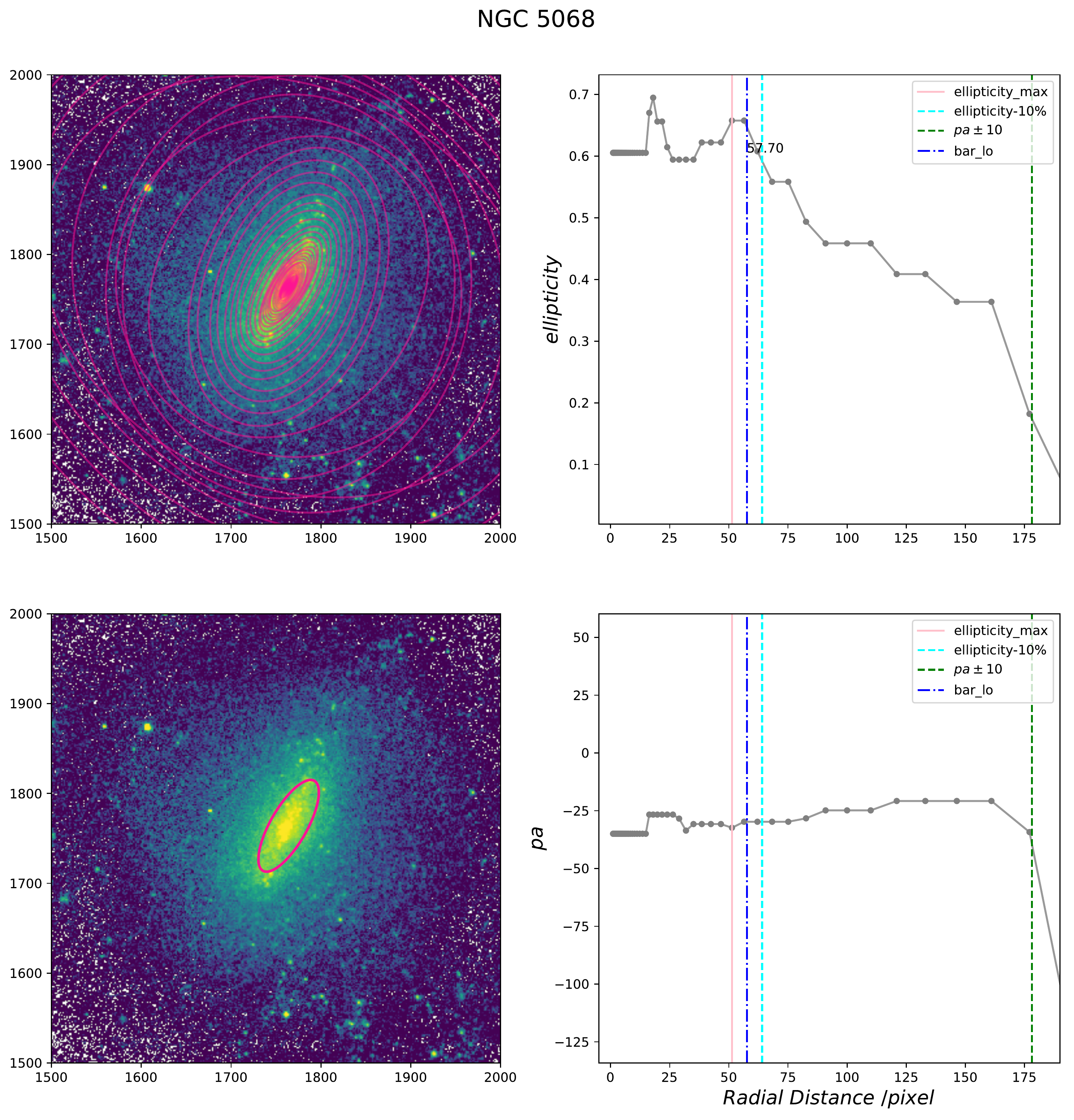}
		}
		\caption{Definition of bar length utilizing ellipticity and P.A. profiles. For each sample, the top-left panel shows the ellipse fitting result from IRAF, overlapping on the K-band image from VHS survey; the lower-left panel shows the defined bar radius; the ellipticity and P.A. radial profiles are presented in the right panels while the vertical lines are four photometric parameters as introduced above.}
		\label{fig:barlen}
	\end{figure*}

	To identify the length of bar in our sample, we follow the method in \citet{Mu_oz_Mateos_2013} based on the ellipticity and P.A. profiles. They define four parameters,
	\begin{enumerate}
	    \item $a_{\mathrm{\epsilon max}}$, the radius where the ellipticity of the bar is maximum.
	    \item $a_{\mathrm{\epsilon min}}$, the radius where there is a local minimum in ellipticity after the previous maximum.
	    \item $a_{\mathrm{\Delta\epsilon} = 0.1}$, the radius where the ellipticity drops by 0.1 with respect to the maximum one.
	    \item $a_{\mathrm{\Delta P.A.= 10^{\circ}}}$, the radius where the P.A. differs by 10\textdegree\ from the one at $a_{\mathrm{\epsilon max}}$.
	\end{enumerate}
	We adopted $a_{\mathrm{\epsilon max}}$ as a lower limit for the length of bar and the minimum of the other three parameters as an upper limit. The defined bar radius of each galaxy is the average of the lower and upper limit. In our work, we use K-band image data from VHS survey\footnote{http://www.eso.org/sci/observing/phase3/data\_releases/vhs\_dr1.html} as infrared images are good tracers to evaluate the gravitational potential and the bar strength \citep{Mu_oz_Mateos_2013}. We use {\sc IRAF-ellipse} to measure the ellipticity and P.A. radial profiles of galaxies, as shown in Fig~\ref{fig:barlen}. The blue ellipse overlapped on the infrared image in the lower left panel of each sample is our final definition of the bar region. The derived bar radius by photometric approach are in line with manual recognition. 
	
	{}
	
	To aid in comparing the length of each bar with the break radius of metallicity, we deproject the bar radius following the method in Section~\ref{sec:deprojection}. The position angles and inclination angles are the same as listed in Tab~\ref{tab:info}.
	
	\begin{table}
		\begin{tabular}{cc}
			\hline
			Galaxy name  & Deprojected bar length (kpc)\\
			\hline
			NGC~1365 & 9.10\\
			NGC~1566 & 3.83\\
			NGC~1744 & 1.07\\
			NGC~2835 & 1.70\\
			NGC~5068 & 0.55\\
			\hline
		\end{tabular}
		\caption{The length of bar after deprojection (Section~\ref{sec:deprojection}) for each galaxy.}
		\label{tab:barlen}
	\end{table}

    \section{Previous work on the metallicity gradients of our galaxies}\label{sec:individual_galaxy}
    In this section, we collect the metallicity gradients of our galaxies whose metallicity gradients have been reported in previous work.
	
	\subsection{NGC~1365}\label{sec:global}
	NGC~1365 is a strong barred galaxy in the Fornax cluster having a mixed ionising source with both star formation and a harder ionising source \citep[AGN and shocks;][]{Ho_2017} as is evident in the BPT diagram (Fig~\ref{fig:bpt}). The high star formation rate (SFR) is consistent with a harder ionising source and indicates a starburst in NGC~1365.
 
	\cite{Sanchez_2008} report the asymmetry in the dust distribution and the kinematic map of NGC~1365. They suggest the anomalous velocities inside the bar region may due to remnants of an infalling cloud in the disk of NGC~1365. However, it remains unclear whether the infalling gas is an isolated intergalactic cloud or a stream of debris from a gas-rich dwarf galaxy.
	
	\citet{Roy_1997} measure a global metallicity gradient of NGC 1365 as $-$0.02dex/kpc (R97) with a break at 0.55 R$_{25}$ using the R$_{23}$ diagnostic, corresponding to 16 kpc. This break radius in the outer disc of the galaxy and a flattening in the metallicity gradient is faintly visible at scales 15$\sim$20 kpc, albeit with significant scatter. In future work, we will investigate a three component fits to dissect the formation history of the TYPHOON galaxies from the presence of an inner and outer break to the metallicity gradients (Kewley et al., in prep). 
	
	\citet{Ho_2017} use the N2S2 diagnostic and measure a metallicity gradient of $-$0.0139 $\pm$ 0.0015 dex/kpc \citep[Appendix A,][]{Ho_2017} from the HII regions. Their reported value is similar to the single linear fit gradient we measure. The \cite{Ho_2017} data have limited HII regions in the central region of the galaxy at galactocentric distances within R $\leq$ 5 kpc, limiting the comparison of the reported gradients for the two studies within the central region where we find a strong steepening in the metallicity gradient. For better comparison with \citet{Ho_2017}, we perform a single linear fit to metallicity gradient outside the break at 5.84~kpc region with a slope of $-$0.0123 $\pm$ 0.0005 dex/kpc, consistent with the gradient as reported in \cite{Ho_2017}.
	
    \subsection{NGC~1566}
	NGC~1566 is a face-on nearby galaxy with grand design spiral arms and a weak-central bar. It is the brightest galaxy of the Dorado group. NGC 1566 hosts a low luminosity AGN in a low level of activity during observations, classified as Seyfert \citep{Vaucouleurs_1961}.
	
	\citet{Zarisky_1994} measured the metallicity gradient with the R$_{23}$ diagnostic, yielding $-$0.060 $\pm$ 0.080 dex/kpc, consistent with our metallicity gradient measurement within the errors. \citet{Sanchez_2018} adopted the O3N2 calibration \citep{PP_2004} and reported a metallicity gradient of 0.00 dex/R$_e$ with a scatter of 0.033 dex; consistent with our single fit metallicity gradient within the errors. We note that due to their limited field of view, their reported gradient only includes the central southern region of NGC~1566 (approximately 1' $\times$ 1') while the TYPHOON data covers the entire disc of the galaxy with a 18' $\times$ 6.765' field of view.
	
	\subsection{NGC~1744}
	NGC~1744 is a late-type barred galaxy with a higher inclination \citep[69.9\textdegree;][]{Makarov_2014} than the other galaxies in the study, thus it is difficult to trace the spiral arms. The HII regions are located across the bar as well as the northern and western spiral arms \citep{Ryder_1993}.

	\cite{Pisano_1998} report that NGC~1744 shows no evidence of recent interaction with an unperturbed velocity field. This indicates that the bar in NGC~1744 is not a newly-formed but long-lived structure. 
	
	\subsection{NGC~2835}
	NGC~2835 is a nearby spiral galaxy with an small-scale bar whose boundary is difficult to define given the diffuse, flocculent spiral arms.
	
	Using observations with MUSE on the Very Large Telescope (VLT), \citet{Kreckel_2019} reports a metallicity gradient of $-$0.096 $\pm$ 0.003 dex/kpc using the N2S2 metallicity diagnostic, a slope that is twice as steep compared to our measured gradient. The difference between their reported gradient and ours may be attributed to \citet{Kreckel_2019} only observing the center region of NGC 2835 (distances out to $\sim$ 7 kpc) compared to our observations which cover the disc out to 16~kpc. The location of the break radius in our measured metallicity profile of 6.78~kpc is nearly the same as the galactocentric distance covered by \citet{Kreckel_2019}. When we limit our spaxels to the same galactocentric distances, we measure a metallicity gradient of $-$0.071 $\pm$ 0.002~dex/kpc, more consistent to the reported gradient of \citet{Kreckel_2019}. We further note that the study of \citet{Kreckel_2019} is on the scale of entire HII regions while our work is a spaxel-by-spaxel analysis.
	
	\subsection{NGC~5068}
	NGC~5068 is a face-on barred spiral galaxy, known to host Wolf–Rayet (WR) stars. \citet{Rosa_1986} detected spectroscopic signatures of WR in bright HII regions, located throughout the bar region \citep{Bibby_2012}. 
	
	\citet{Ryder_1995} reports a metallicity gradient of $-$0.35 $\pm$ 0.26 dex/R$_{25}$ using the R$_{23}$ diagnostic \citep{Pagel+1981}. \citet{Bibby_2012} reports a metallicity gradient as $-$0.61 $\pm$ 0.22 dex/R$_{25}$ using N2 and O3N2 diagnostics. Assuming a distance of 5.16~Mpc (Table~\ref{tab:info}), we convert our metallicity gradients to $-$0.397 $\pm$ 0.011 dex/R$_{\mathrm{25}}$. Despite the different metallicity calibrations used, our derived single fit metallicity gradient is consistent with prior measurements from both Ryder and Bibby et al., within the errors.
	
	Based on the bright HII regions throughout the bar region, we hypothesise that the bar of NGC~5068 induces gas inflows, fueling the central region. The inflows mix ISM inside the bar region ($\sim$ 1~kpc) which result in the flattened metallicity gradient within the break radius (0.82~kpc). The nearly flat ($-$0.005 dex/kpc) inner metallicity gradient is strong evidence for efficient radial migration and material mixing driven by the bar.
	
\end{CJK} 	
\end{document}